%% file: v5.tex
\documentclass[useAMS,usenatbib]{mn2e}
\usepackage{graphicx}
\usepackage{amssymb}
\usepackage{natbib,times}

\title[Abnormal emission events of PSR B0919+06]
      {Jiamusi Pulsar Observations: I. Abnormal emission events of PSR B0919+06}
\author[Han et al. ]
       {Jun Han$^{1,4}$, J. L. Han$^{1}$ \thanks{E-mail: hjl@nao.cas.cn},
         Ling-Xiang Peng$^{2,3}$, De-Yu Tang$^{2}$, Jun Wang$^2$, \and Jun-Qiang Li$^2$,
         Chen Wang$^{1,3}$, Ye-Zhao Yu$^{1,4}$, Bin Dong$^{1,3}$ \\
         \\
         $^1$National Astronomical Observatories, Chinese Academy of Sciences,
         Jia-20 Datun Road, ChaoYang District, Beijing 100012, China\\
         $^2$Jiamusi Deep Space Station, China Xi'an Satellite Control Center,
         Jiamusi, Heilongjiang 154002, China\\
         $^3$The State Key Laboratory of Astronautic Dynamics, Xi'an,
         Shaanxi 710043, China\\
         $^4$School of Astronomy, University of Chinese Academy of Sciences, 
                Beijing 100049, China
       }

\begin{document}

\date{Accepted 2015 December 03. Received 2015 December 02; in original form 2015 October 30}

\pagerange{\pageref{firstpage}--\pageref{lastpage}}
\pubyear{2015}

\maketitle

\label{firstpage}

\begin{abstract}
  PSR B0919+06 generally radiates radio pulses in a normal phase
  range. It has been known for its occasional perplexing abnormal
  emission events wherein individual pulses come to an earlier phase
  range for a few tens of periods and then returns to its usual phase.
  Heretofore, only a few such events have been available for study. We
  observed PSR B0919+06 for about 30 hours using the Jiamusi 66-m
  telescope at Jiamusi Deep Space Station at S-band, and detected 92
  abnormal emission events. We identify four types of events based on
  the abrupted or gradual phase-shifting of individual pulses. The
  abnormal emission events are seen to occur randomly some every
  1000 to 3000 periods, and they affect the leading edge of the mean profile
  by up to 2\% in amplitude. The abnormal emission events are probably
  related to gradual changes of emission processing in the pulsar
  magnetosphere.
\end{abstract}

\begin{keywords}
pulsars: general -- pulsars: individual: PSR B0919+06.
\end{keywords}

\section{Introduction}

Pulsars radiate the pulsed emission periodically, though individual
pulses vary in shape and in intensity. Sub-pulses of individual pulses
can appear in different ranges of pulsar rotation phases. Various
phenomena such as nulling, subpulse drifting and mode-changing have
been observed from many pulsars \citep[see e.g.][]{syh+15}. When
individual pulses are averaged over many periods, the mean pulse
profile is very stable for a pulsar in general, which is believed to
be a cut of the emission window of pulsar magnetosphere \citep{man95},
and their polarization features are closely related to the emission
geometry \citep[e.g.][]{lm88,gan01,whw13}.  Sub-pulses of some pulsars
drift within the pulsar emission window \citep[e.g.][]{ran86,rankin+13}.
Some pulsars exhibit different emission modes in a sequence of
individual pulses. For example, the precursor of the main pulse and
the interpulse of PSR B1822$-$09 are relevantly switched on and off
for several hundreds of periods \citep{gil+94,lmr12}.
The extreme case for emission mode changes is the nulling phenomena,
in which case the emission process appears to be fully stopped for
many periods \citep[e.g.][]{wmj07,yhw14,syh+15}.

PSR B0919+06 is a nearby bright pulsar with a period of 0.43062~s and
a small dispersion measure of 27.27 pc~cm$^{-3}$ \citep{hhc+09}. Its
narrow mean pulse profile at high frequencies (e.g. above 1.4~GHz) is
highly polarized \citep{scr+84,jkmg08,hr10}, and is very asymmetric
with a long gradual leading edge and a sharp trailing edge. Such a
profile was classified as a partial cone by \citet{lm88}, and the
emission at high frequencies originates mainly from the trailing conal
component. At lower frequencies, two or three profile components
emerge \citep{pw92,sse+06,jkmg08,hr10}, and the central core and
leading conal components become stronger. Orthogonal polarization
modes have been observed mainly in the central weak component
\citep{scr+84,rrw06,sse+06}.
  
The abnormal emission events of PSR B0919+06 were first discovered by
\citet{rrw06}, who noticed five occurrences of a gradual shifting of
individual pulses about $5^{\circ}$ in longitude towards early pulse
phases from four pulse squences of 10763 pulses in total. From their
observations made on MJD 52854 at 1425~MHz, two abnormal emission
events are seen in the pulse series of 480~s (1115 pulses), each
lasting for 30-40 periods with a separation of some 650
periods. \citet{rrw06} observed PSR B0919+06 at 327~MHz for 30 minutes
(4180 pulses) on MJD 52916, and detected one event lasting for 60
periods. They also found two more abnormal emission events in archive
data of \citet{scr+84} of 3582 pulses observed on MJD 44857 at
1.4~GHz, but no event from 1886 pulses observed on MJD 44859. The
great advantage of the observations made by \citet{rrw06} is that
polarization data are recorded. The orthogonal polarization modes and
the integrated polarization profiles for the normal emission mode and
the abnormal emission events of PSR B0919+16 obtained from these
observations clearly show that the leading conal component and the
core component emerge during the abnormal emission events at 1425~MHz
and during the normal the abnormal modes at 327~MHz. The orthogonal
polarization modes were detected from not only the core component at
1425~MHz but also both the leading conal component and the core
component at 327~MHz. A very important conclusion derived from the
observation of \citet{rrw06} is that the polarization angles during
the abnormal emission events follow the same ``S''shaped-curve as the
trailing conal component (shown in their Figure~3) after the
orthogonal polarization modes are considered, suggesting that the
abnormal emission events occur as if the two profile components at
early pulse phases are lightened and the trailing conal component is
switched off.

Such abnormal emission events of PSR B0919+06 for a gradual shifting
of individual pulses were recently confirmed by \citet{psw+15} from
their single pulses observed at MJD 55751. In addition to such a
so-called ``main flare state'' of $5^{\circ}$ shifting \citet{psw+15}
found a ``small flare state'', characterized by a smaller gradual
shifting of about 3$^{\circ}$ in longitude of pulse phases for a few
seconds.

Very intriguing is the fact that the profiles of PSR B0919+06 have
also shown two states \citep{lyne10, psw+15} with the most distinctive
difference exactly in the range of pulse phases where abnormal
emission events stand out. Long term timing observations of PSR
B0919+06 \citep{lyne10, psw+15} show that its spin-down rate
$\dot{\nu}$ varies between ``low and high states'' with a double-peak
structure and a quasi-periodicity of 630 days or 550 days, which was
seen to be very closely correlated with variations between the two
profile states. \citet{psw+15} removed the identified ``flare states''
shown in one or two subintegrations (10~seconds or 1~minute each) from
their timing data of normally 20~minutes and found that the
correlation still exists between the slow $\dot{\nu}$ variations and
the profile-shape parameters. It is possible that variations of
profiles arised from either some un-identified ``flare states'' or the
leakages of flares to nearby subintegrations or ``small flares'' in
their timing data.

At the present, it is unclear how often and how long the abnormal
emission events occur in PSR B0919+06 and how much they contribute to
the profile variations. Here we report our observations of PSR
B0919+06 for about 30 hours by using the Jiamusi 66-m telescope at
Jiamusi Deep Space Station, in which we find 92 abnormal emission
events.
The observation system is briefly described in Sect.~2, the results
are presented and analyzed in Sect.~3, and discussions and conclusions
are given in Sect.~4.

\begin{figure}
  \centering
  \includegraphics[width=0.47\textwidth]{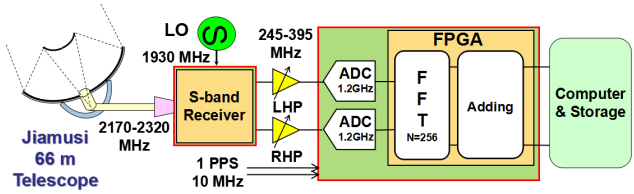}
  \caption{The system used for observations. See text
    for explanations.}
  \label{fig1}
\end{figure}

\begin{table}
\centering
\caption{Observational parameters for PSR B0919+06.}
\setlength{\tabcolsep}{1mm}
\begin{tabular}{cccccc}
\hline
Obs. Date &MJD  &Start Time  &Obs. length &Channel   &Sample \\ 
 UTC  &   &UTC   &min. / Pulse No.  & No.    &(ms)  \\
\hline
\hline
2015.04.18 &57130  &13:29   & 15.0 / 2088    &256  &0.19968 \\           
2015.07.12 &57215  &04:09   &381.0 / 53053   &256  &0.19968 \\           
2015.07.14 &57217  &01:07   &485.2 / 67611   &256  &0.19968 \\           
2015.08.17 &57251  &22:31   &466.5 / 65003   &128  &0.09984 \\           
2015.08.18 &57252  &22:15   &253.4 / 35306   &256  &0.19968 \\ 
2015.08.19 &57253  &03:20   &179.4 / 25001   &256  &0.19968 \\ 
\hline
\end{tabular}
\label{tab1}
\end{table}

\begin{figure}
  \centering
  \includegraphics[angle=-90,width=0.45\textwidth]{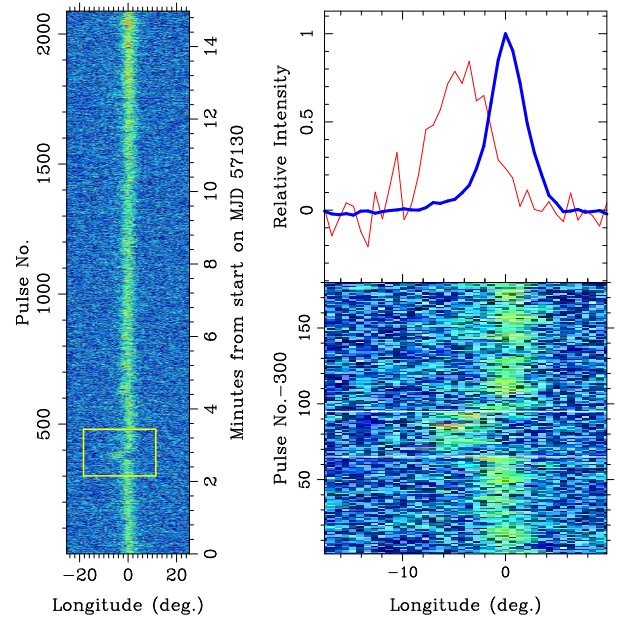}
  \caption{Results from a test observation of PSR B0919+06 on MJD
    57130. The left panel shows a plot of 2088 individual pulses
    observed in 15 minutes. An enlarged plot for 180 pulses, including
    an abnormal emission event between the pulses numbers 365 and 395,
    is shown in the right-bottom panel. The comparison of mean
    profiles for the normal emission mode (thick line) and abnormal
    emission event (thin line) are shown in the top-right panel.}
  \label{fig2}
\end{figure}

\input tab2.tex

\section{The observation system and data taking}

The Jiamusi 66-m telescope is located at Jiamusi in Heilongjiang
province, China, which is operated by the Jiamusi Deep Space Station
of the China Xi'an Satellite Control Center. A cryogenically cooled
dual-channel S-band receiver was used for the observations, which has
a bandwidth of about 140~MHz at the radio frequency around
2.25~GHz. After adjusting the signal amplitude of the intermediate
frequency (IF) output with two variable-gain amplifiers, we connected
a digital backend to the receiver (see Figure~\ref{fig1} for
illustration). The backend samples the IF signals and then channelizes
the signals via a FFT module in a Field-Programmable Gate Array (FPGA)
board. The total radio power is obtained by summing the detected power
from the left hand and right hand polarization receivers for each of
the 256 or 128 frequency channels, and then accumulated for duration
of the sampling time. The accumulated radio powers of 256 or 128
channels are saved to a computer with a time resolution of 0.2~ms or
0.1~ms.

A test observation was made for 15 minutes on 2015 April 18 (MJD
57130), and long observations of PSR B0919+06 were made in 2015 July
and August (MJD 57215 -- 57253, see Table~\ref{tab1}).

\begin{figure}
  \includegraphics[height=229mm]{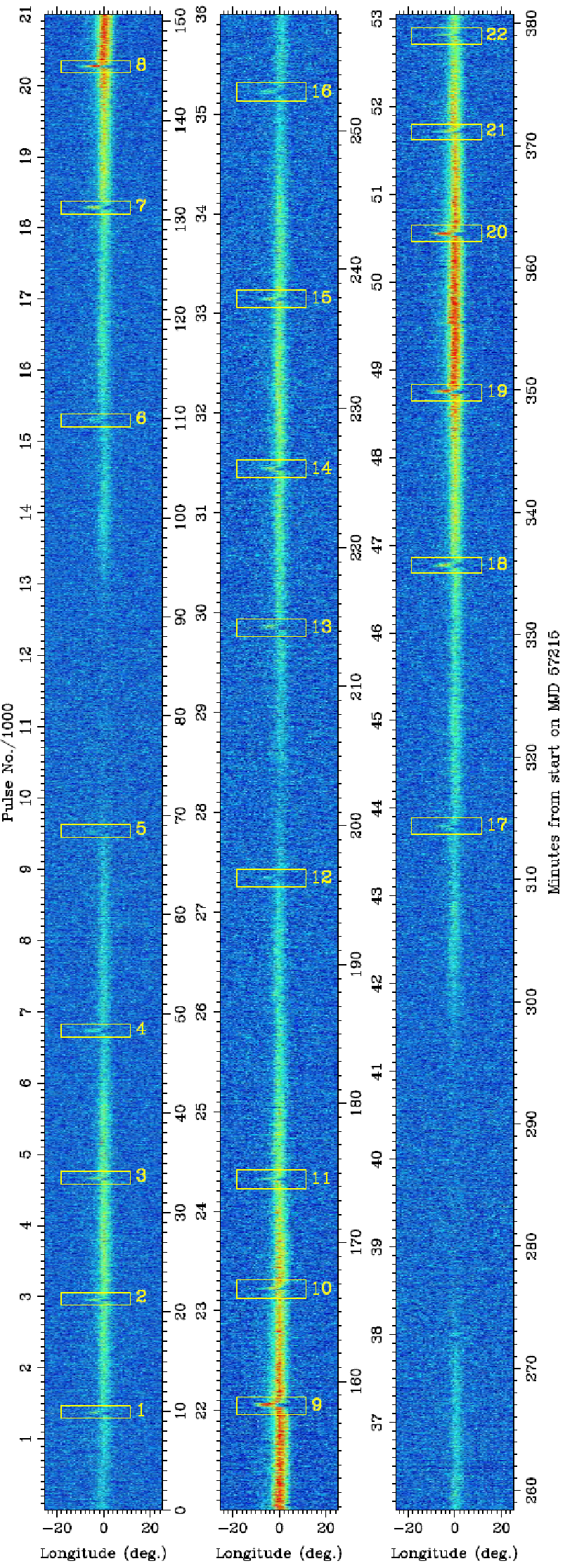}
  \caption{The 53,053 individual pulses observed on MJD 57215, in which 22
    abnormal emission events are seen as indicated by the boxes.}
  \label{fig3}
\end{figure}

\begin{figure}
  \includegraphics[height=229mm]{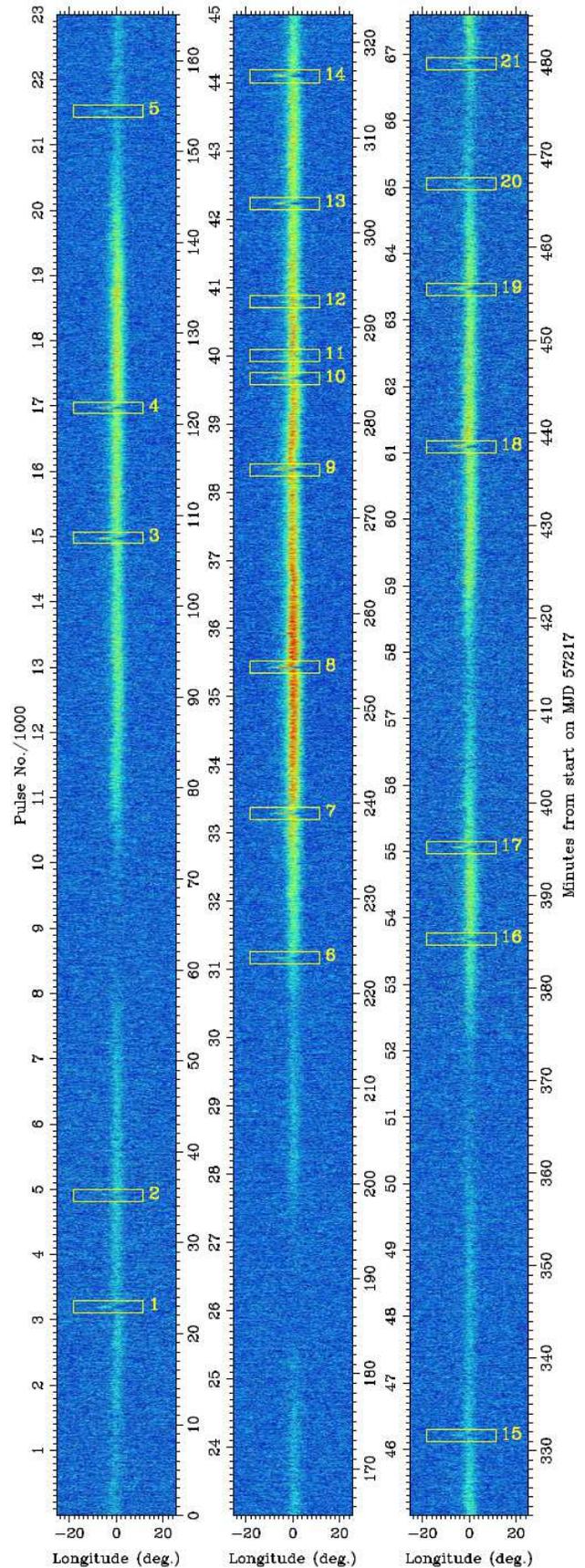}
  \caption{The same as Fig.\ref{fig3} but for 67,611 individual pulses
    observed on MJD 57217 and the 21 abnormal emission events.}
  \label{fig4}
\end{figure}

\begin{figure}
  \includegraphics[height=229mm]{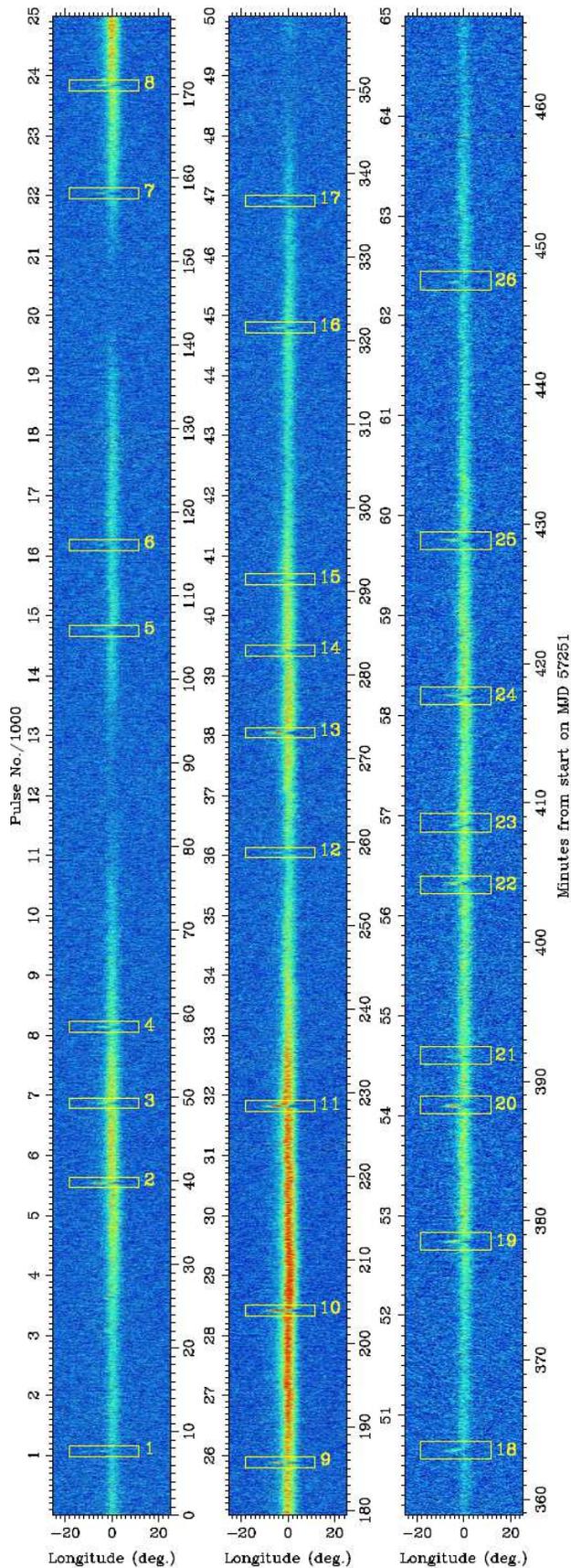}
  \caption{The same as Fig.\ref{fig3} but for 65,003 individual pulses
    observed on MJD 57251 and 26 abnormal emission events.}
  \label{fig5}
\end{figure}

\begin{figure}
  \includegraphics[height=229mm]{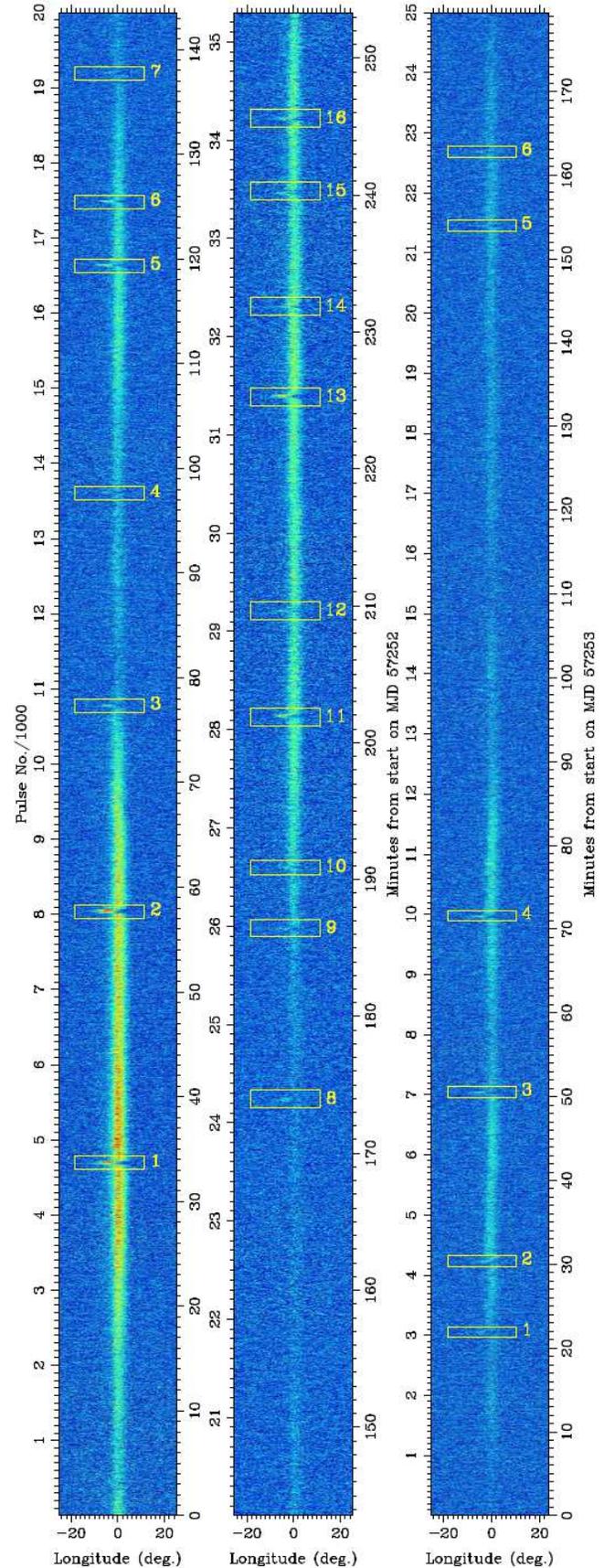}
  \caption{The same as Fig.\ref{fig3} but for 35,306 individual pulses observed on MJD 57252 and 25,001 pulses observed on MJD 57253, in which 16 and 6
    abnormal emission events are found, respectively.}
  \label{fig6}
\end{figure}

\begin{figure}
  \centering
  \includegraphics[angle=-90,width=0.38\textwidth]{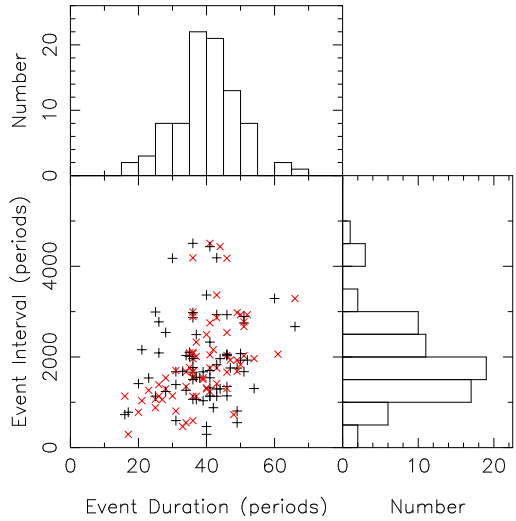}
  \caption{Duration statistics for the normal emission mode and
    abnormal emission events of PSR B0919+06. When the signal-to-noise
    ratio is good enough to count individual pulses, the period number
    of an abnormal emission event is taken as the horizontal value in
    the main panel, and the period number of the normal emission mode
    ahead of (or after) the event is taken as the vertical value, so
    that a point is plotted as ``+'' (or ``$\times$''). The histograms
    on the top panel and the right panel show pulse number
    distribution of abnormal emission events and the event intervals,
    respectively.}
  \label{fig7}
\end{figure}

\section{Data Analysis and results}

A few steps were needed to analyze the recorded data. First, the radio
frequency interference (RFI) was cleaned. RFI appears in a
few channels for a long time or in many channels for a some short
time. The polluted data were identified and then replaced with
randomly selected samples from the same channel or nearby channels to
keep the statistical properties. The total power data from all
channels were dedispersed with the known dispersion measure value of
PSR B0919+06, and then summed with proper weights due to different
gains of the 256 or 128 channels to get a long time series. Finally,
the individual pulses were aligned with the best period which we
found out around the formal period value.

Results from a test observation of PSR B0919+06 on MJD 57130 are shown
in Figure~\ref{fig2}. One abnormal emission event was identified
between pulses 365 and 395, as illuminated in the right
panels, in which pulses show up at earlier phases of about $5^{\circ}$
in longitude. This event is very similar to the two abnormal emission
events observed by \citet[][shown in their Figure~1]{rrw06} and the
one ``main flare state'' observed by \citet[][shown in their
  Figure~10]{psw+15}, during which individual pulses have a gradual
shifting in the pulse rotation phase for 30-40 periods.

\input event.tex

To investigate how often such events occur, we have carried out
observations in 5 days, several hours in each day. The aligned plots
of individual pulses\footnote{Dedispersed data series are available at
  http://zmtt.bao.ac.cn/psr-jms/} are shown in
Figures~\ref{fig3}--\ref{fig6}. Due to interstellar scintillation, the
flux intensity of PSR B0919+06 clearly shows variations on a time
scale of one to two hours. From these long series of individual
pulses, we found 92 abnormal emission events through the
visual-checking of plots. See Table~\ref{tab2} for their start and end
pulse numbers. Based on the total number of events over the entire
timespan of observations, we can estimate that one event, on average,
occurs about every 20 minutes. However, there are possibilities that
events were missing when pulsar signals are weaked by
scintillation. The total number of events should therefore be taken as
the lower limit, and hence the waiting time of events are expected to
be shorter. As shown in Figure~\ref{fig7}, we made the duration
statistics for the pulsar normal mode emission and abnormal emission
events. The abnormal emission events were seen to last for 15--70
periods with a peak of about 40 periods (i.e. $\le$20 seconds,
remembering the pulsar period of 0.43062~s), and normal emission
lasted for 1000 to 3000 periods with a peak about 1500 periods
(i.e. $\sim$10 minutes).

The abnormal emission events of PSR B0919+06 were obviously detected
by looking at plots for individual pulses if with a good
signal-to-noise ratio. For events with weak signals, the data samples
of neighbor phase bins and from a few periods could be averaged to
enhance the signal-to-noise radio. Based on their details of
phase-shifting of individual pulses, we classified abnormal emission
events as following:
\begin{itemize}
\item {\bf $\Pi$ type:} Individual pulses of abnormal emission events
  abruptly come a few degrees early in longitude, and then stay in this
  phase range (i.e. the ``high state'') for a few tens of periods,
  eventually come back abruptly to the normal phase range. There
  are seven such cases detected during our observations, one shown in
  Figure~\ref{fig2} and six shown in Figure~\ref{fig8}. Such a quick
  switching is almost the same as the mode-changing phenomena of other
  pulsars. Note also that subpulse drifting was observed before and
  during the abnormal emission events No.~8 and No.~19 observed on MJD
  57215, which can be seen from very enlarged plots of Figure~8.
\item {\bf M type}: It is the original type of abnormal emission event
  that was discovered by \citet{rrw06}. Individual pulses gradually
  ``go up'' on the longitude over a few periods, stay on the ``high
  state'' for more than 20 periods perhaps with some fluctuations in
  phases, and finally gradually come back to the normal phase over a
  few periods. We identified 33 events of this type (see
  Figure~\ref{fig9}).
\item {\bf $\Lambda$ type}: Individual pulses of this type of events
  rise and fall gradually without staying in the ``high state''. We
  identified 22 events of this type (shown in
  Figure~\ref{fig10}). Some of the events were the ``small flares'' as
  these seen by \citet{psw+15}.
\item {\bf $\lambda$ type}: A few abnormal emission events show a
  rapid ``rise'' but a slow ``decay'', which appears as a longer tail
  in Figure~\ref{fig11}.
\end{itemize}
In addition to these classified events, 26 events (see
Figure~\ref{fig12}) have a very weak signal-to-noise ratio, which can
not be certainly classified even after their ``images'' are enhanced
by a ``Gaussian smoothing''. These events are marked with a ``?'' for
the type in Table~\ref{tab2}.

\begin{figure}
  \centering
  \includegraphics[angle=-90,width=0.35\textwidth]{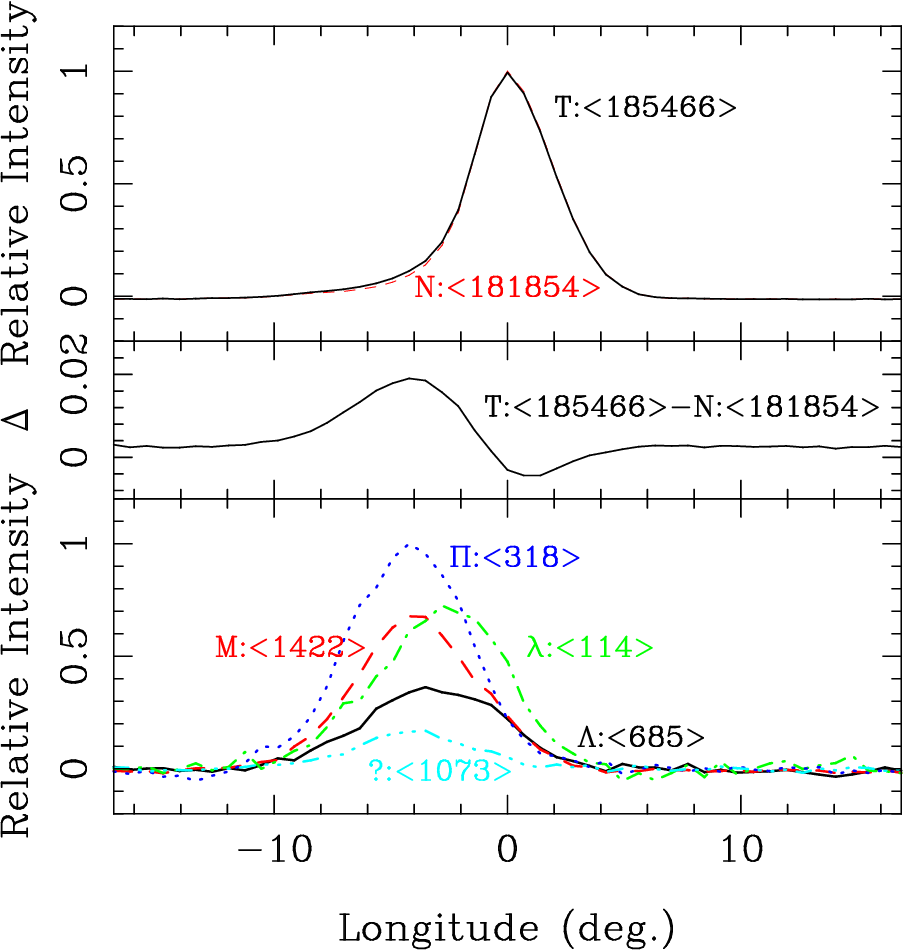}
  \caption{Comparison of profiles for the normal emission mode and
    abnormal emission events. On the top panel the mean profile of
    181854 normal emission pulses are compared with all 185466 pulses
    including 3612 pulses in the abnormal emission events. The
    difference of their normalized profiles is shown in the middle
    panel. In the lower panel, the mean profiles are presented for 318
    individual pulses from the $\Pi$ type, 1422 pulses from the M
    type, 685 pulses from the $\Lambda$ type, 114 pulses from the
    $\lambda$ type of abnormal events are shown, together with the
    mean profile from the 1073 weak pulses from the unclassified
    events, which are all offset from the normal pulses.}
  \label{fig13}
\end{figure}

The short duration and the small total number of individual pulses of
abnormal emission events indicates that they contribute very little
($\sim2\%$) to the mean intensity. In Figure~\ref{fig13}, we compare
the mean pulsar profile from all 185466 bright pulses including
abnormal emission events with that from 181854 bright pulses excluding
the abnormal emission events in between. The difference between the
above two profiles are shown in the middle panel. In the bottom panel
the mean profiles of all types of abnormal emission events are shown,
and clearly they are different. The abnormal emission events mostly
affect the mean profile in the phase window of $-2^{\circ}$ to
$-8^{\circ}$.

\section{Discussions and Conclusions}

We observed PSR B0919+06 for about 30 hours and detected 92 abnormal
emission events during which individual pulses are shifted to early
pulse phases by a few degrees in longitude. Such events last for a few
tens of periods, and are seen to happen randomly every 1000 to 3000
periods. We have classified the abnormal emission events into four
types based on the phase-shift features of individual pulses.

Although the pulse number during the abnormal emission events are small (3612
vs. 185466 normal pulses, i.e. about 2\%), it is clear that the
abnormal emission events should lead to changes of mean pulse profiles
of PSR B0919+06 up to 2\%. This amplitude is roughly equal to the
difference of the two profile modes shown by \citet{psw+15}. Timing
observations for 20 minutes or less may or may not include one
abnormal emission event, since the events occur randomly with an
interval of 500 to 3500 periods (i.e. 200 to 1400 seconds). Therefore
it is not clear if the long-term variations of profiles are related to
such abnormal emission events. In addition, during timing, it seems to
be important to get the time of arrival of pulses by fitting the
standard profile to the profiles only in the later phase range of
$>-1^{\circ}$, rather than the all phase ranges of pulsed emission.

Can the precession be the causal reason for the abnormal emission
events?  Free precession has been suggested to cause long-term
variations of pulsar profiles and $\dot{\nu}$ \citep{sls00,khjs15},
especially their correlation \citep{alw06,jon12}. However, a small
wobbling of the rotation axis should lead a gradual variation of
profiles quasi-periodically, not just in short duration of 20 seconds
of every 10 to 15 minutes. Therefore, it is very unlikely that the
precession is the intrinsic cause for such abnormal emission events
lasting only for a few tens of periods.

We emphasize that PSR B0919+06 is not the only pulsar with such a
phase shift. PSR B1859+07 was also found \citep{rrw06,psw+16} to
exhibit similar events, though the events were seen to be much more
frequent. Other pulsars might have similar emission features
uncovered, simply due to lack of long observations.

Based on the features of abnormal emission events of PSR B0919+06 we
observed, together with the same mean ``S''-shaped curve of
polarization angles of the abnormal emission events as the normal
trailing conal component observed by \citet[][as described in
  Section~1]{rrw06}, we believe that the abnormal emission events of
PSR B0919+06, as the phenomena of mode-changing and subpulse drifting
observed in other pulsars, probably originate from emission processes
in the pulsar magnetosphere.

\section*{Acknowledgments}
We thank Prof. JinXin Hao of NAOC for his support of this work,
Prof. Biping Gong from Huazhong University of Science and Technology
for stimulative discussions, and Ms. SuSu Shan, Mr. Fan Yang and the
operation team of Jiamusi 66~m Deep Space Station of the China Xi'an
Satellite Control Center for the assistance of observations.
We also thank the referee for helpful comments.
The authors are partially supported by the National Natural Science
Foundation of China through grants No. 11473034 and 11273029, the
Strategic Priority Research Program ``The Emergence of Cosmological
Structures'' of the Chinese Academy of Sciences, Grant
No. XDB09010200”, and the Open Fund of the State Key Laboratory of
Astronautic Dynamics of China.

\bibliographystyle{mn2e}
\bibliography{journals,psr}


\label{lastpage}
\end{document}

%% file: tab2.tex
\begin{table}
\caption{A list of abnormal emission events for detailed plots}
\label{tab2}
\setlength{\tabcolsep}{1.5mm}
\begin{tabular}{ccccll}
\hline
MJD       & Event No.  & Start--End pulse No. & Duration & Type  & Details \\
\hline                                                                    
57130      & 1    & 365    -- 395   &31 & $\Pi$       & Fig.2   \\[1.5mm]
57215      & 1    & 1360   -- 1400  &41 & ?           & Fig.12  \\
57215      & 2    & 2940   -- 2975  &36 & M           & Fig.9   \\   
57215      & 3    & 4638   -- 4683  &46 & M           & Fig.9   \\   
57215      & 4    & 6710?  -- 6760  &51 & M           & Fig.9   \\   
57215      & 5    & 9510   -- 9550  &41 & ?           & Fig.12  \\
57215      & 6    & 15280? -- 15325?&46 & ?           & Fig.12  \\
57215      & 7    & 18258  -- 18309 &52 & $\Pi$       & Fig.8   \\    
57215      & 8    & 20238  -- 20284 &47 & $\Pi$       & Fig.8   \\   
57215      & 9    & 22040  -- 22080 &41 & $\Pi$       & Fig.8   \\   
57215      & 10   & 23213  -- 23248 &36 & $\lambda$   & Fig.11  \\   
57215      & 11   & 24315  -- 24339 &25 & $\Lambda$   & Fig.10  \\   
57215      & 12   & 27330? -- 27365?&36?& ?           & Fig.12  \\
57215      & 13   & 29845  -- 29882?&38?& M           & Fig.9   \\   
57215      & 14   & 31422  -- 31460 &39 & M           & Fig.9   \\   
57215      & 15   & 33130  -- 33163 &34 & M           & Fig.9   \\   
57215      & 16   & 35195  -- 35240 &46 & M           & Fig.9   \\   
57215      & 17   & 43770  -- 43812 &43 & ?           & Fig.12  \\
57215      & 18   & 46745  -- 46794 &50 & $\Pi$       & Fig.8   \\   
57215      & 19   & 48722  -- 48770 &49 & $\Pi$       & Fig.8   \\   
57215      & 20   & 50525  -- 50567 &43 & $\Pi$       & Fig.8   \\   
57215      & 21   & 51700  -- 51736 &37 & $\lambda$   & Fig.11  \\   
57215      & 22   & 52796  -- 52822 &27 & $\Lambda$   & Fig.10  \\[1.5mm]
57217      & 1    & 3175   -- 3225? &51?& M           & Fig.9   \\   
57217      & 2    & 4900?  -- 4935? &36?& ?           & Fig.12  \\
57217      & 3    & 14950  -- 14993 &44 & M           & Fig.9   \\   
57217      & 4    & 16955  -- 16990 &36 & M           & Fig.9   \\   
57217      & 5    & 21495  -- 21535 &41 & ?           & Fig.12  \\
57217      & 6    & 31165? -- 31190?&26?& $\Lambda$   & Fig.10  \\   
57217      & 7    & 33278  -- 33312 &35 & M           & Fig.9   \\   
57217      & 8    & 35414  -- 35449 &36 & M           & Fig.9   \\   
57217      & 9    & 38315  -- 38357 &43 & M           & Fig.9   \\   
57217      & 10   & 39651  -- 39690 &40 & M           & Fig.9   \\   
57217      & 11   & 39982  -- 39998 &17 & $\Lambda$   & Fig.10  \\   
57217      & 12   & 40777  -- 40796 &20 & $\Lambda$   & Fig.10  \\   
57217      & 13   & 42208  -- 42250 &43 & M           & Fig.9   \\   
57217      & 14   & 44078  -- 44127 &50 & M           & Fig.9   \\   
57217      & 15   & 46200  -- 46235 &36 & ?           & Fig.12  \\
57217      & 16   & 53655  -- 53700 &46 & M           & Fig.9   \\   
57217      & 17   & 55047  -- 55080 &34 & M           & Fig.9   \\   
57217      & 18   & 61080  -- 61120 &41 & $\lambda$   & Fig.11  \\   
57217      & 19   & 63446  -- 63482 &37 & M           & Fig.9   \\   
57217      & 20   & 65035? -- 65070?&36?& ?           & Fig.12  \\
57217      & 21   & 66840  -- 66875 &36 & ?           & Fig.12  \\[1.5mm]
57251      & 1    & 1050?  -- 1090? &41?& ?           & Fig.12  \\
57251       & 2   & 5522   -- 5565  &44 & M          & Fig.9    \\   
57251       & 3   & 6860   -- 6887  &28 & $\Lambda$  & Fig.10   \\    
57251       & 4   & 8125   -- 8165? &41?& M          & Fig.9    \\   
57251       & 5   & 14750? -- 14780?&31?& ?          & Fig.12   \\
57251       & 6   & 16170  -- 16195 &26 & ?          & Fig.12   \\
57251       & 7   & 22020? -- 22055 &36?& M          & Fig.9   \\    
57251       & 8   & 23822  -- 23856 &35 & $\Lambda$  & Fig.10   \\   
57251       & 9   & 25867  -- 25903 &37 & M          & Fig.9    \\   
57251       & 10  & 28397  -- 28436 &40 & M          & Fig.9    \\   
57251       & 11  & 31802  -- 31844 &43 & M          & Fig.9    \\   
57251       & 12  & 36030? -- 36065?&36?& $\Lambda$  & Fig.10   \\   
57251       & 13  & 38027  -- 38080 &54 & M          & Fig.9    \\   
57251       & 14  & 39385  -- 39430 &46 & $\Lambda$  & Fig.10   \\   
57251       & 15  & 40571  -- 40600 &30 & $\Lambda$  & Fig.10   \\   
57251       & 16  & 44775  -- 44820 &46 & M          & Fig.9    \\   
57251       & 17  & 46880  -- 46940 &61 & ?          & Fig.12   \\
57251       & 18  & 50630? -- 50675 &46?& $\Lambda$  & Fig.10   \\   
\end{tabular}
\end{table}
\begin{table}
  \addtocounter{table}{-1}
  \center
 \setlength{\tabcolsep}{1.5mm}

\caption{ -- continued}\center

\begin{tabular}{ccccll}
MJD       & Event No.  & Start--End pulse No. & Duration & Type  & Details \\
\hline
57251       & 19  & 52720  -- 52760 &41 & $\Lambda$  & Fig.10   \\   
57251       & 20  & 54076  -- 54115  &40 & M          & Fig.9    \\   
57251       & 21  & 54580  -- 54612  &33 & $\Lambda$  & Fig.10   \\  
57251       & 22  & 56295  -- 56343  &49 & M          & Fig.9    \\  
57251       & 23  & 56892  -- 56925  &34 & $\Lambda$  & Fig.10   \\  
57251       & 24  & 58188  -- 58210  &23 & $\Lambda$  & Fig.10   \\  
57251       & 25  & 59745  -- 59772  &28 & $\Lambda$  & Fig.10   \\  
57251       & 26  & 62310  -- 62355  &46 & ?          & Fig.12   \\[1.5mm]
57252       & 1   & 4667   -- 4726   &60 & M          & Fig.9    \\  
57252       & 2   & 8015   -- 8080   &66 & M          & Fig.9    \\  
57252       & 3   & 10750  -- 10800  &51 & ?          & Fig.12   \\
57252       & 4   & 13605? -- 13640? &36?& ?          & Fig.12   \\
57252       & 5   & 16618  -- 16666  &49 & $\Lambda$  & Fig.10   \\  
57252       & 6   & 17472  -- 17502  &31 & $\Lambda$  & Fig.10   \\  
57252       & 7   & 19175? -- 19210? &36?& ?          & Fig.12   \\
57252       & 8   & 24220  -- 24260  &41 & ?          & Fig.12   \\
57252       & 9   & 25960  -- 25990  &31 & ?          & Fig.12   \\
57252       & 10  & 26585  -- 26620  &36 & ?          & Fig.12   \\
57252       & 11  & 28122  -- 28160  &39 & M          & Fig.9    \\  
57252       & 12  & 29195  -- 29215  &21 & $\Lambda$  & Fig.10   \\  
57252       & 13  & 31375  -- 31416  &42 & M          & Fig.9    \\  
57252       & 14  & 32296  -- 32320  &25 & $\Lambda$  & Fig.10   \\  
57252       & 15  & 33455? -- 33470? &16?& $\Lambda$  & Fig.10   \\  
57252       & 16  & 34198  -- 34245  &48 & $\Lambda$  & Fig.10   \\[1.5mm]
57253       & 1   & 3035?  -- 3075?  &41?& ?          & Fig.12   \\
57253       & 2   & 4210   -- 4235   &26 & ?          & Fig.12   \\
57253       & 3   & 7010   -- 7060   &51 & ?          & Fig.12   \\
57253       & 4   & 9950?  -- 9985?  &36?& ?          & Fig.12   \\
57253       & 5   & 21426  -- 21480? &55?& ?          & Fig.12   \\
57253       & 6   & 22665? -- 22700? &36?& ?          & Fig.12   \\
\hline
\end{tabular}
\\ Note: ``?'' indicates the uncertainty in the start--end pulse
numbers, duration and classification. 
\end{table}

%% file: event.tex
\begin{figure}
 \includegraphics[bb = 60 45 505 685,clip,width=0.114\textwidth,angle=-90]{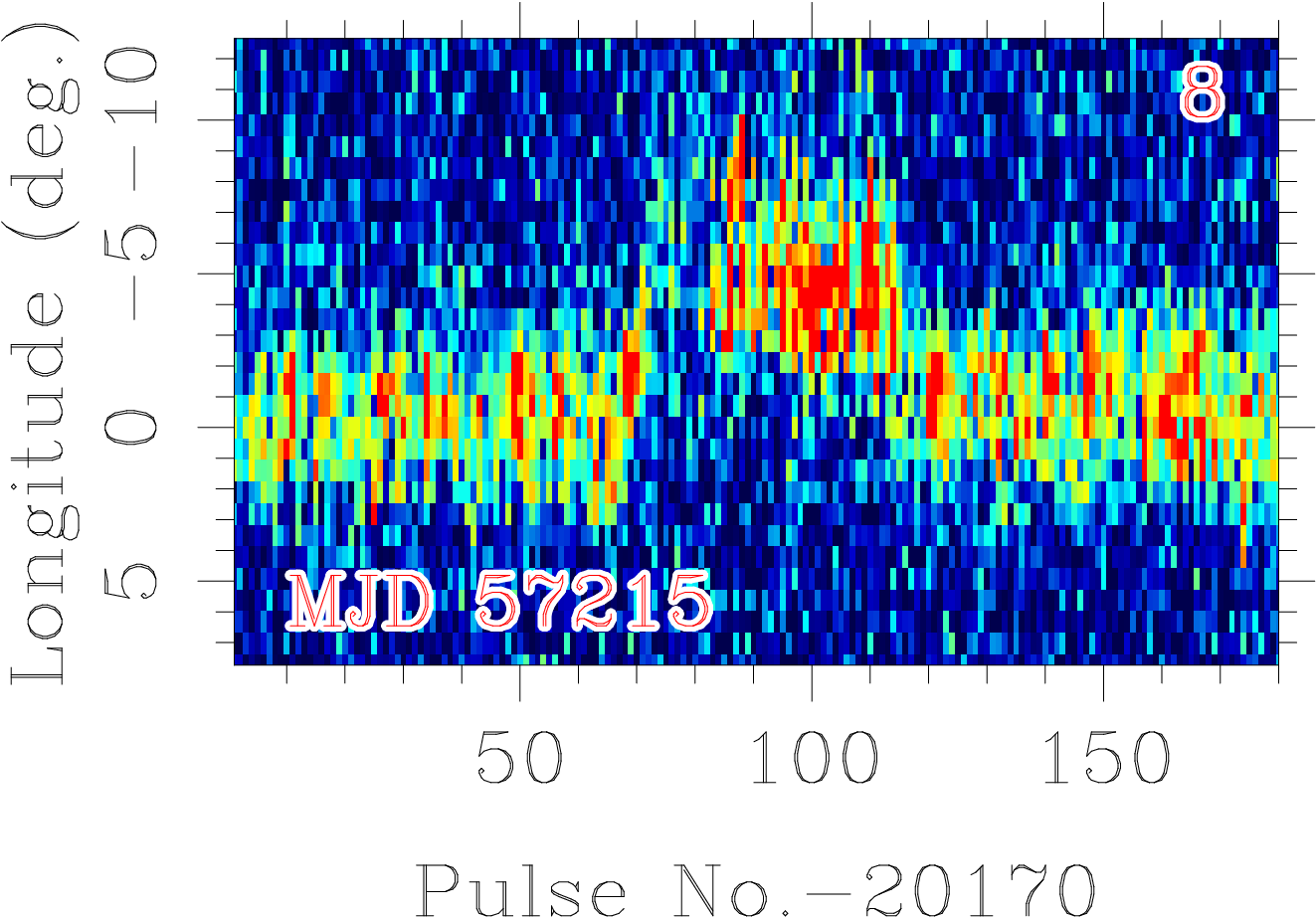}
 \includegraphics[bb = 60 95 505 685,clip,width=0.114\textwidth,angle=-90]{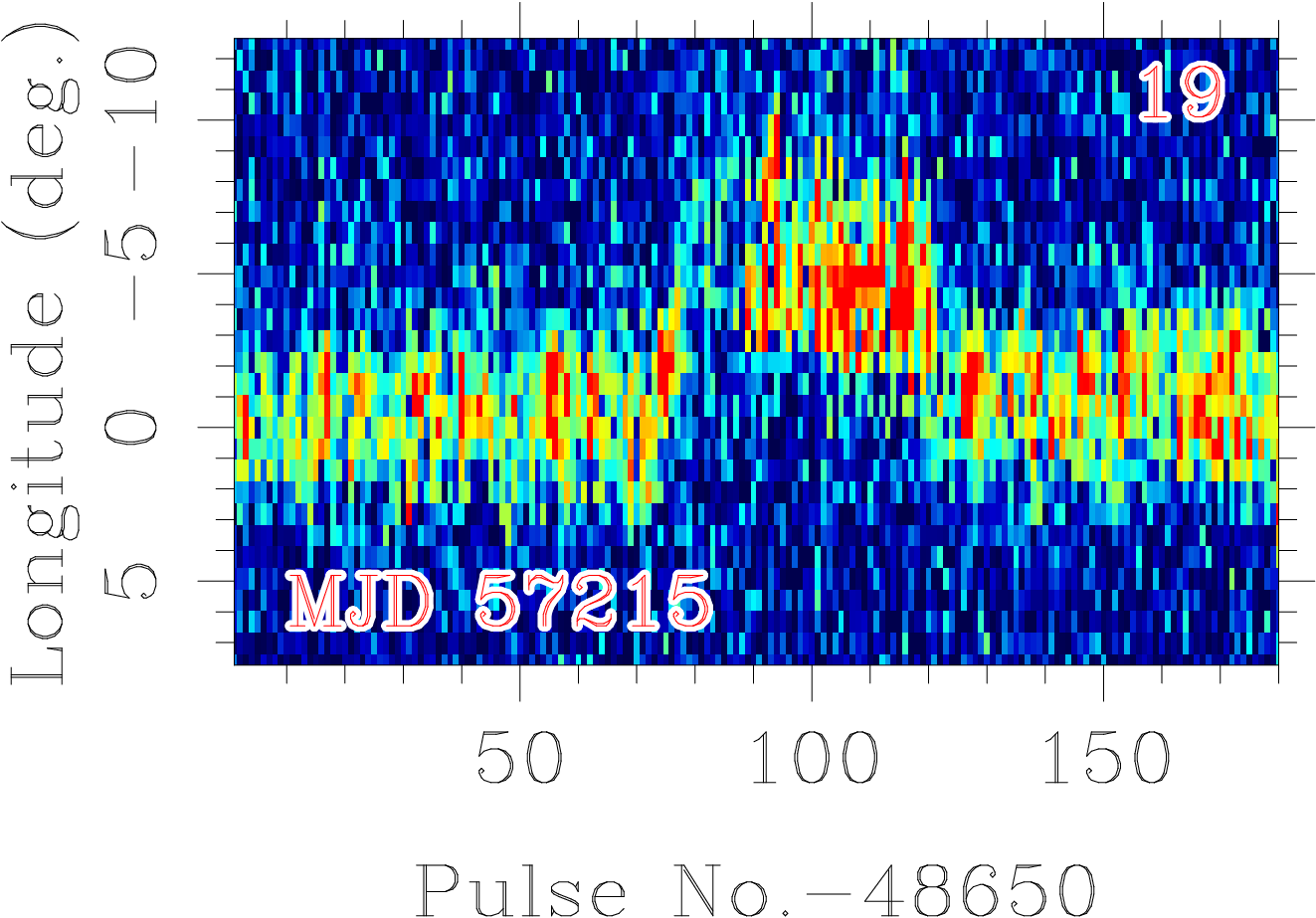} 
 \includegraphics[bb = 60 95 505 685,clip,width=0.114\textwidth,angle=-90]{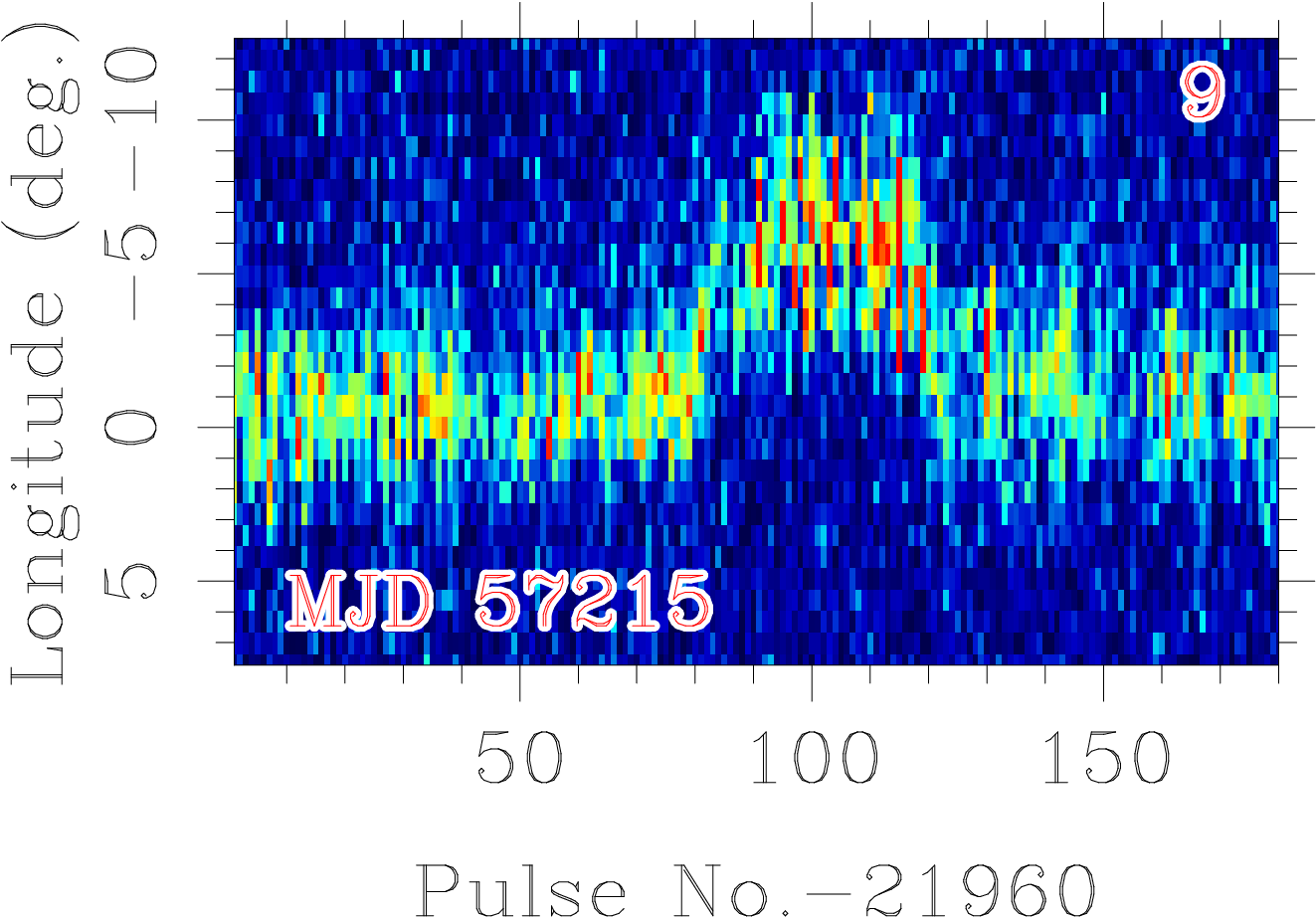}
 \includegraphics[bb = 60 45 505 685,clip,width=0.114\textwidth,angle=-90]{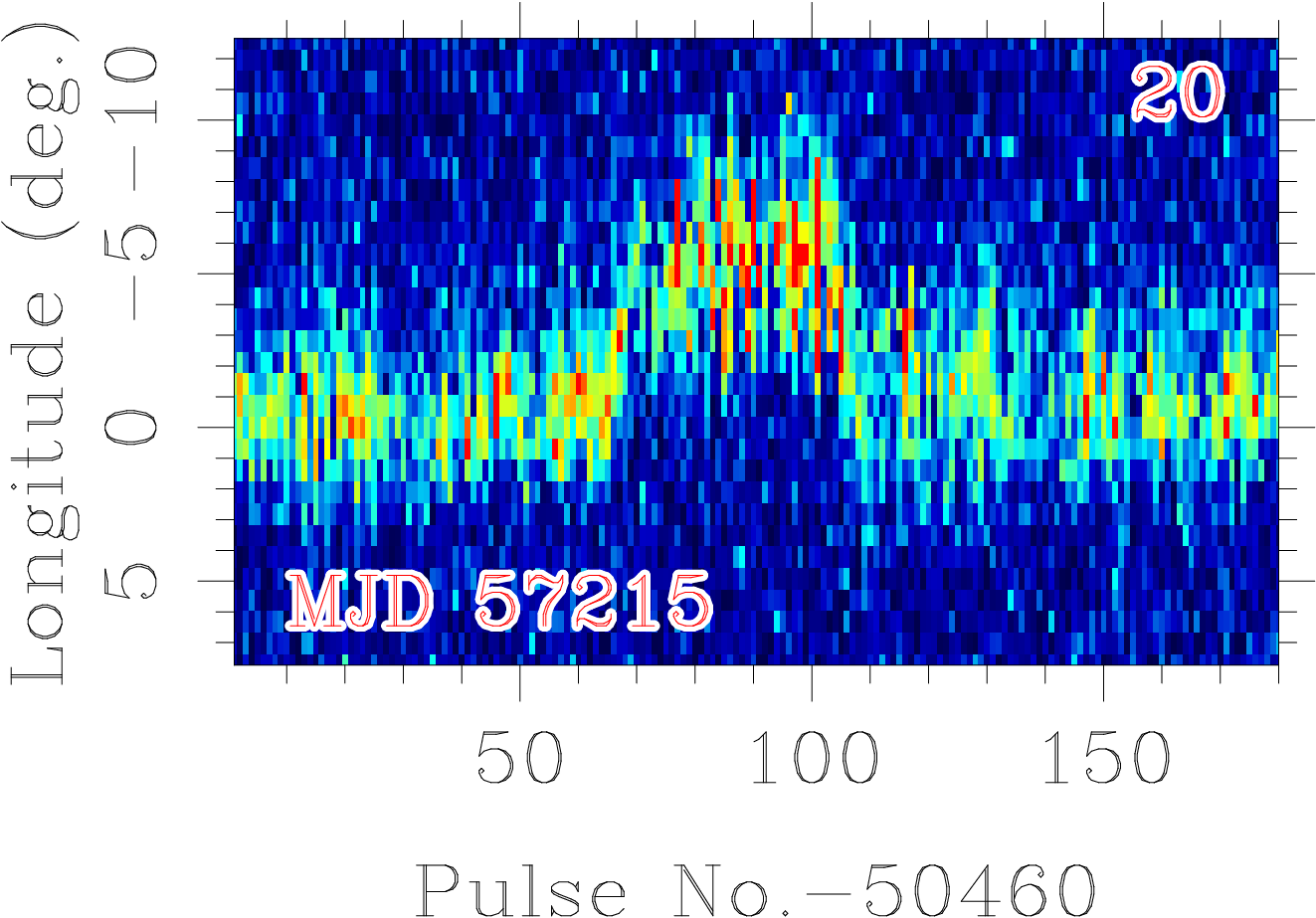}
 \includegraphics[bb = 60 95 505 685,clip,width=0.114\textwidth,angle=-90]{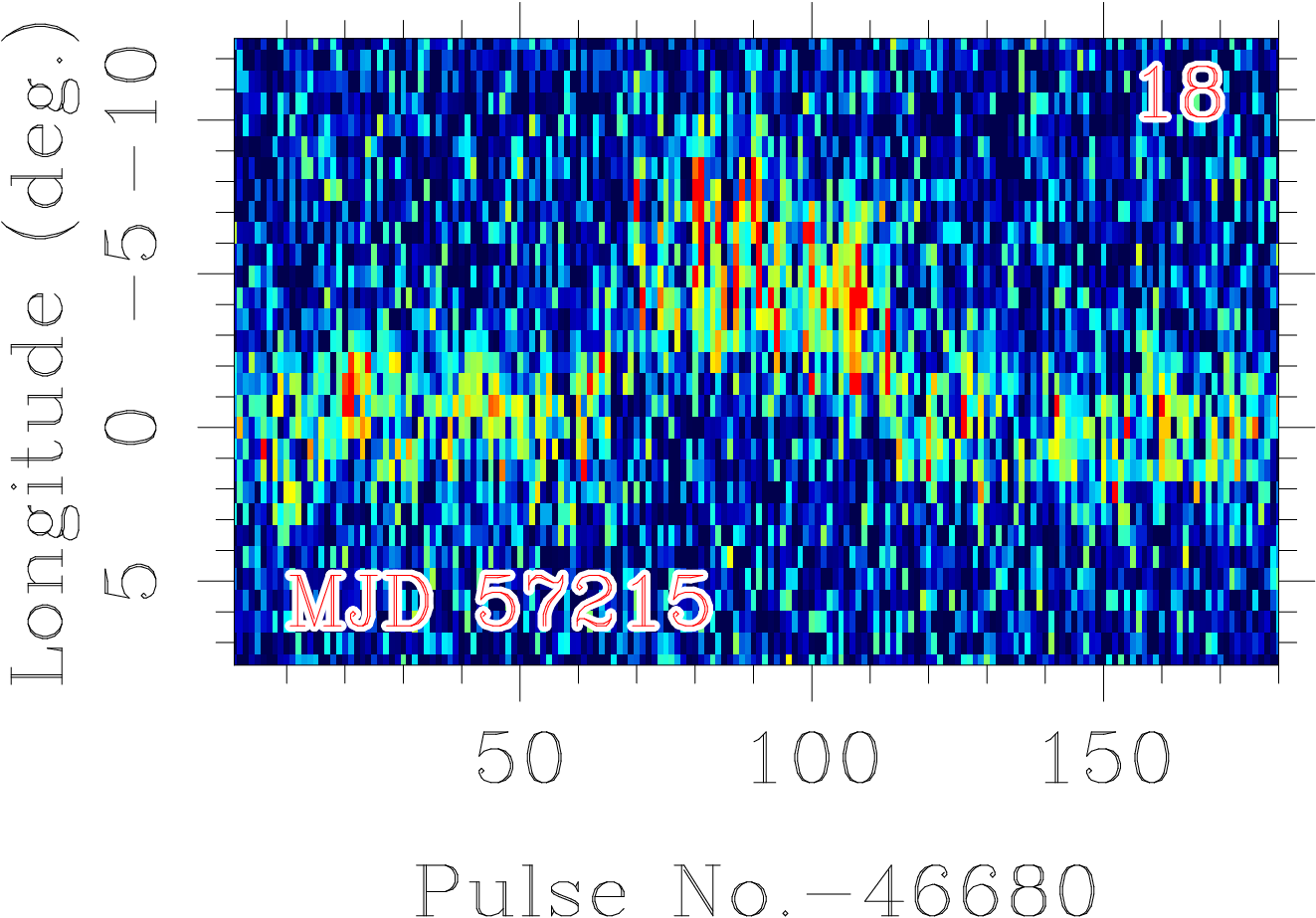}
 \includegraphics[bb = 60 95 505 685,clip,width=0.114\textwidth,angle=-90]{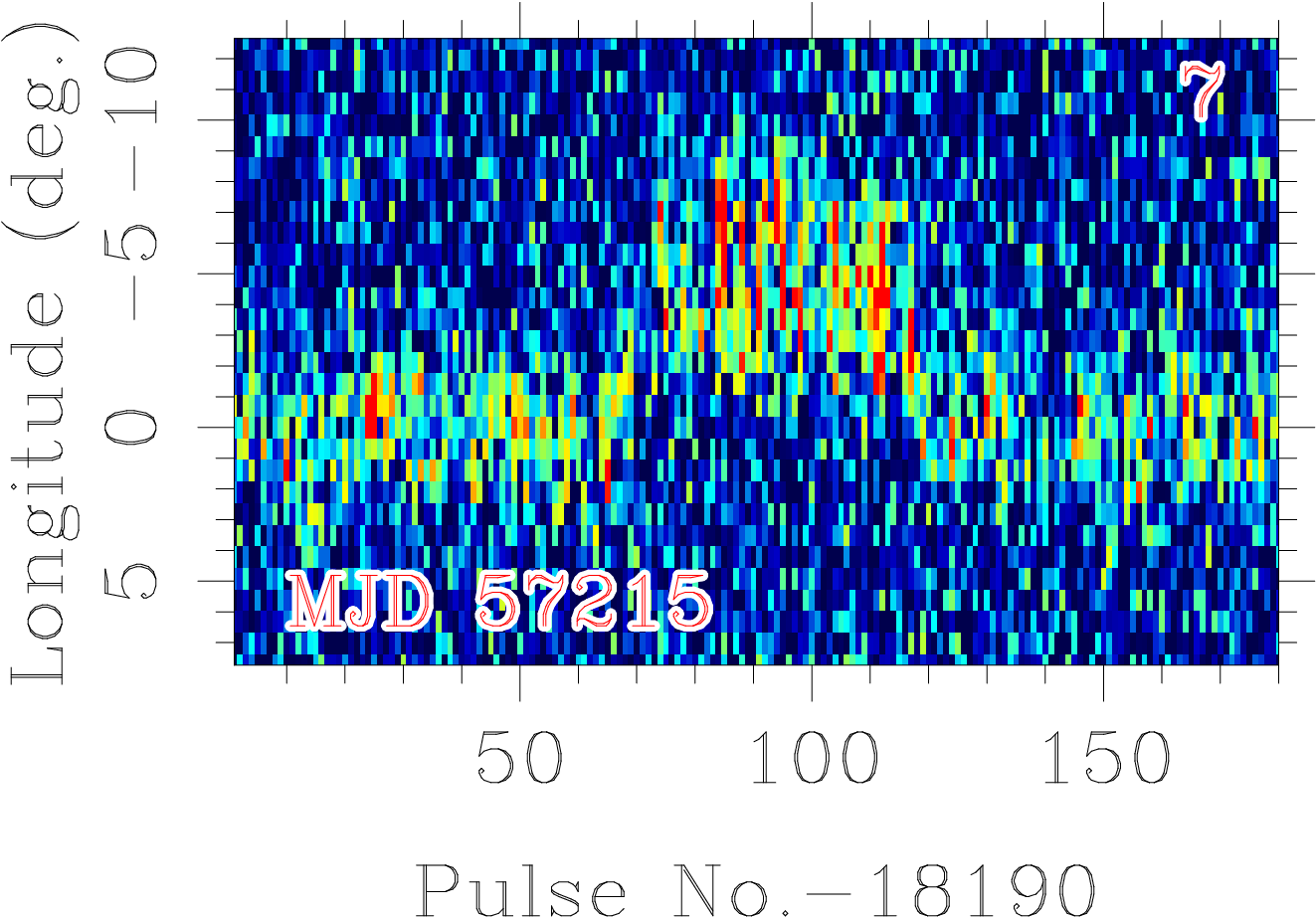}
 \caption{The detailed plots for 6 abnormal emission events of ``$\Pi$ type''. The pulse
   number is plotted on the horizontal axis, and the longitude for the pulse phase
   on the vertical axis. The flux intensity of pulses is
   represented in colors with the dark color for zero and bright color 
   for high intensity. MJD and the event number are indicated in each plot.}
  \label{fig8}
\end{figure}
\begin{figure}
 \includegraphics[bb = 60 45 505 685,clip,width=0.114\textwidth,angle=-90]{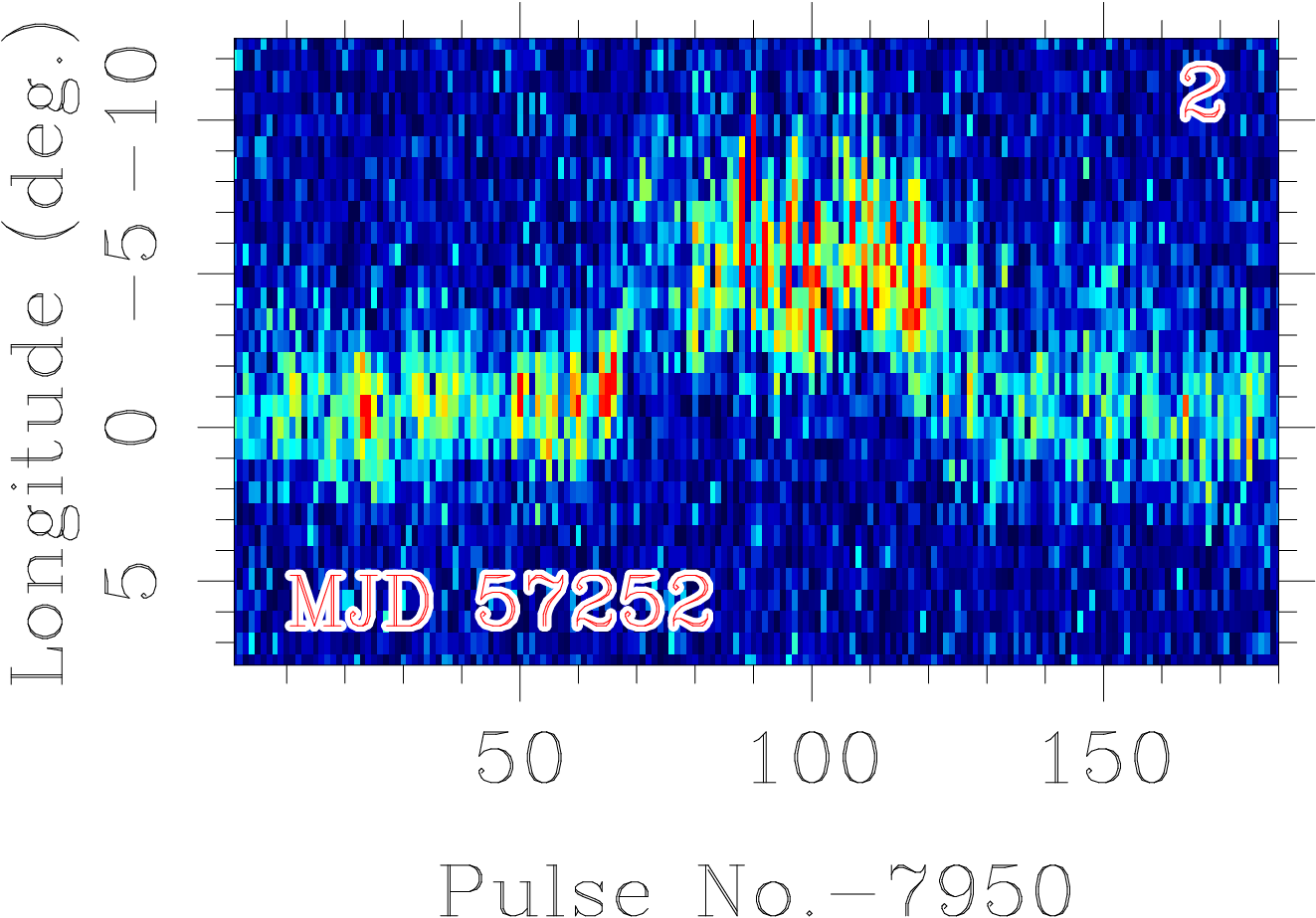}
 \includegraphics[bb = 60 95 505 685,clip,width=0.114\textwidth,angle=-90]{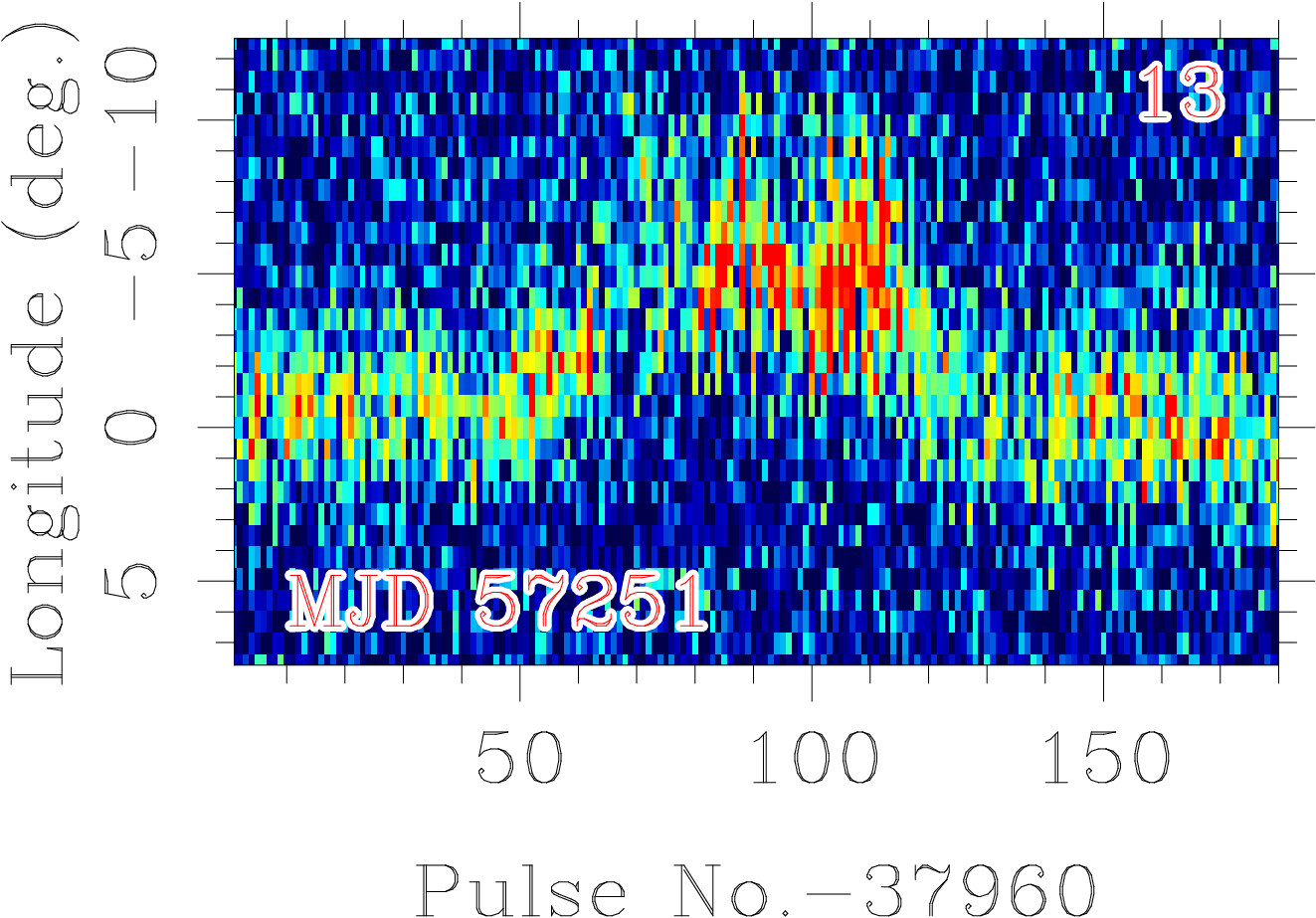}
 \includegraphics[bb = 60 95 505 685,clip,width=0.114\textwidth,angle=-90]{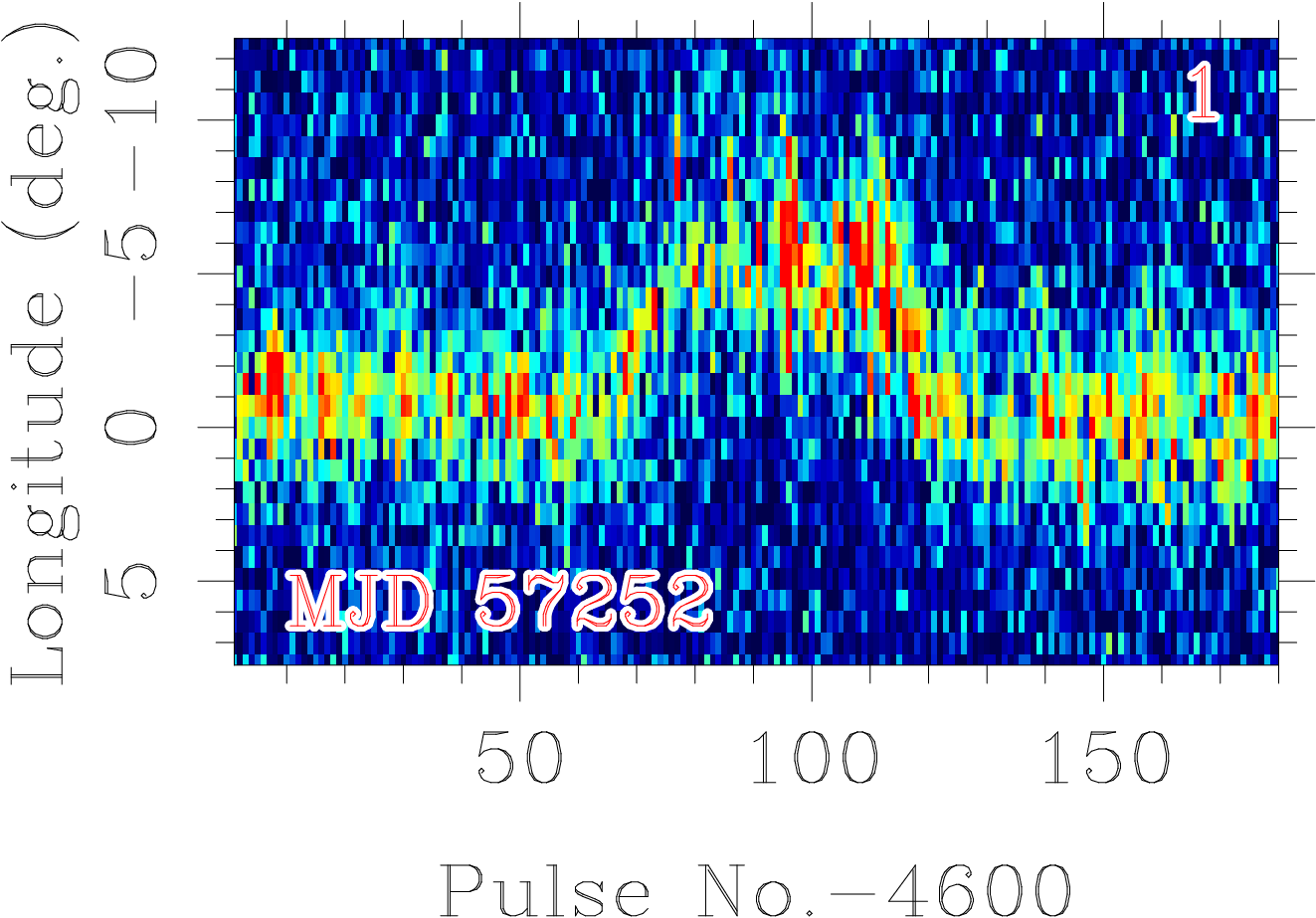}
 \includegraphics[bb = 60 45 505 685,clip,width=0.114\textwidth,angle=-90]{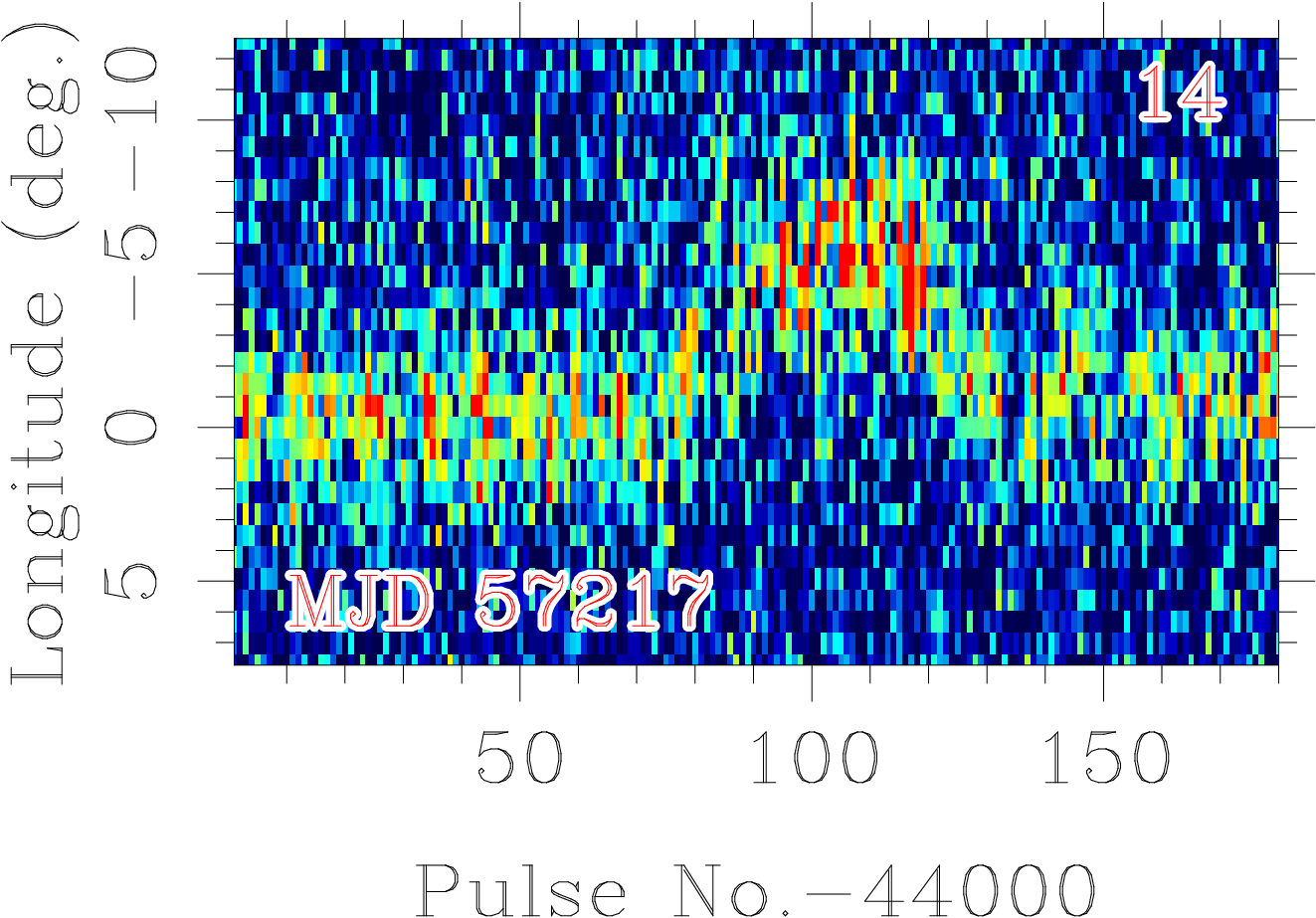}
 \includegraphics[bb = 60 95 505 685,clip,width=0.114\textwidth,angle=-90]{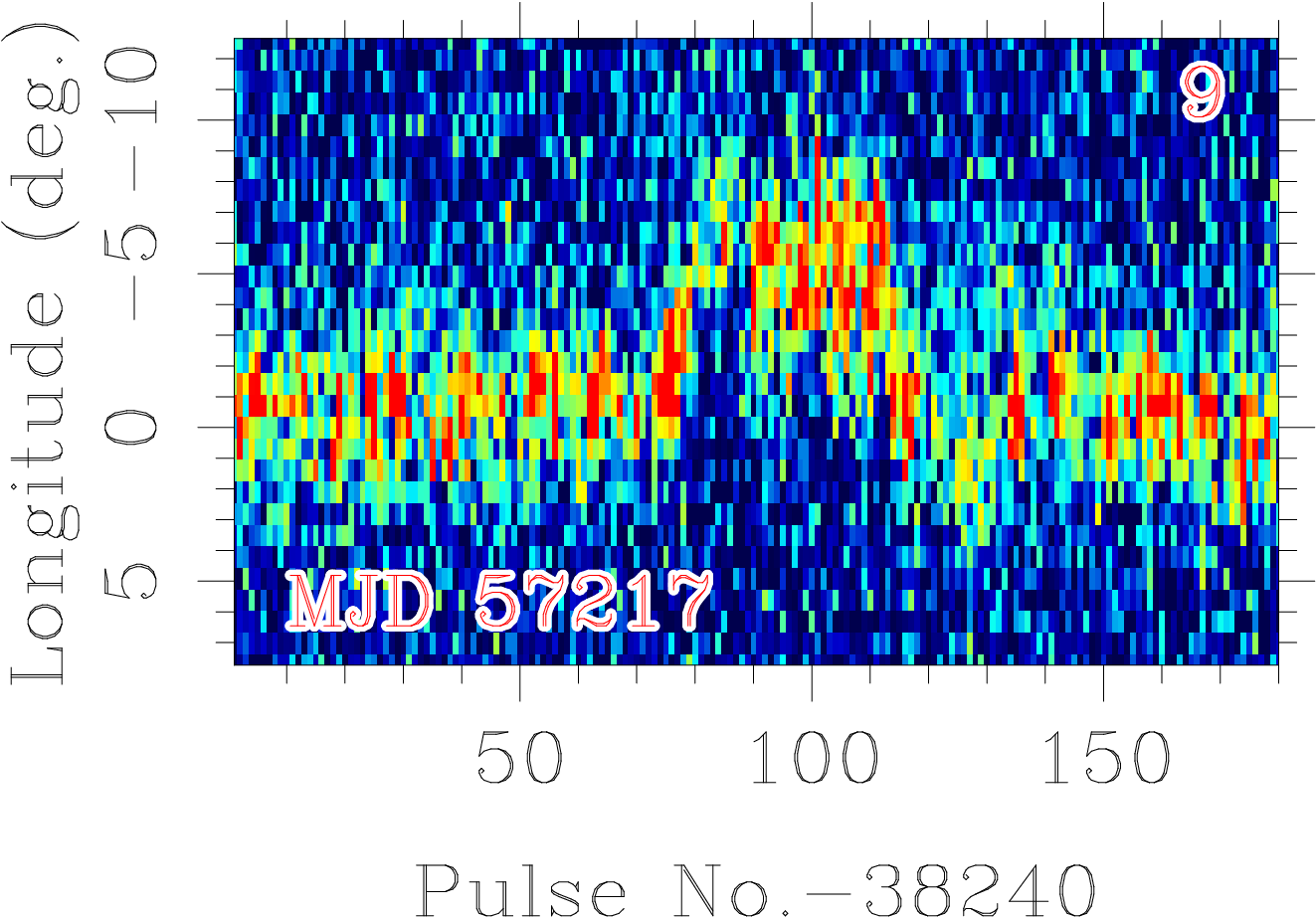}
 \includegraphics[bb = 60 95 505 685,clip,width=0.114\textwidth,angle=-90]{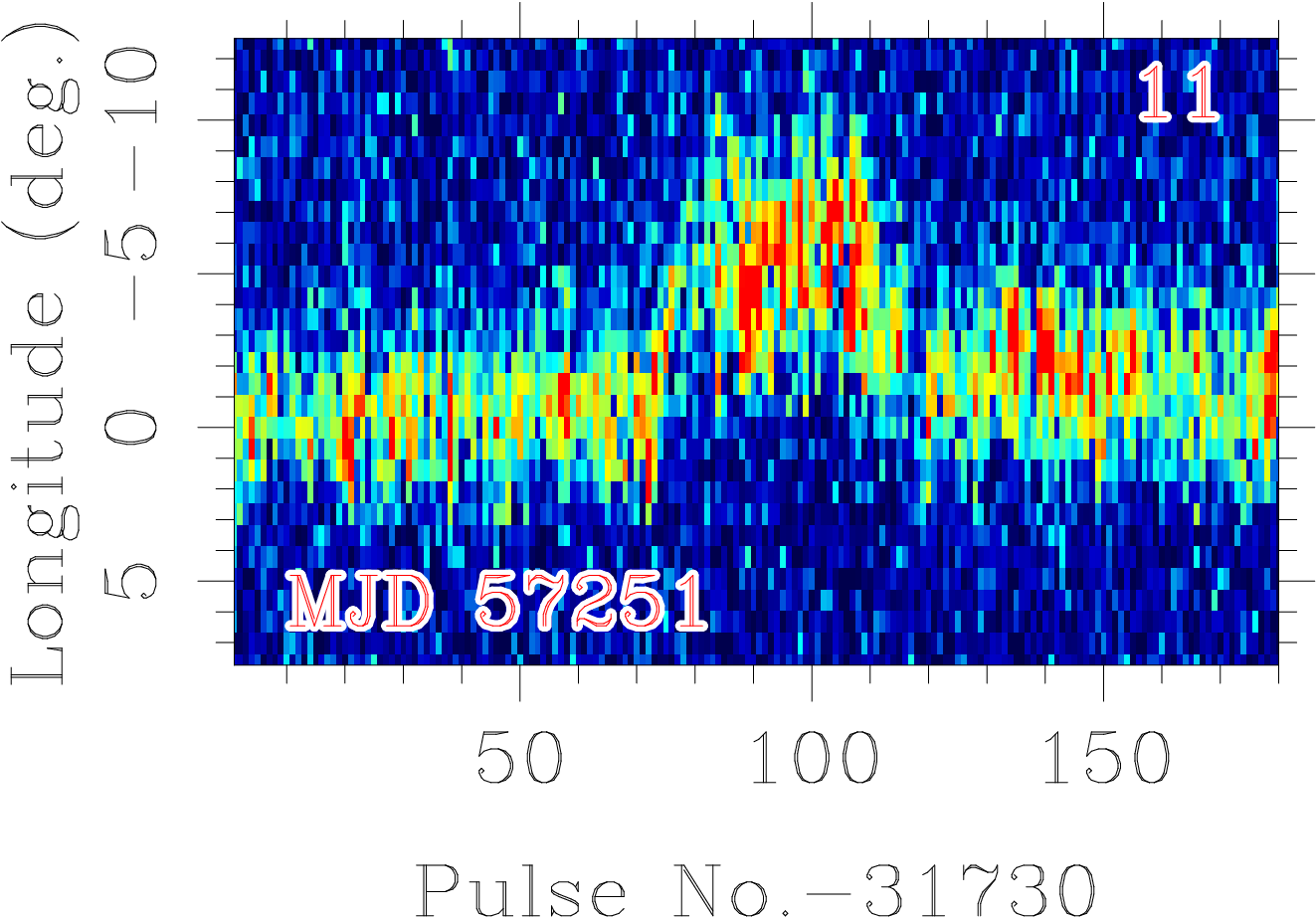}
 \includegraphics[bb = 60 45 505 685,clip,width=0.114\textwidth,angle=-90]{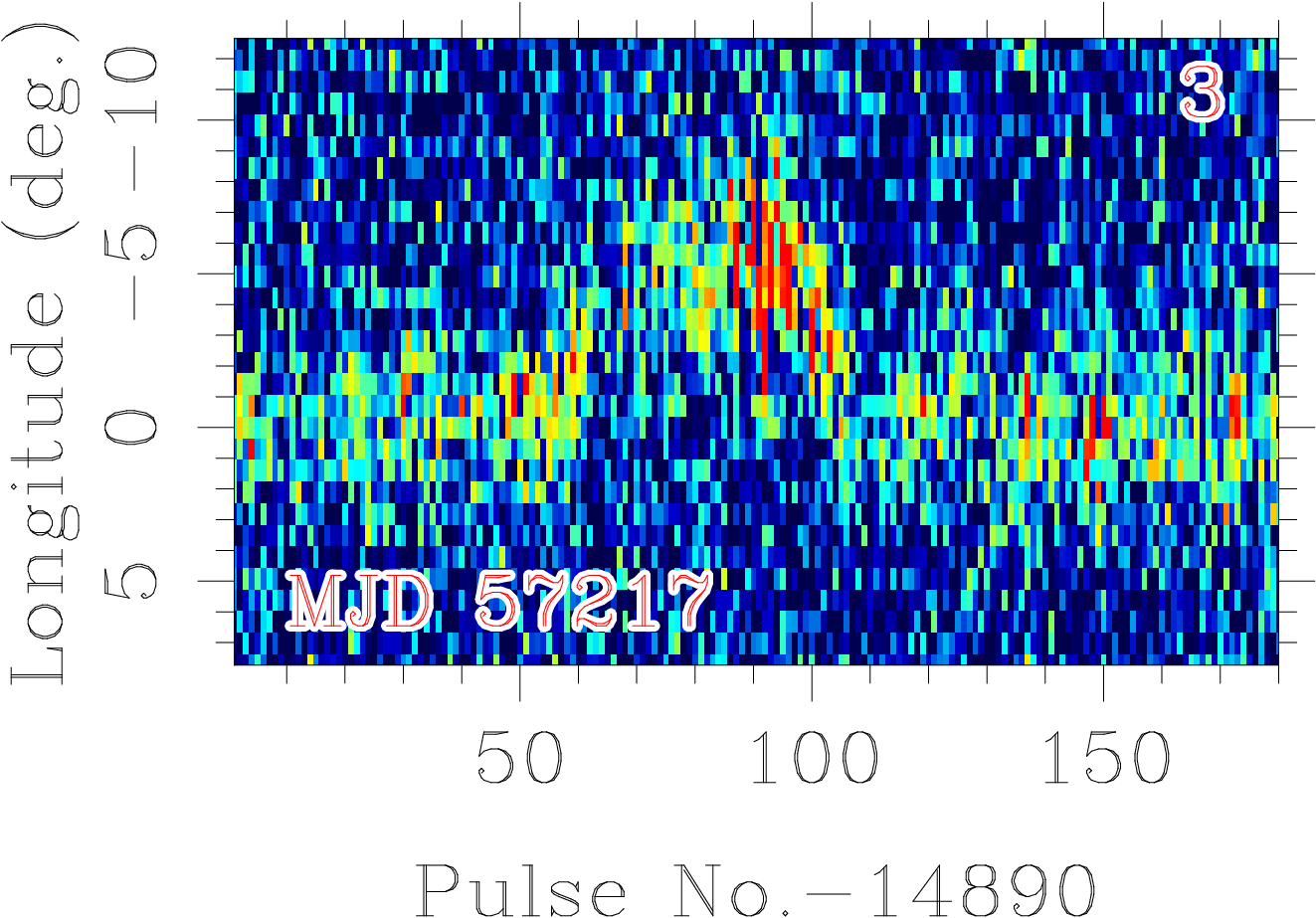}
 \includegraphics[bb = 60 95 505 685,clip,width=0.114\textwidth,angle=-90]{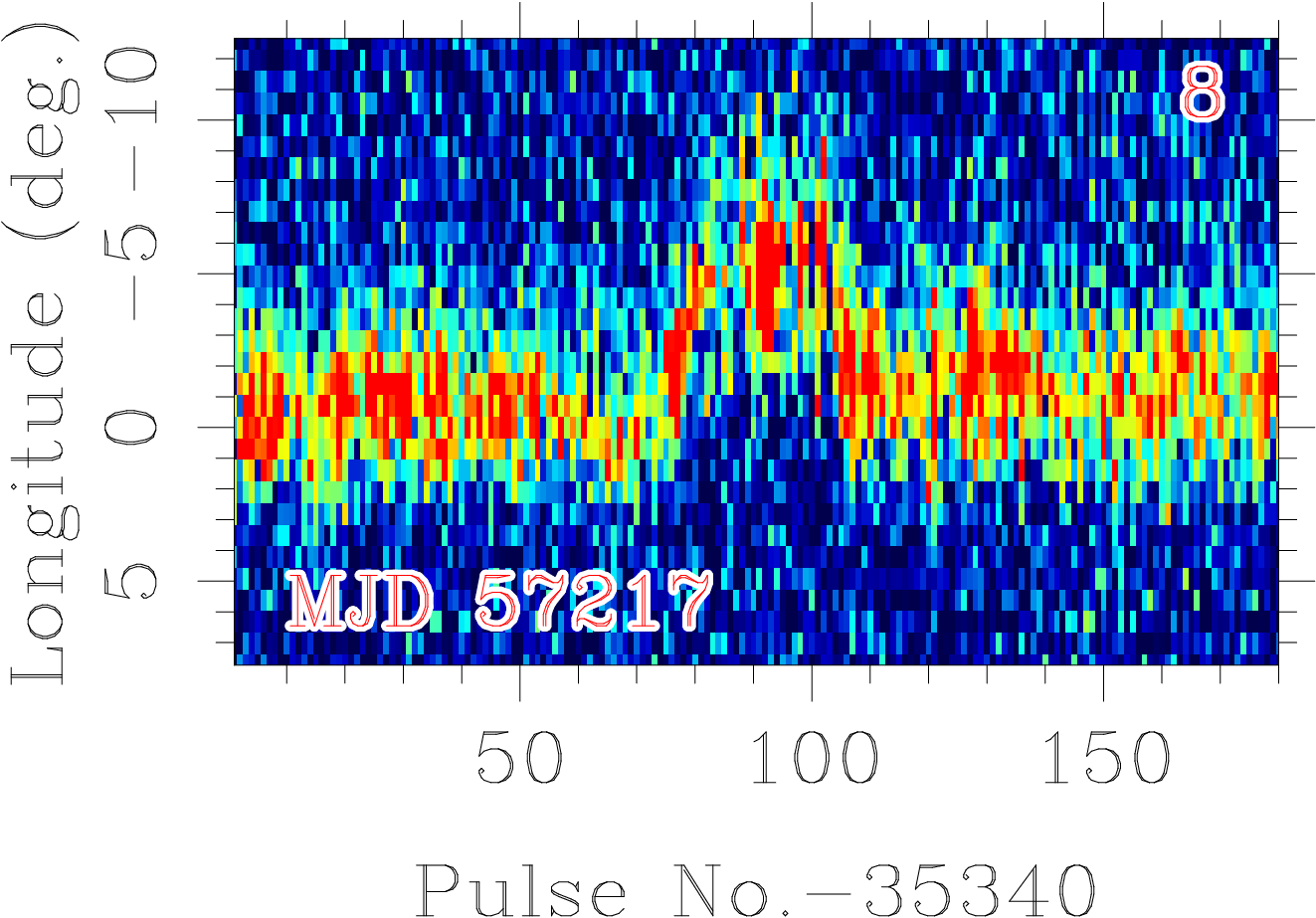}
 \includegraphics[bb = 60 95 505 685,clip,width=0.114\textwidth,angle=-90]{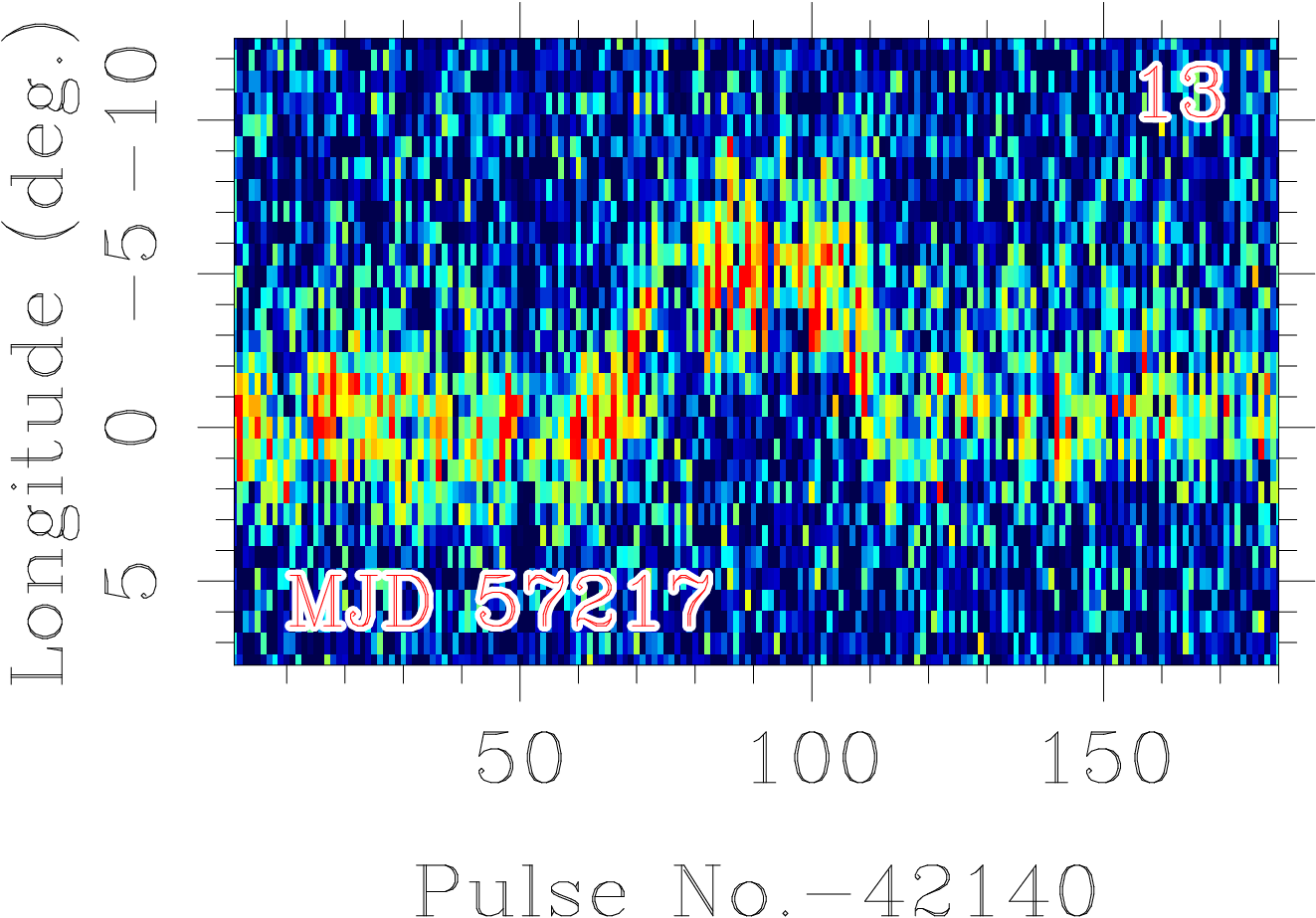}
 \includegraphics[bb = 60 45 505 685,clip,width=0.114\textwidth,angle=-90]{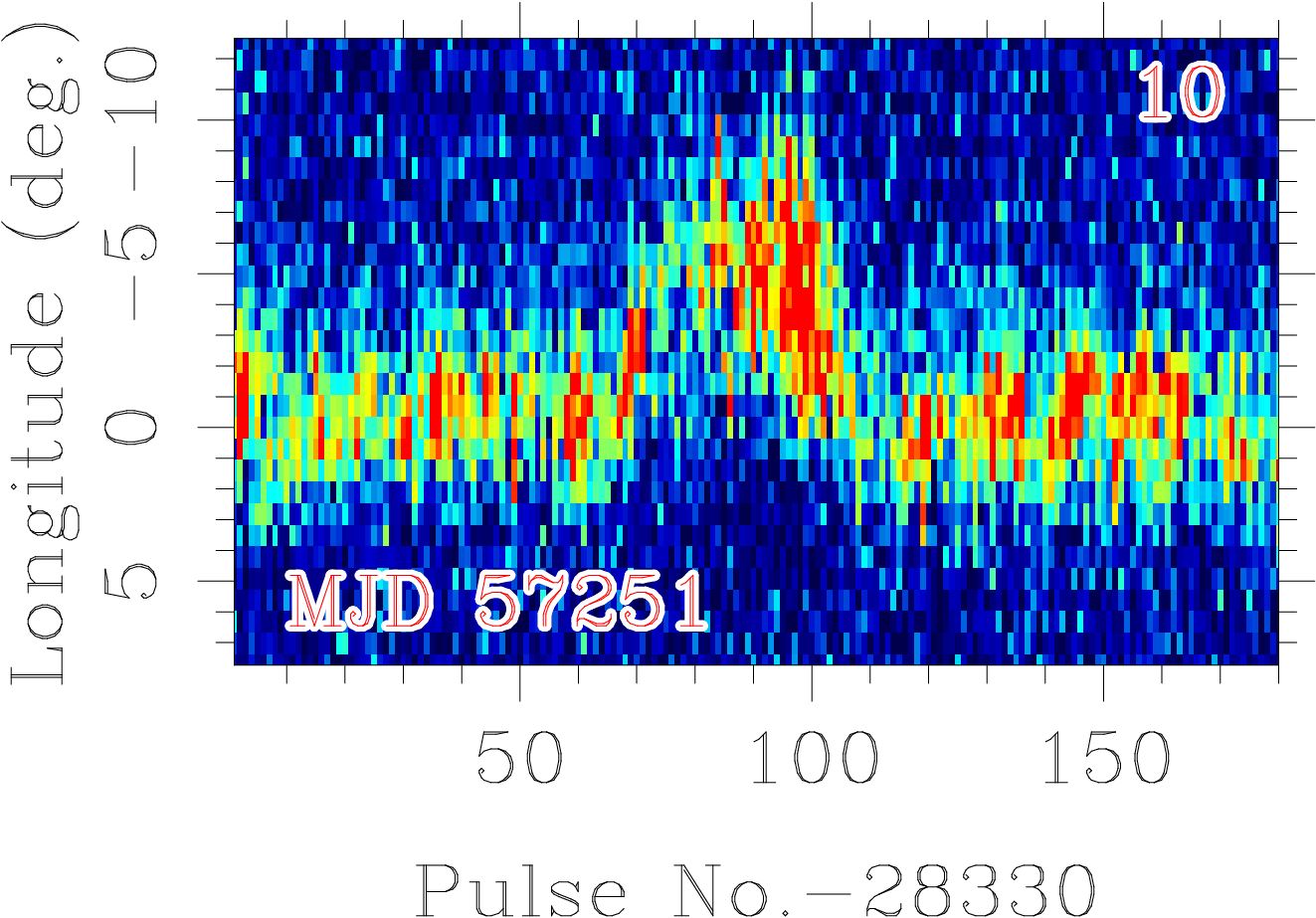}
 \includegraphics[bb = 60 95 505 685,clip,width=0.114\textwidth,angle=-90]{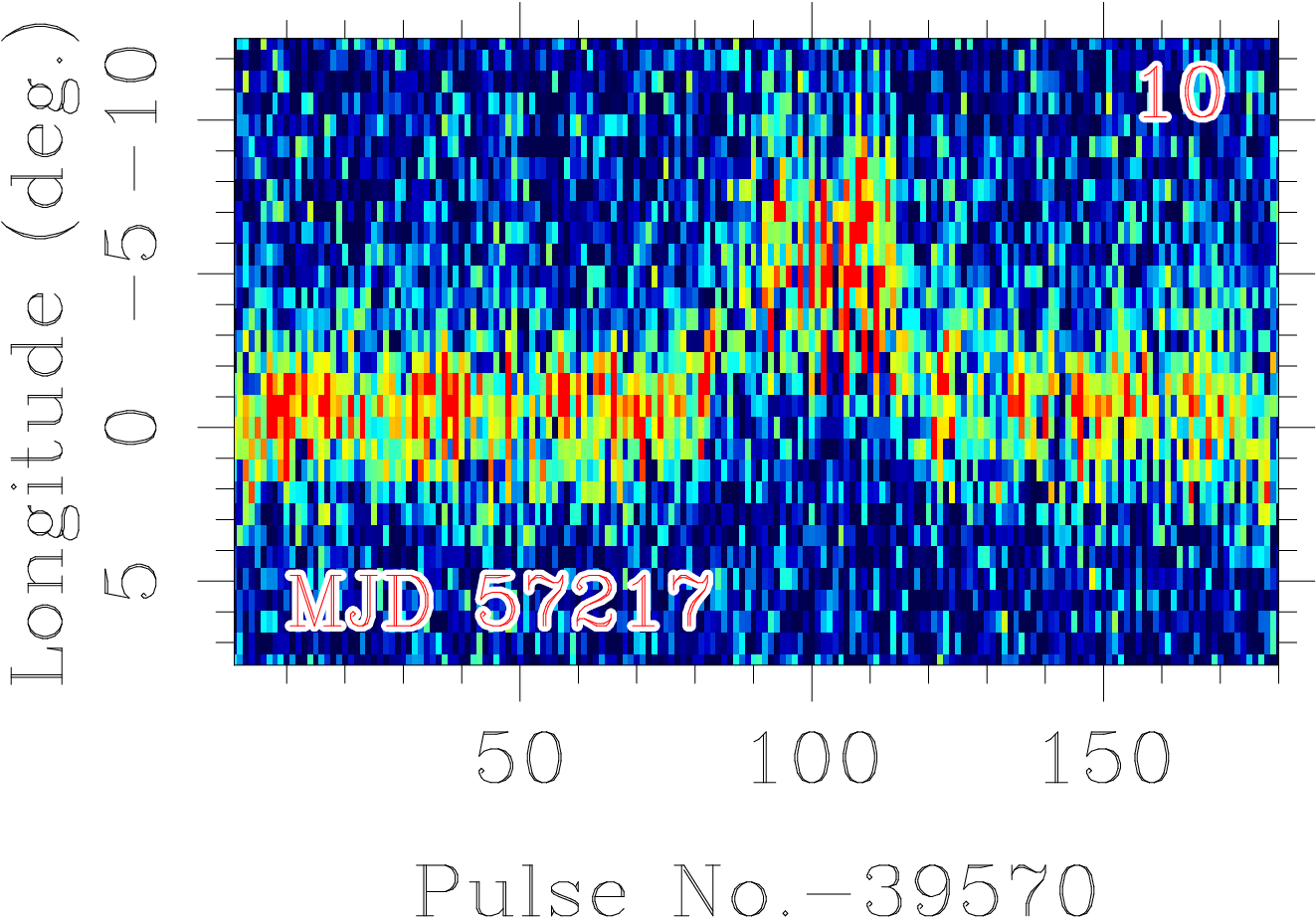}
 \includegraphics[bb = 60 95 505 685,clip,width=0.114\textwidth,angle=-90]{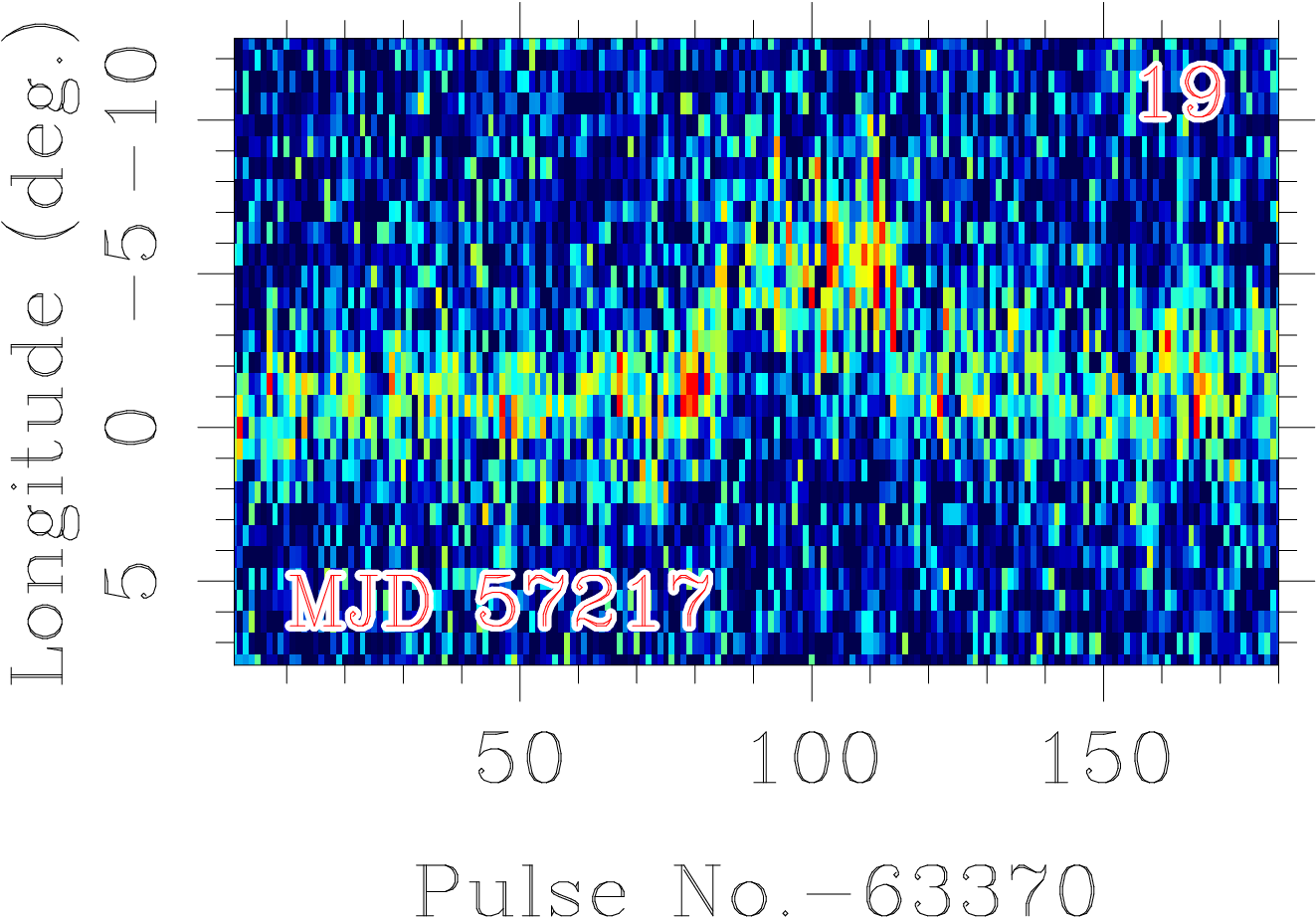}
 \includegraphics[bb = 60 45 505 685,clip,width=0.114\textwidth,angle=-90]{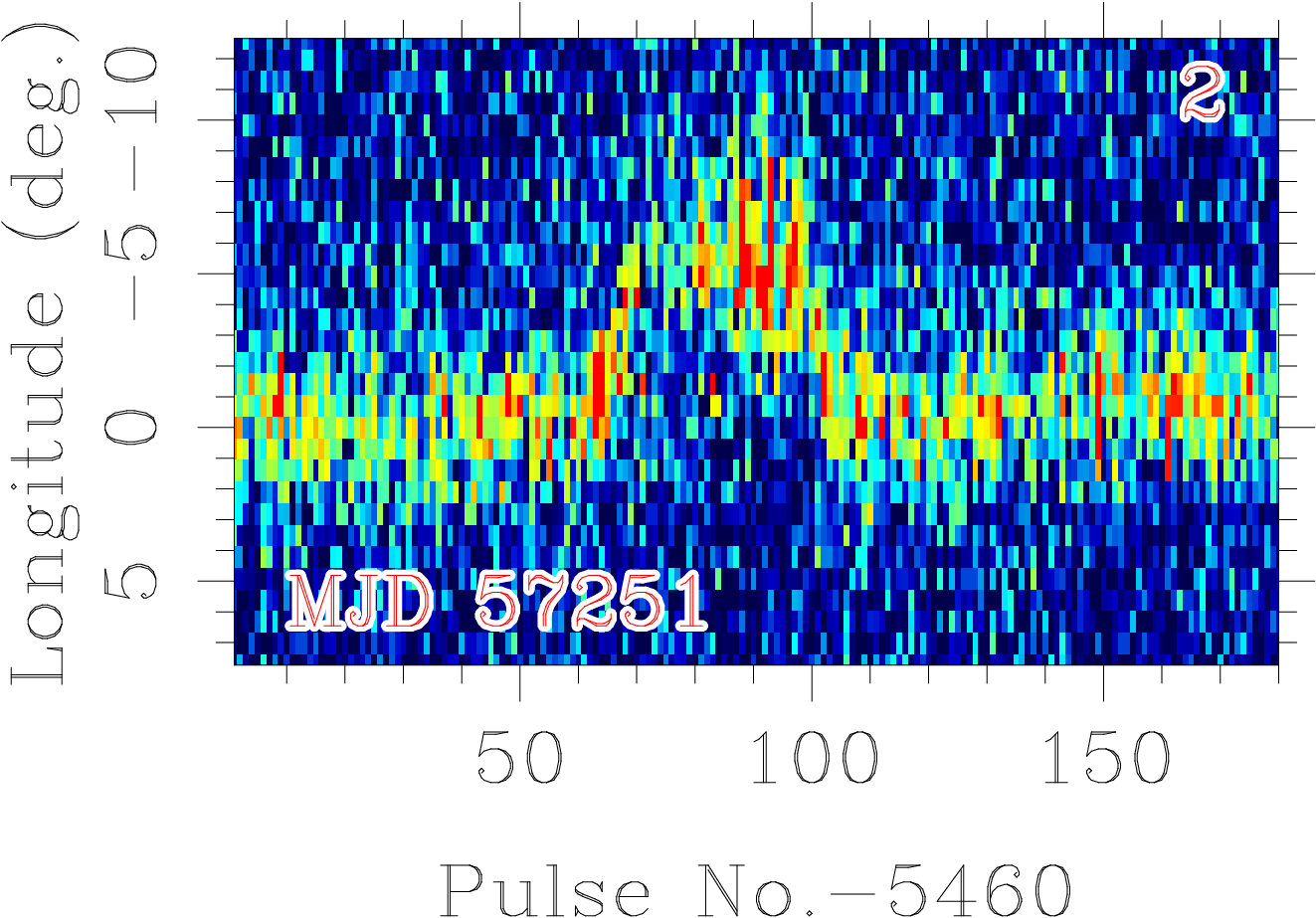}
 \includegraphics[bb = 60 95 505 685,clip,width=0.114\textwidth,angle=-90]{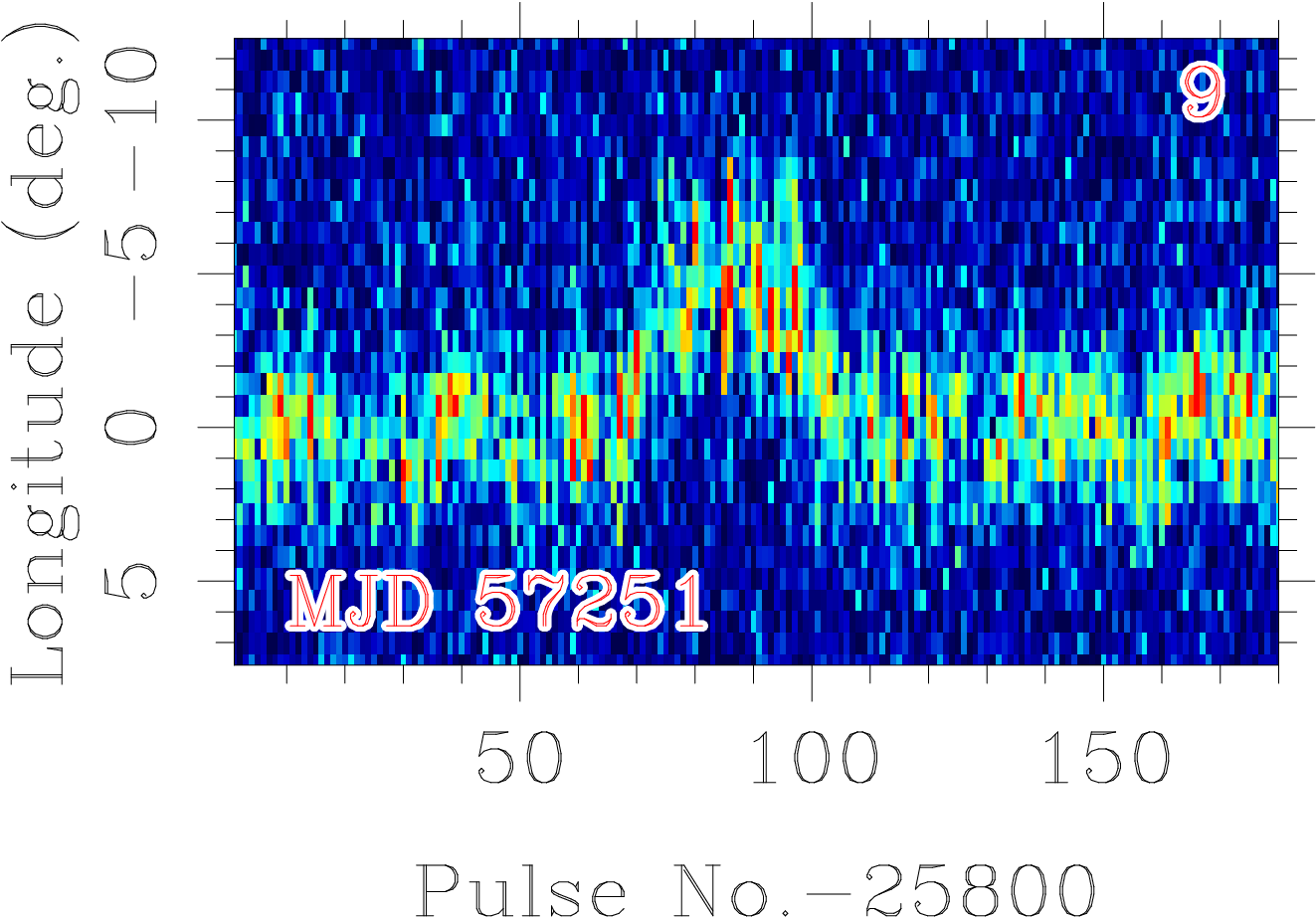}
 \includegraphics[bb = 60 95 505 685,clip,width=0.114\textwidth,angle=-90]{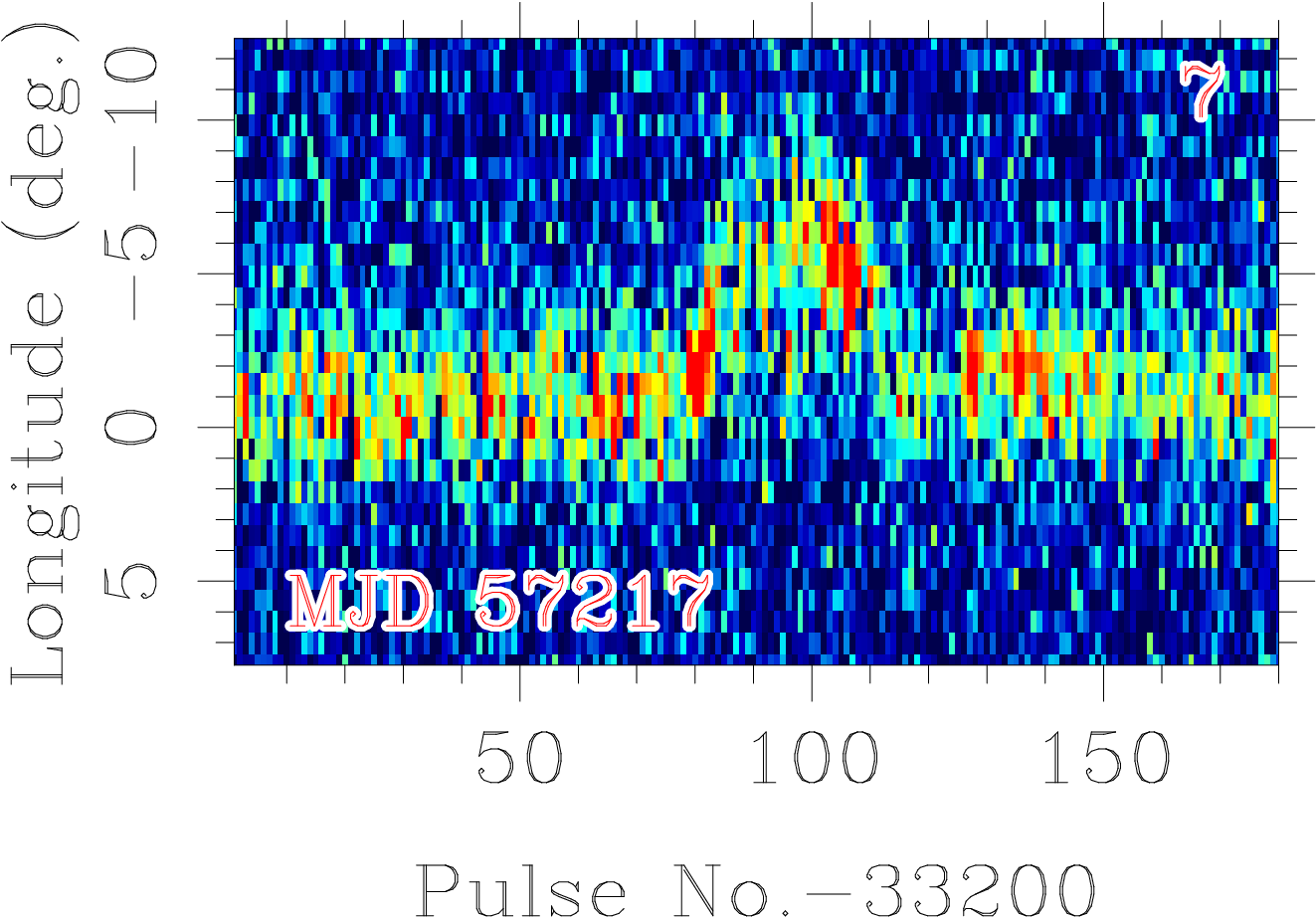}
 \includegraphics[bb = 60 45 505 685,clip,width=0.114\textwidth,angle=-90]{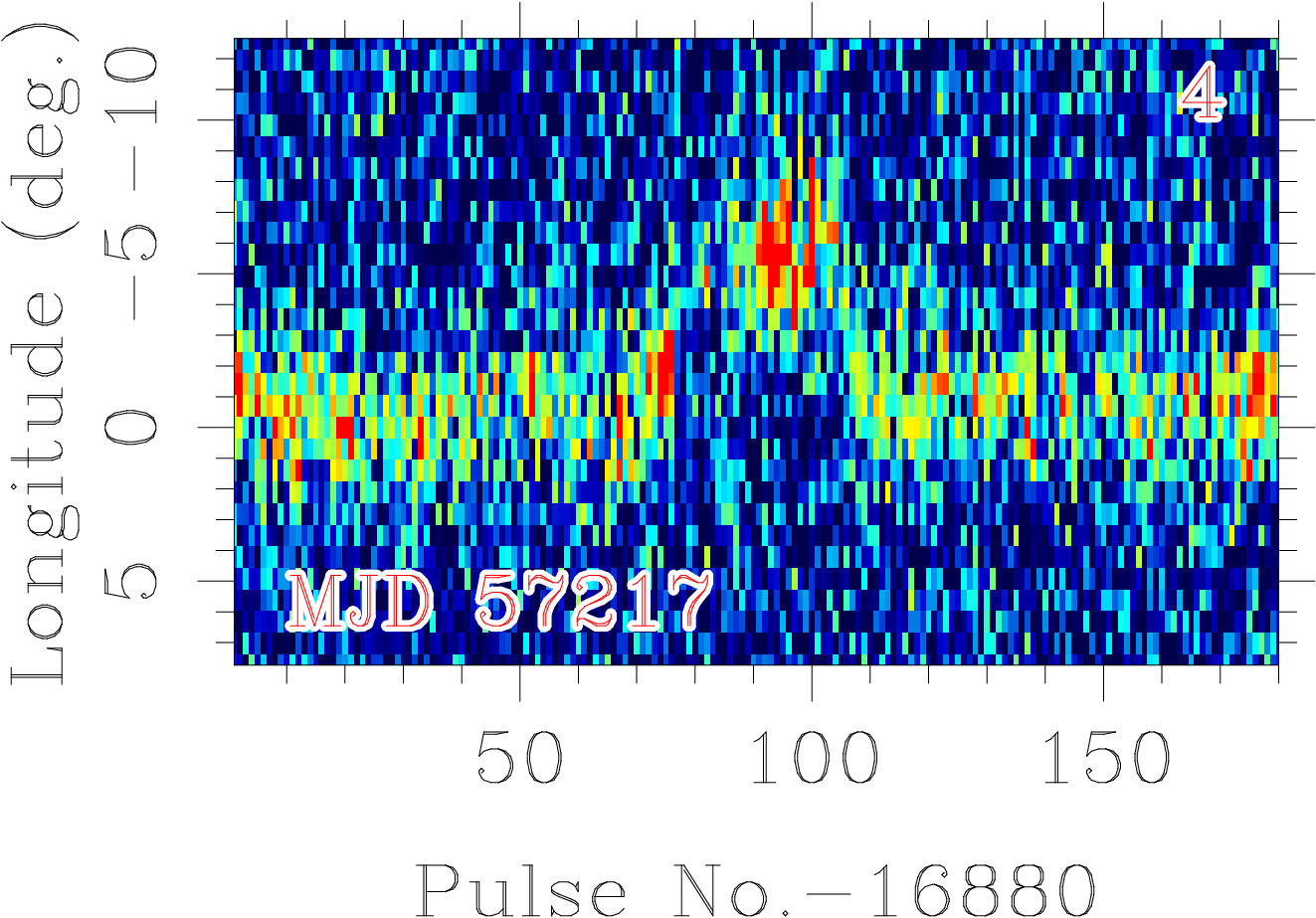}
 \includegraphics[bb = 60 95 505 685,clip,width=0.114\textwidth,angle=-90]{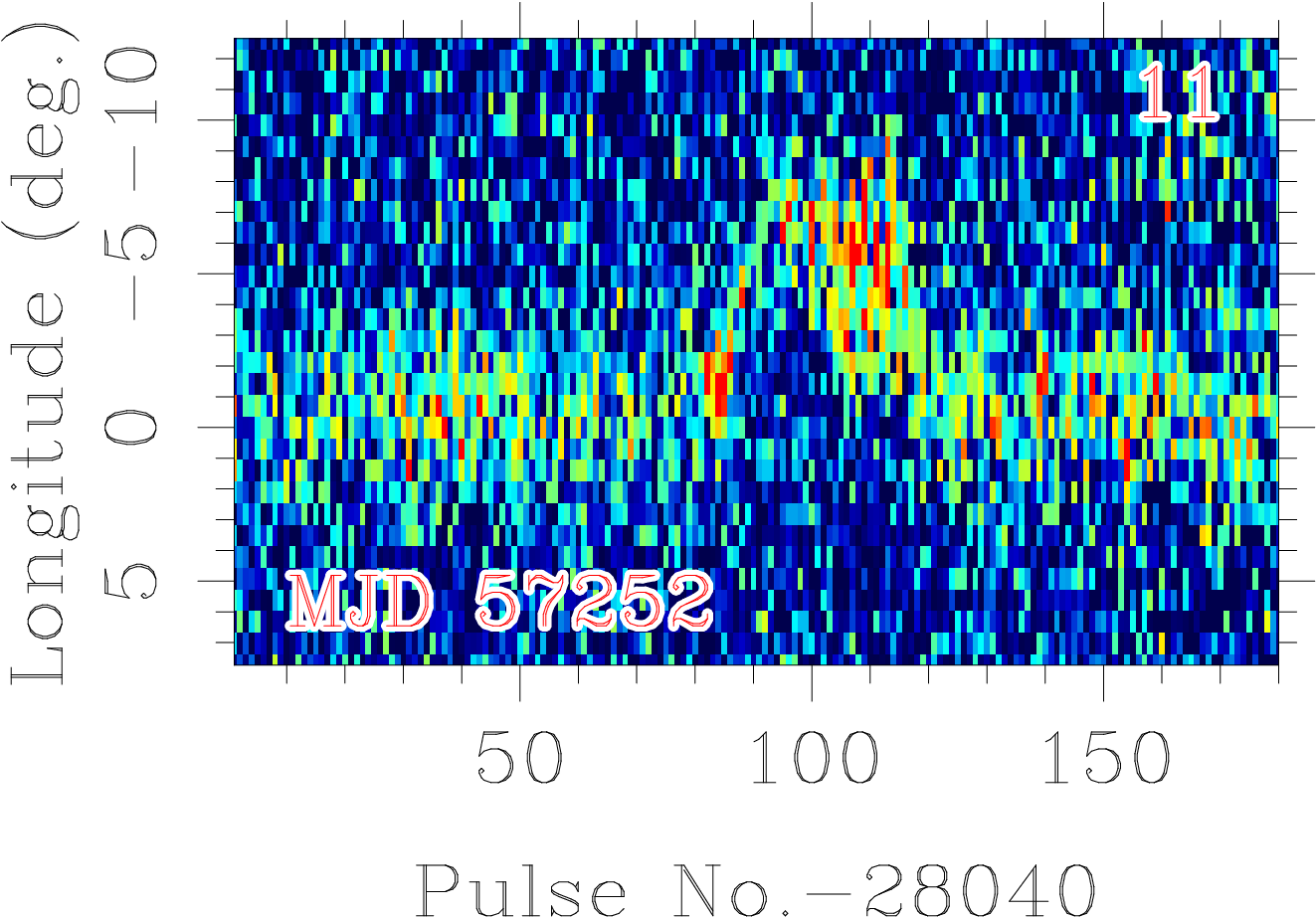} 
 \includegraphics[bb = 60 95 505 685,clip,width=0.114\textwidth,angle=-90]{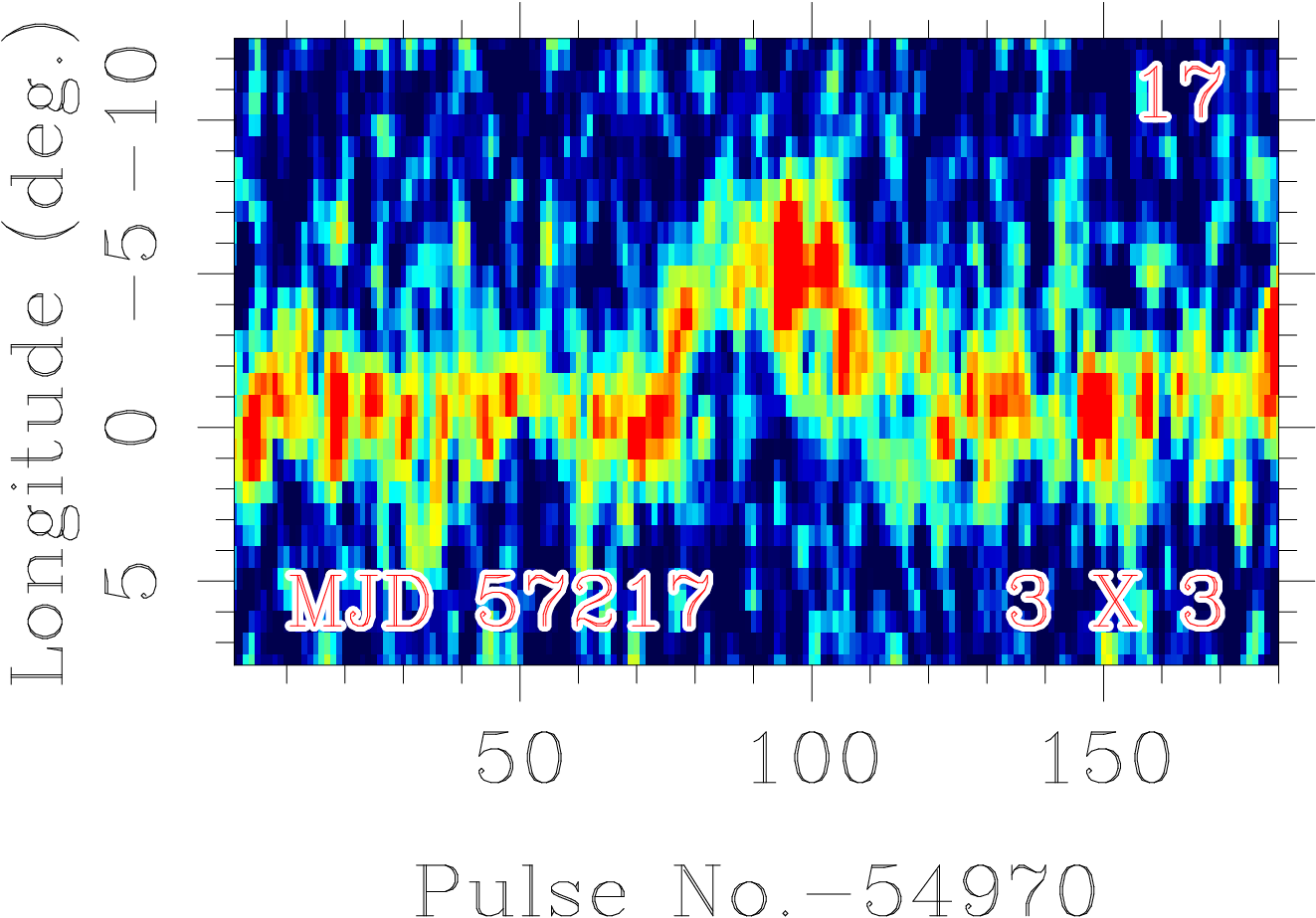}
 \includegraphics[bb = 60 45 505 685,clip,width=0.114\textwidth,angle=-90]{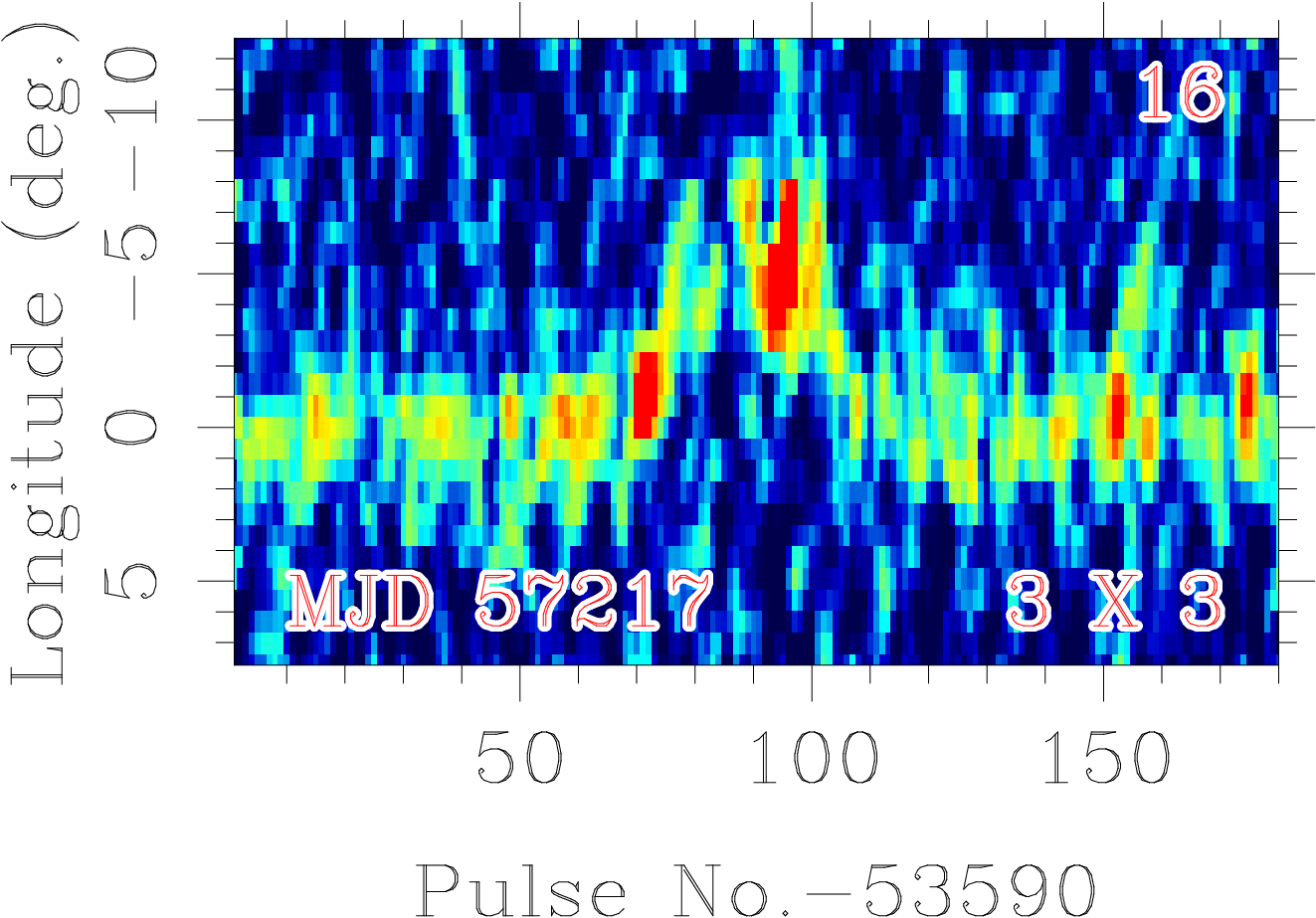}
 \includegraphics[bb = 60 95 505 685,clip,width=0.114\textwidth,angle=-90]{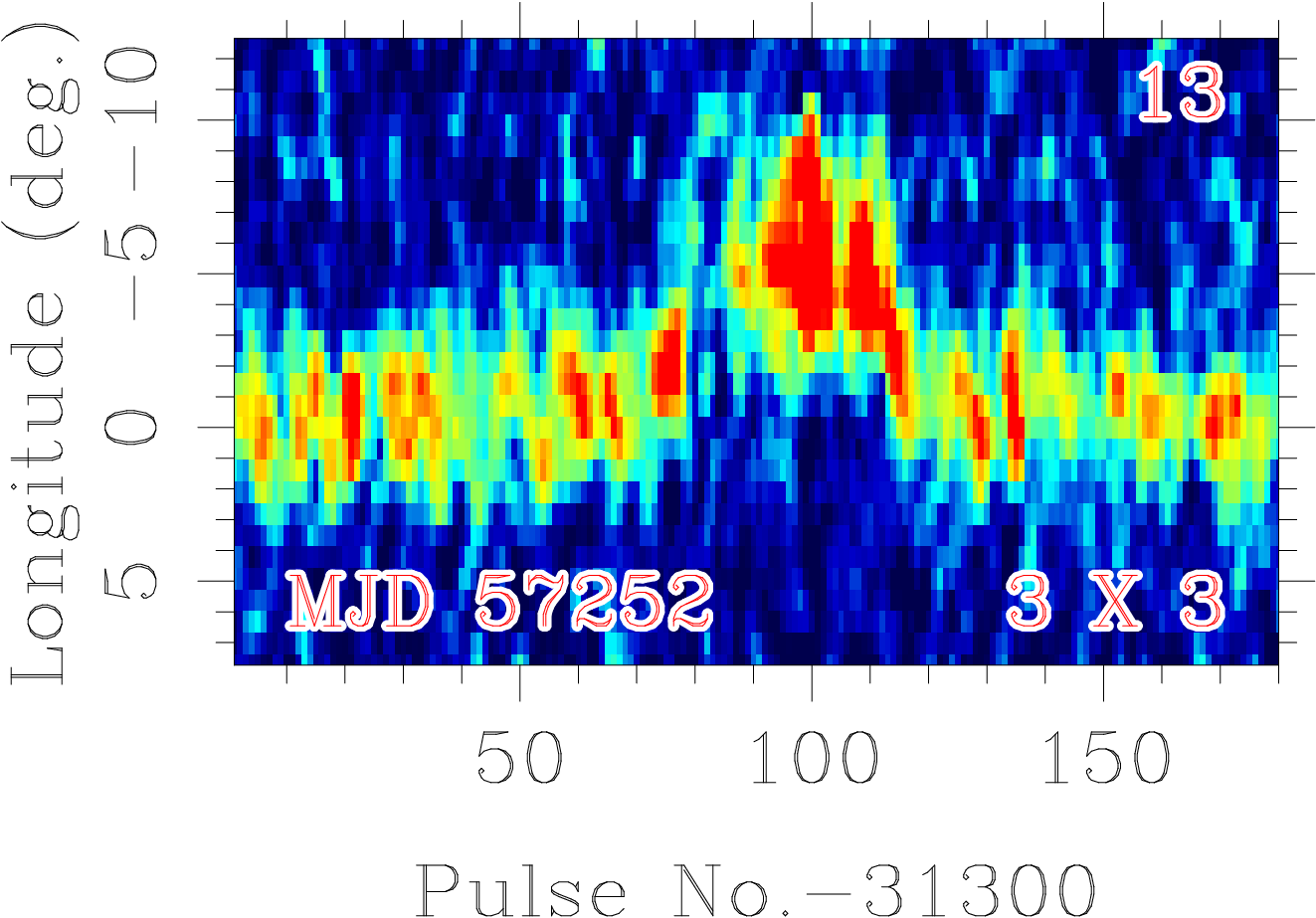} 
 \includegraphics[bb = 60 95 505 685,clip,width=0.114\textwidth,angle=-90]{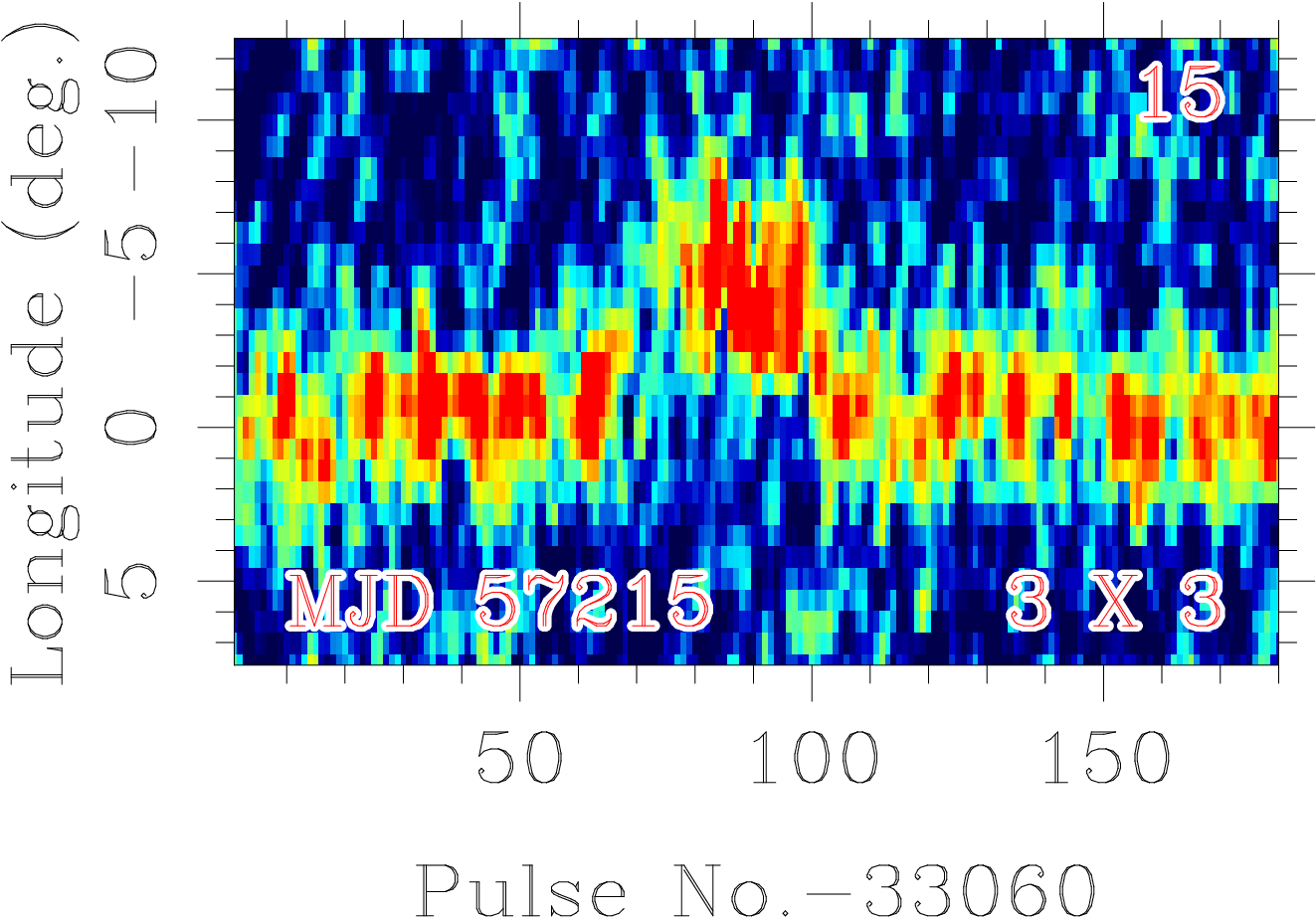}
 \includegraphics[bb = 60 45 505 685,clip,width=0.114\textwidth,angle=-90]{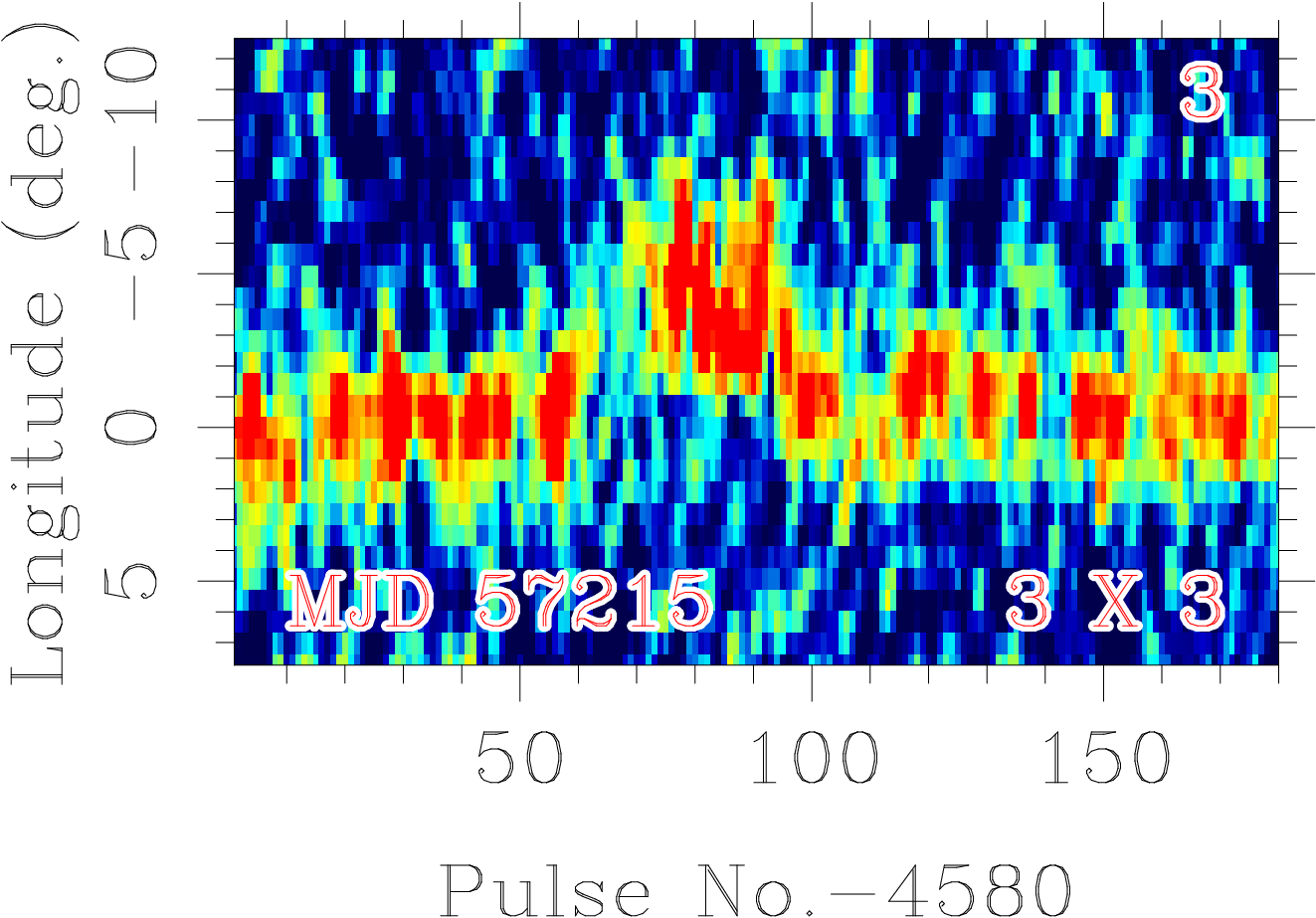}
 \includegraphics[bb = 60 95 505 685,clip,width=0.114\textwidth,angle=-90]{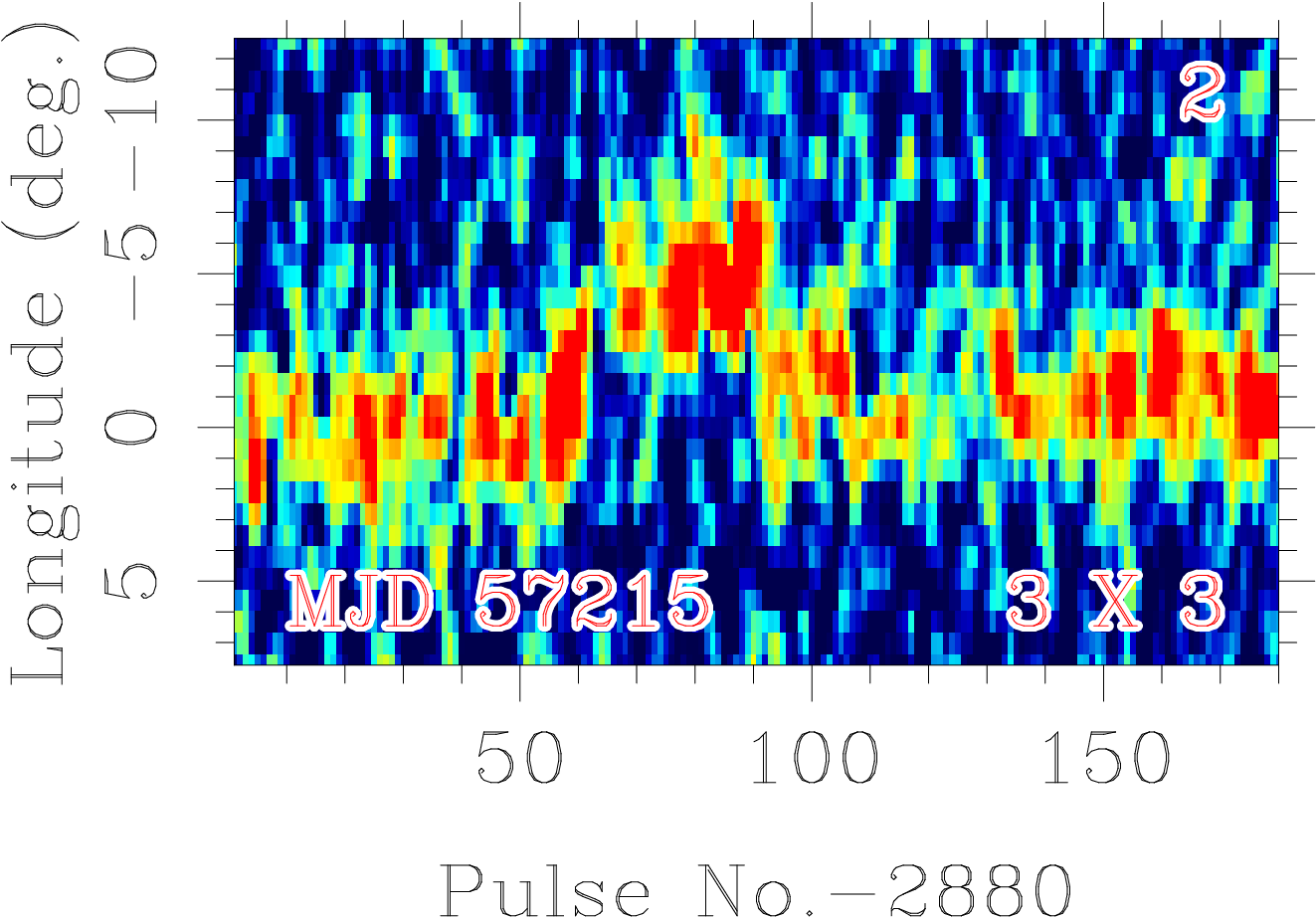}
 \includegraphics[bb = 60 95 505 685,clip,width=0.114\textwidth,angle=-90]{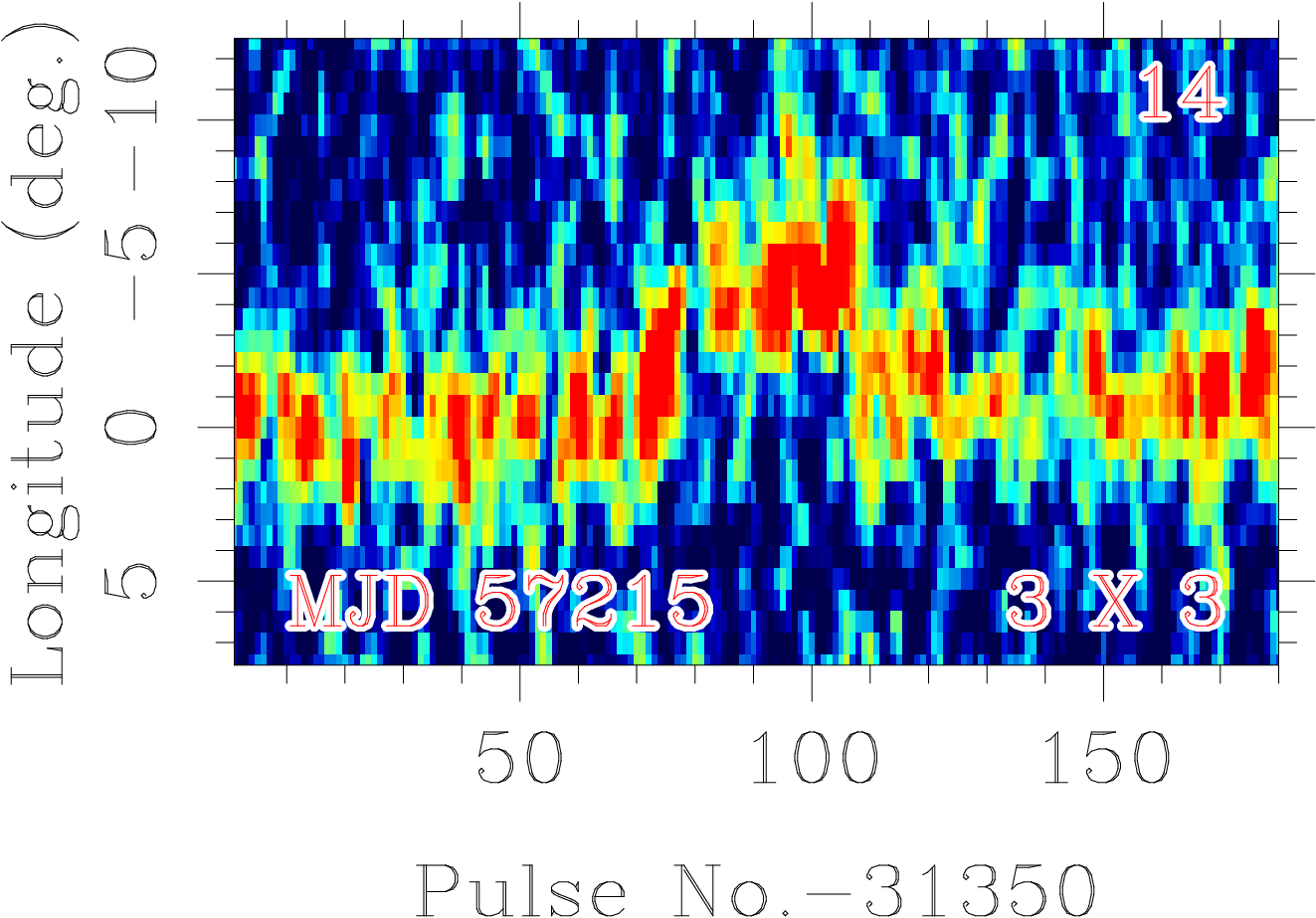}
 \includegraphics[bb = 60 45 505 685,clip,width=0.114\textwidth,angle=-90]{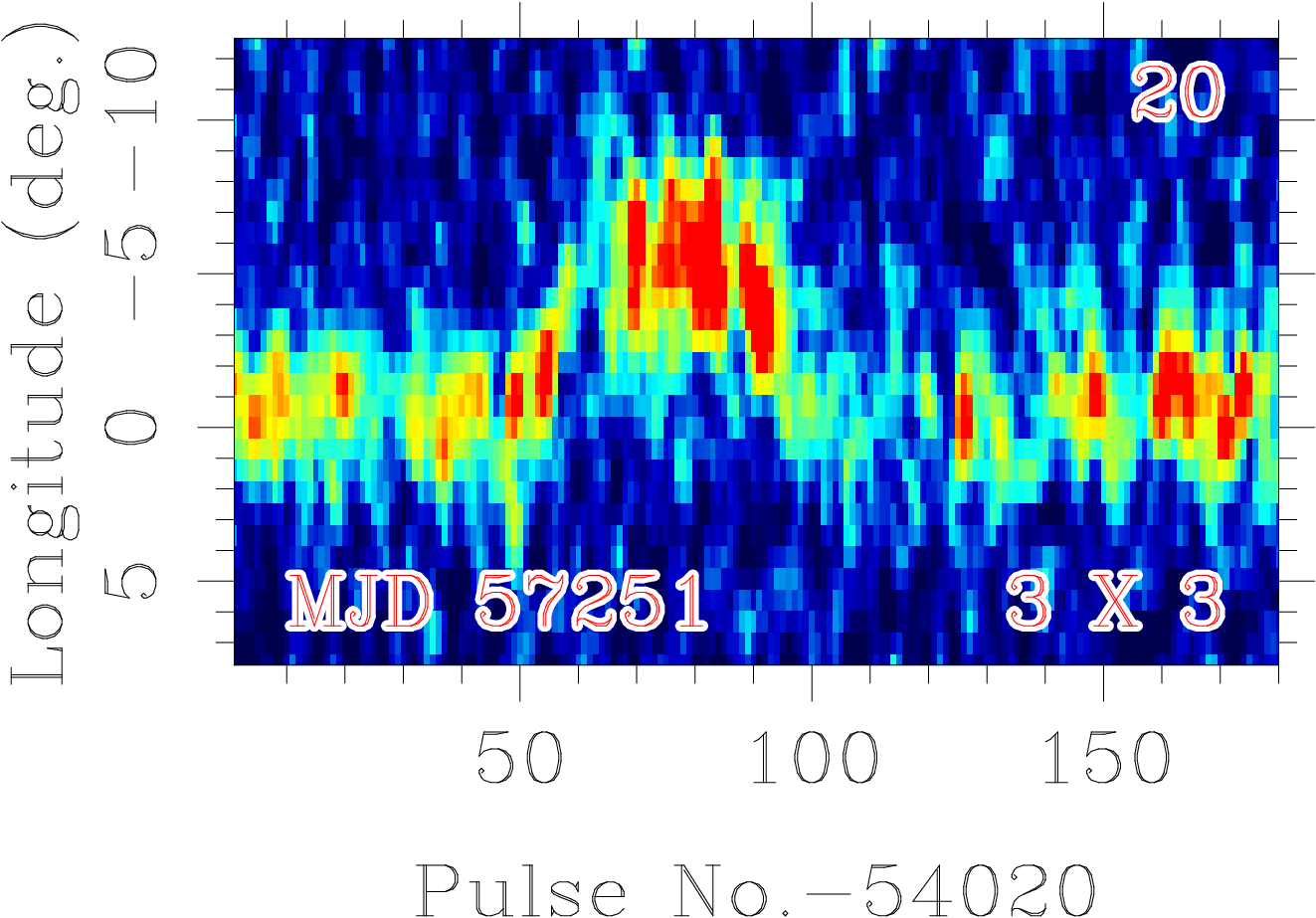}
 \includegraphics[bb = 60 95 505 685,clip,width=0.114\textwidth,angle=-90]{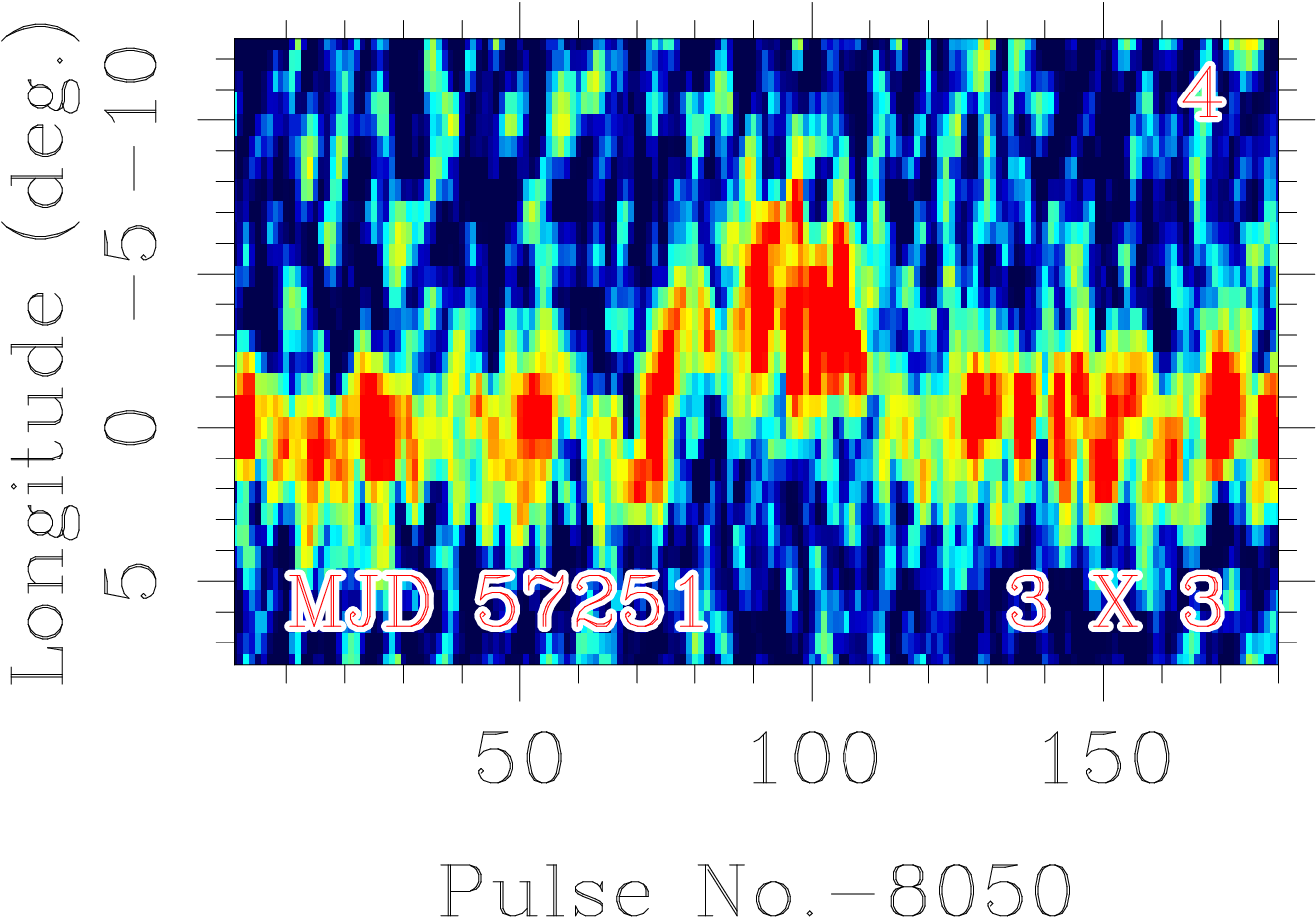}
 \includegraphics[bb = 60 95 505 685,clip,width=0.114\textwidth,angle=-90]{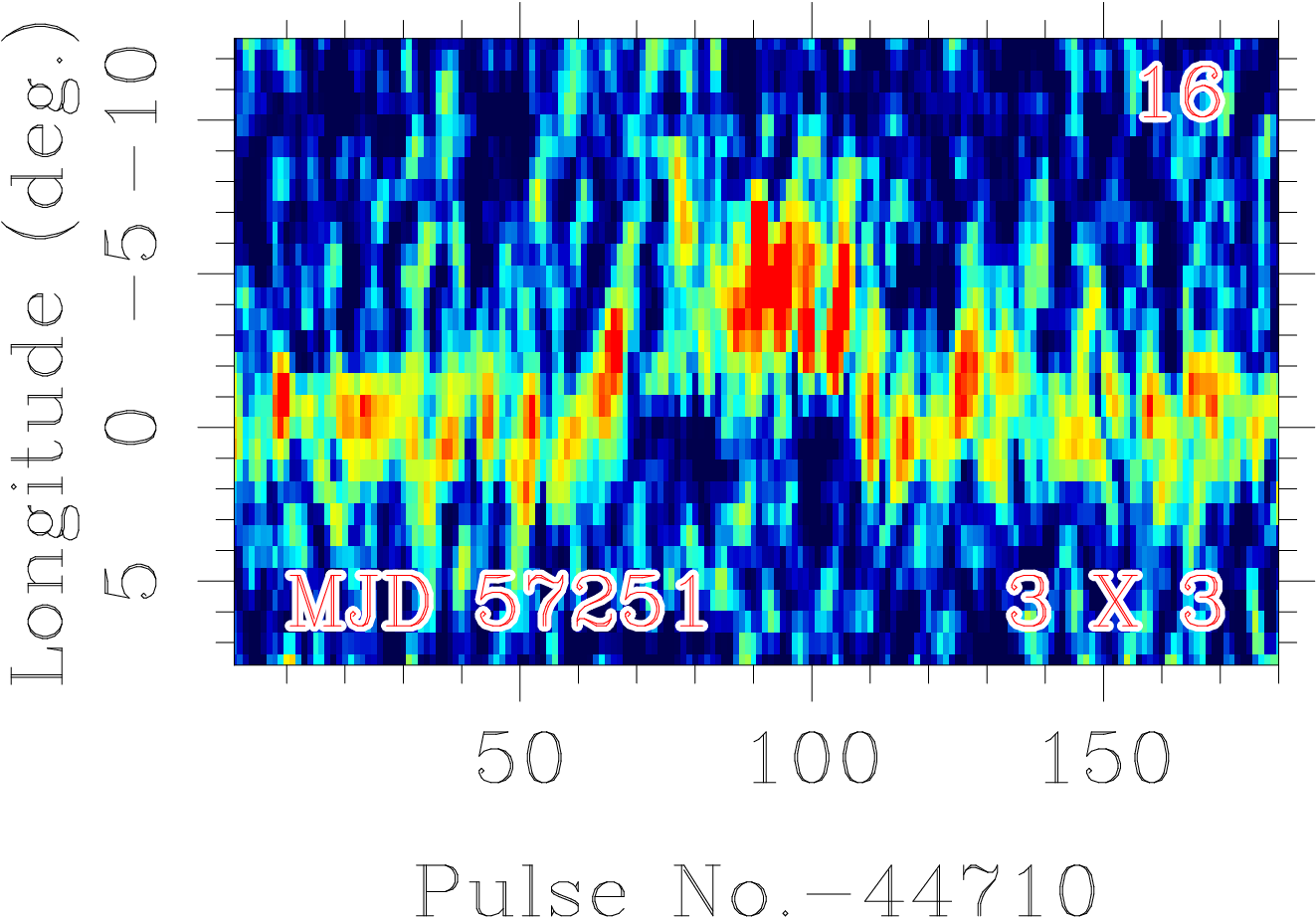}
 \includegraphics[bb = 60 45 505 685,clip,width=0.114\textwidth,angle=-90]{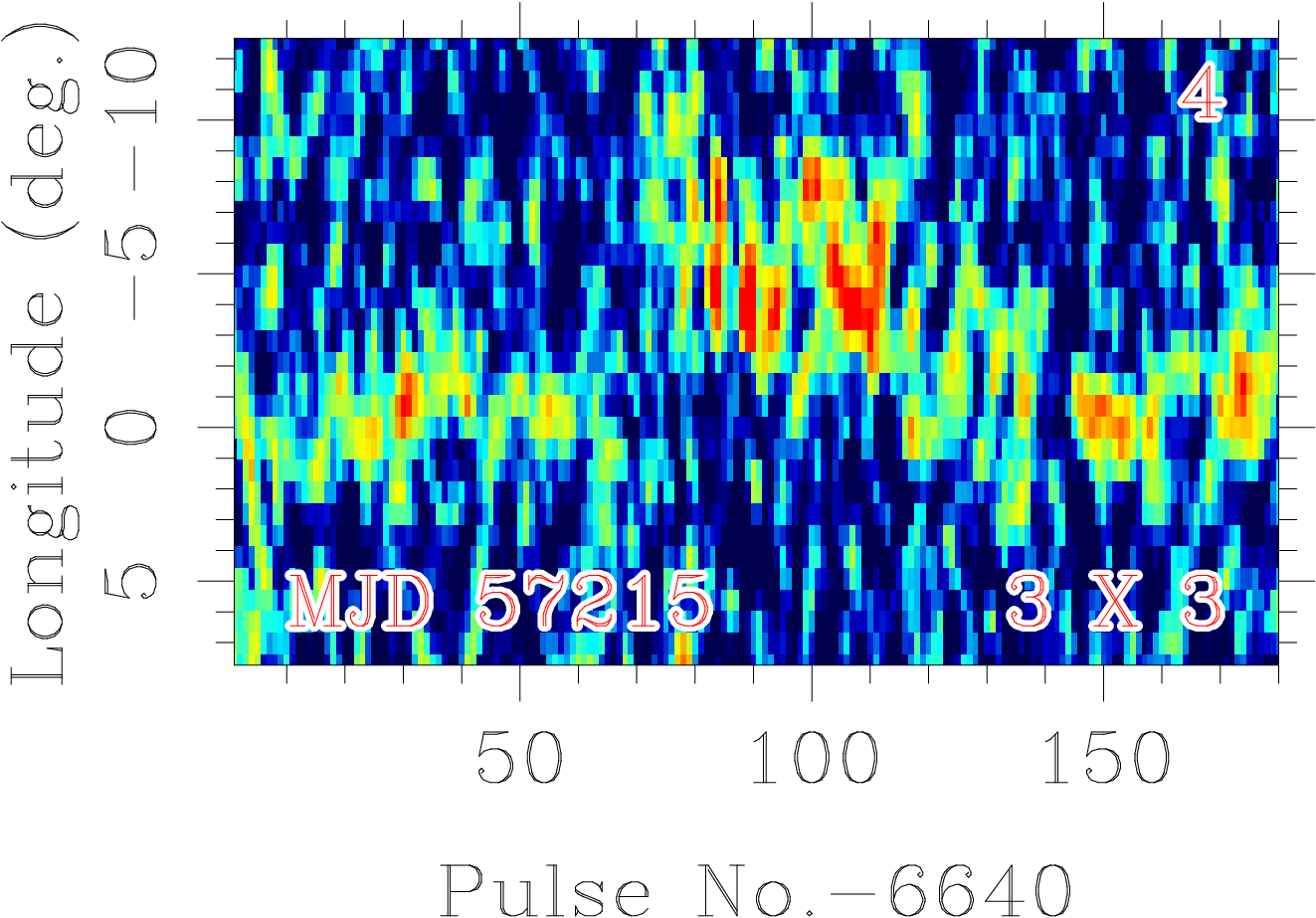}
 \includegraphics[bb = 60 95 505 685,clip,width=0.114\textwidth,angle=-90]{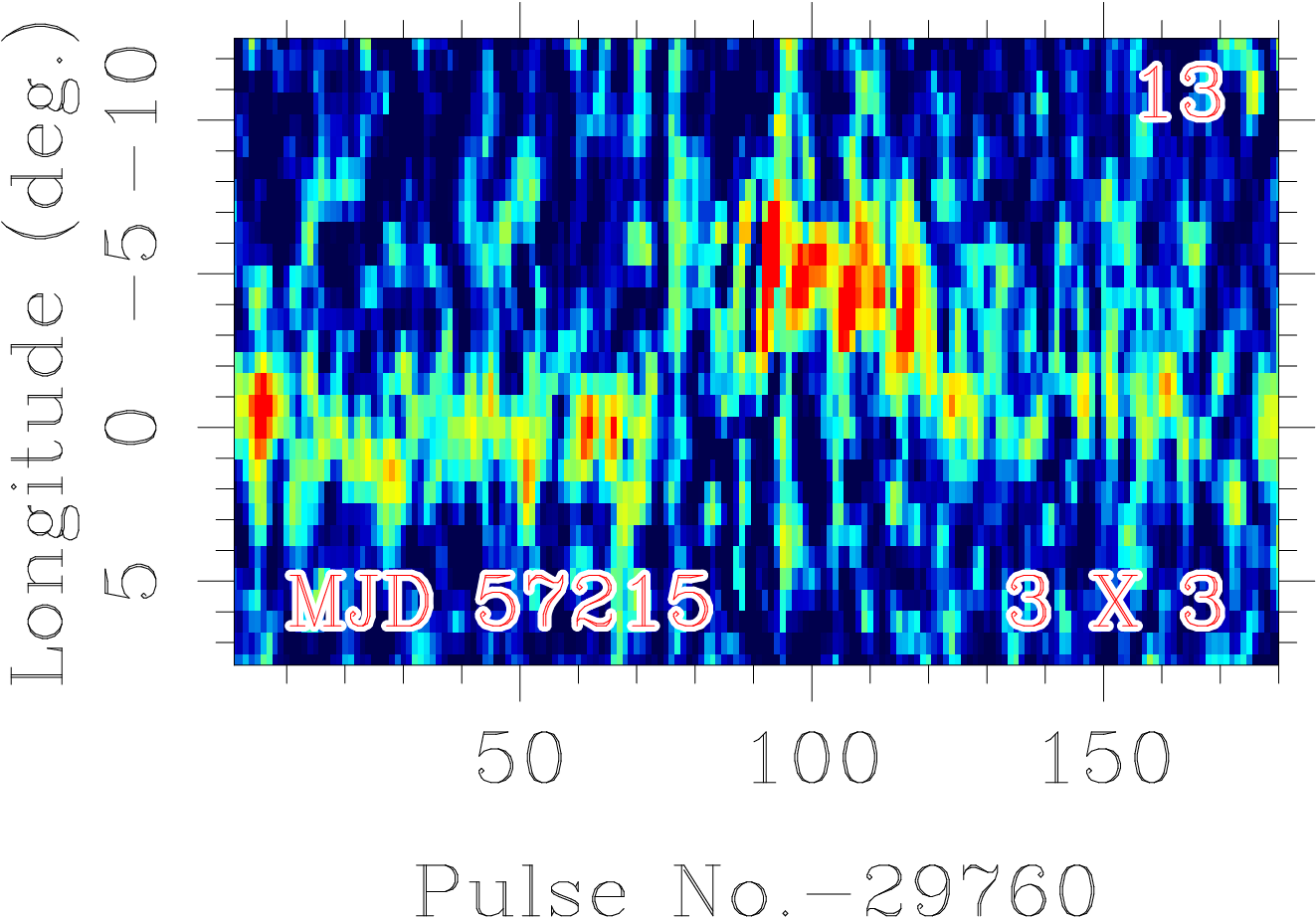}
 \includegraphics[bb = 60 95 505 685,clip,width=0.114\textwidth,angle=-90]{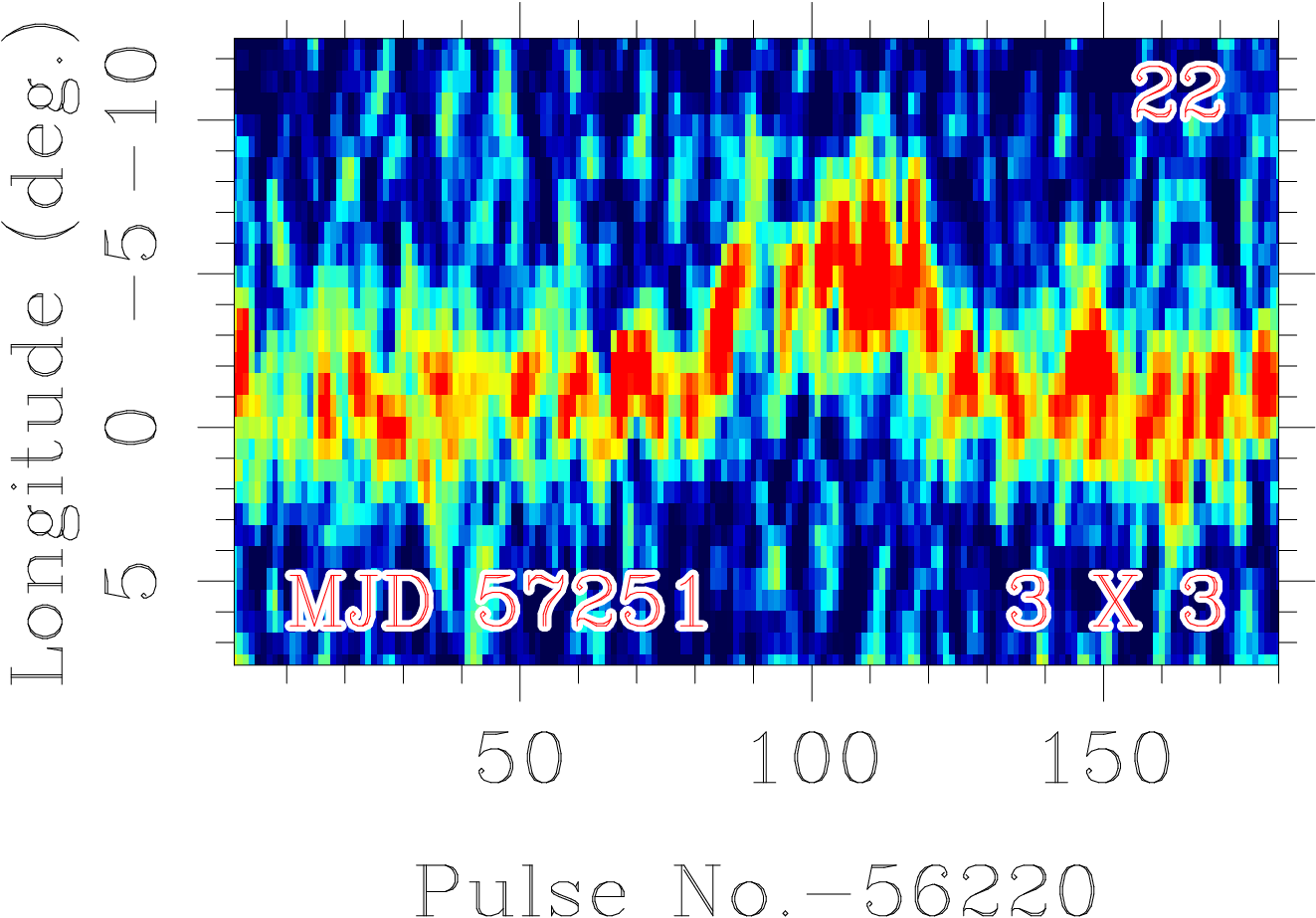}
 \includegraphics[bb = 60 45 505 685,clip,width=0.114\textwidth,angle=-90]{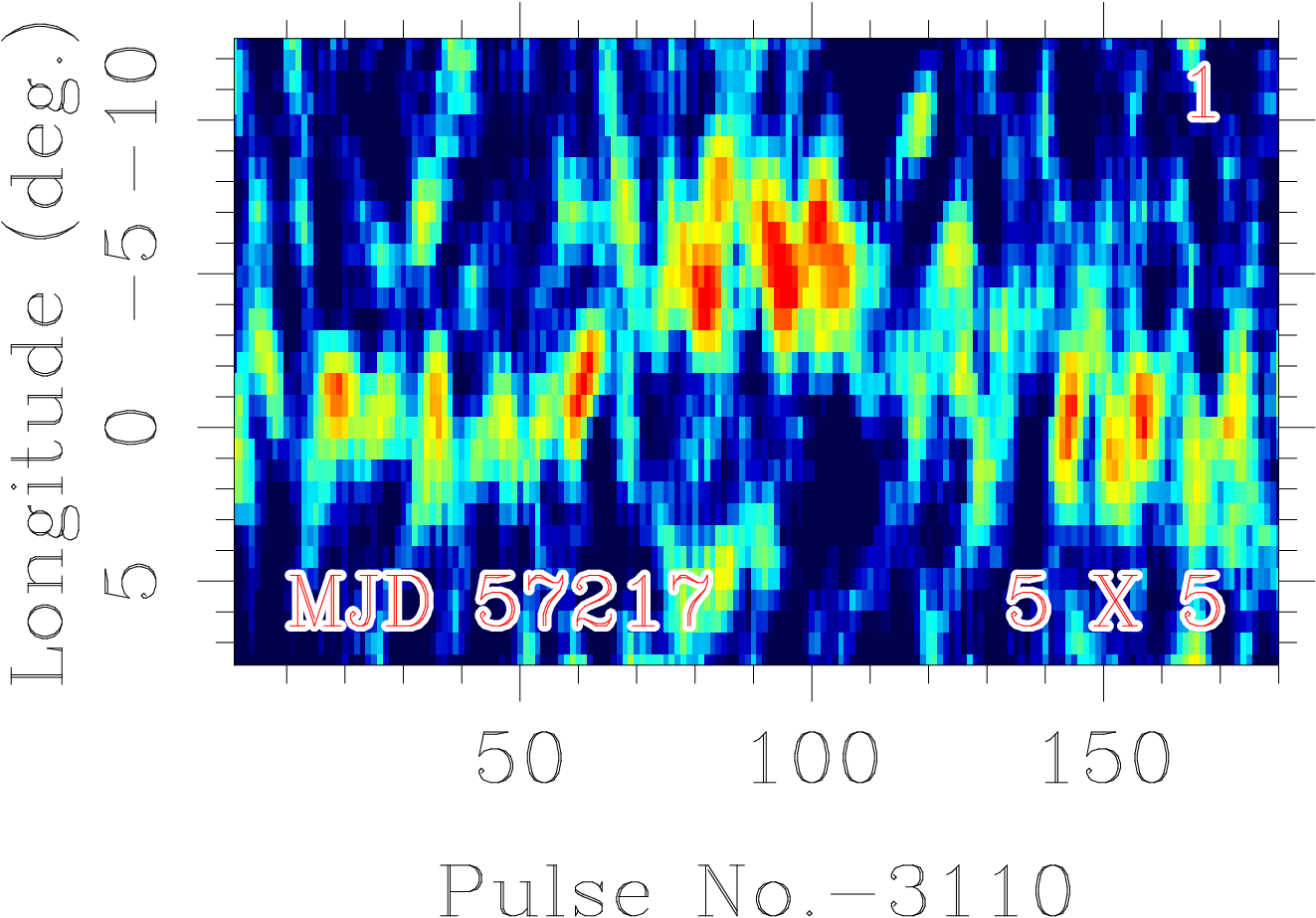}
 \includegraphics[bb = 60 95 505 685,clip,width=0.114\textwidth,angle=-90]{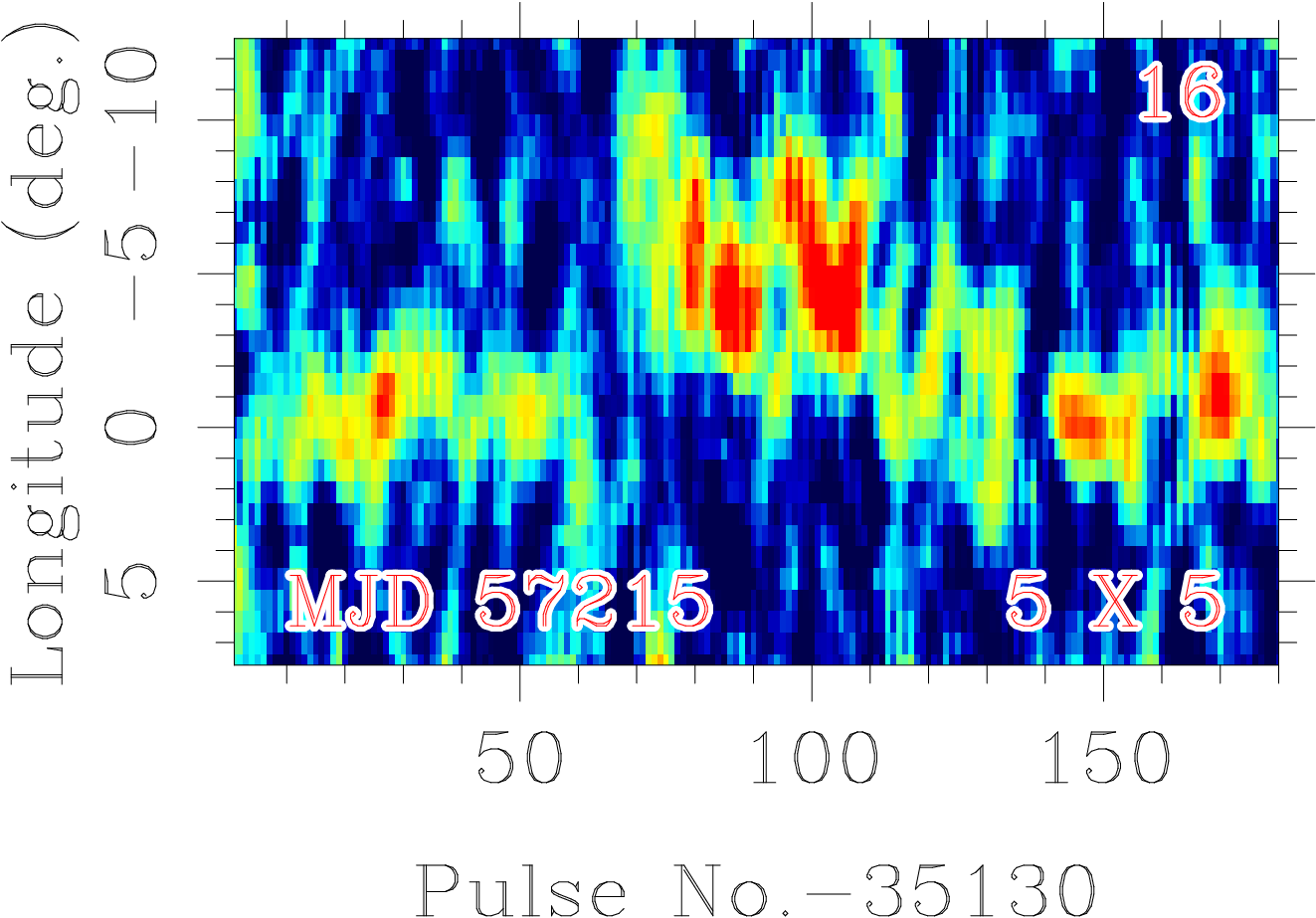}
 \includegraphics[bb = 60 95 505 685,clip,width=0.114\textwidth,angle=-90]{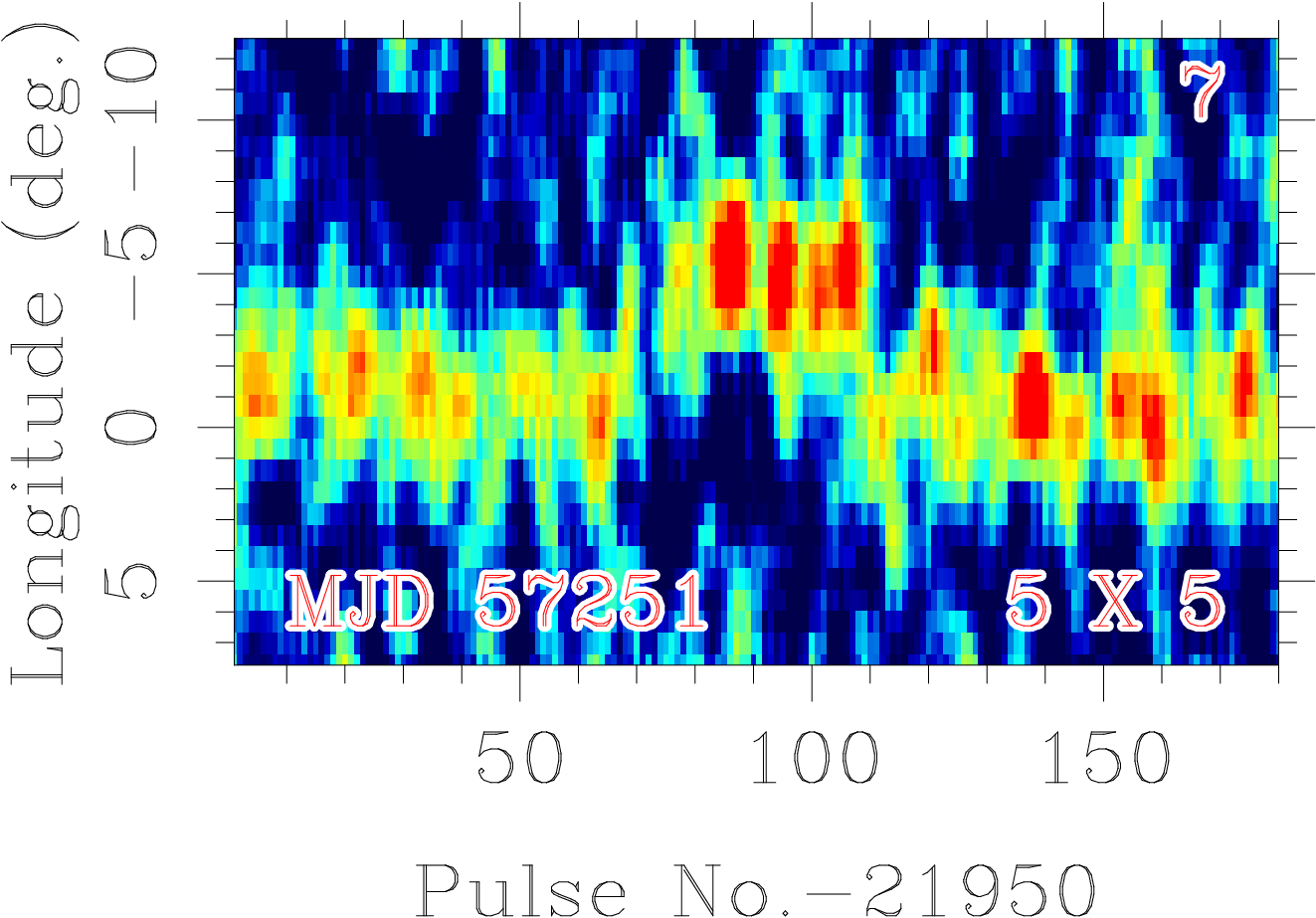}
 \caption{The detailed plots for abnormal emission events of ``M
   type''. For events with a weak signal-to-noise ratio, the
   ``images'' are enhanced by a Gaussian smooth of $3\times3$ pixels
   or $5\times5$ pixels.}
  \label{fig9}
\end{figure}
\begin{figure}
 \includegraphics[bb = 60 45 505 685,clip,width=0.114\textwidth,angle=-90]{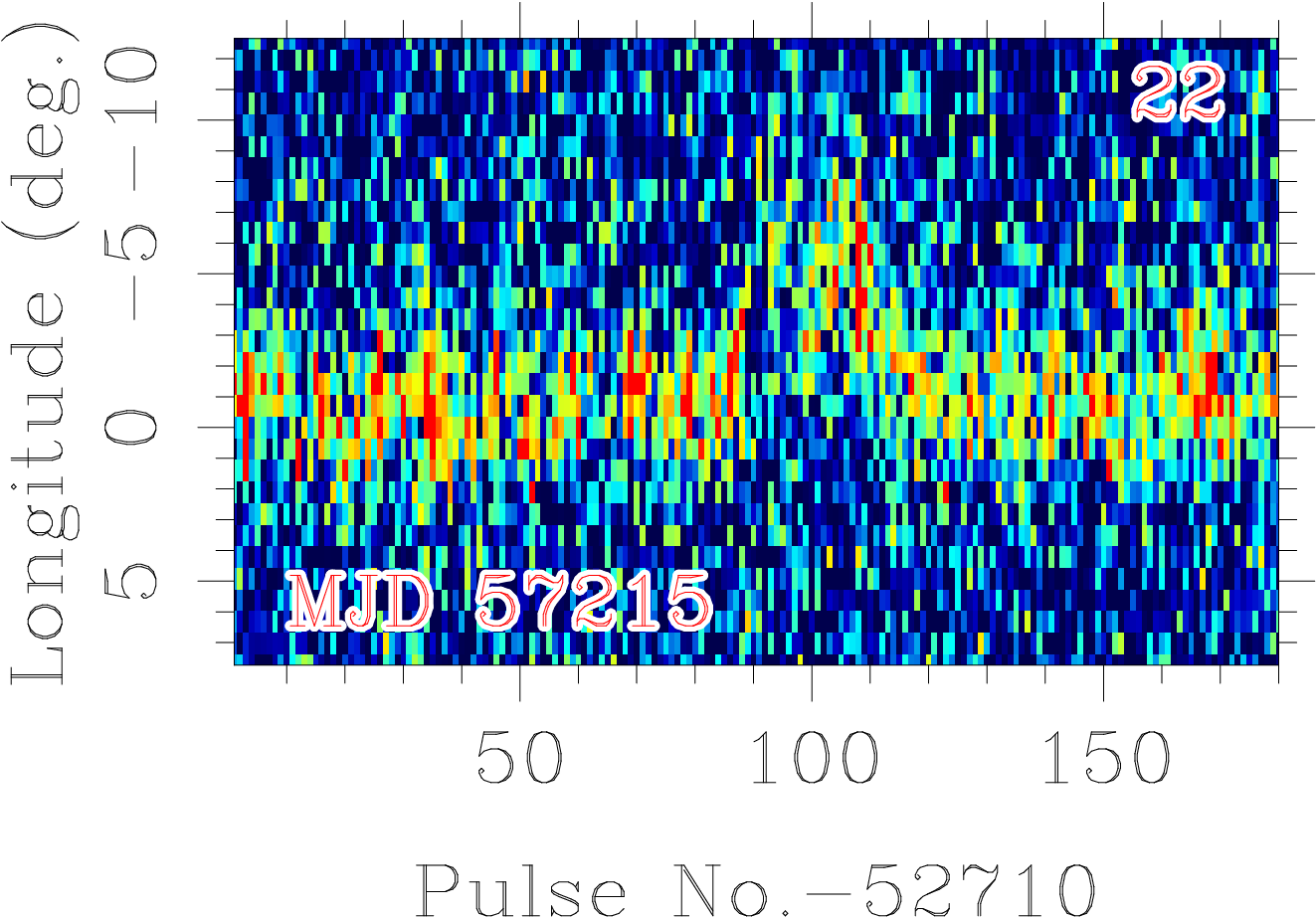}
 \includegraphics[bb = 60 95 505 685,clip,width=0.114\textwidth,angle=-90]{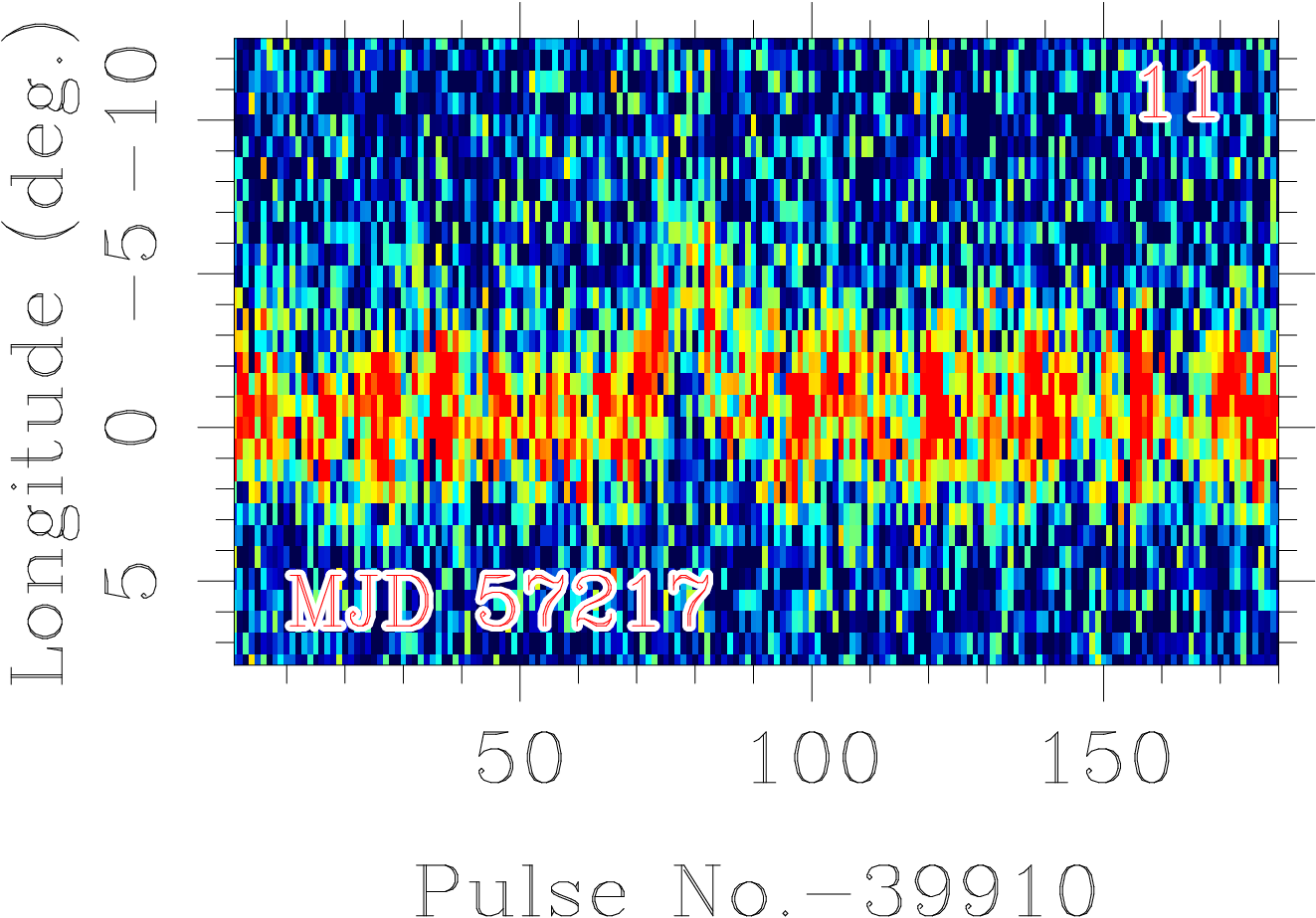}
 \includegraphics[bb = 60 95 505 685,clip,width=0.114\textwidth,angle=-90]{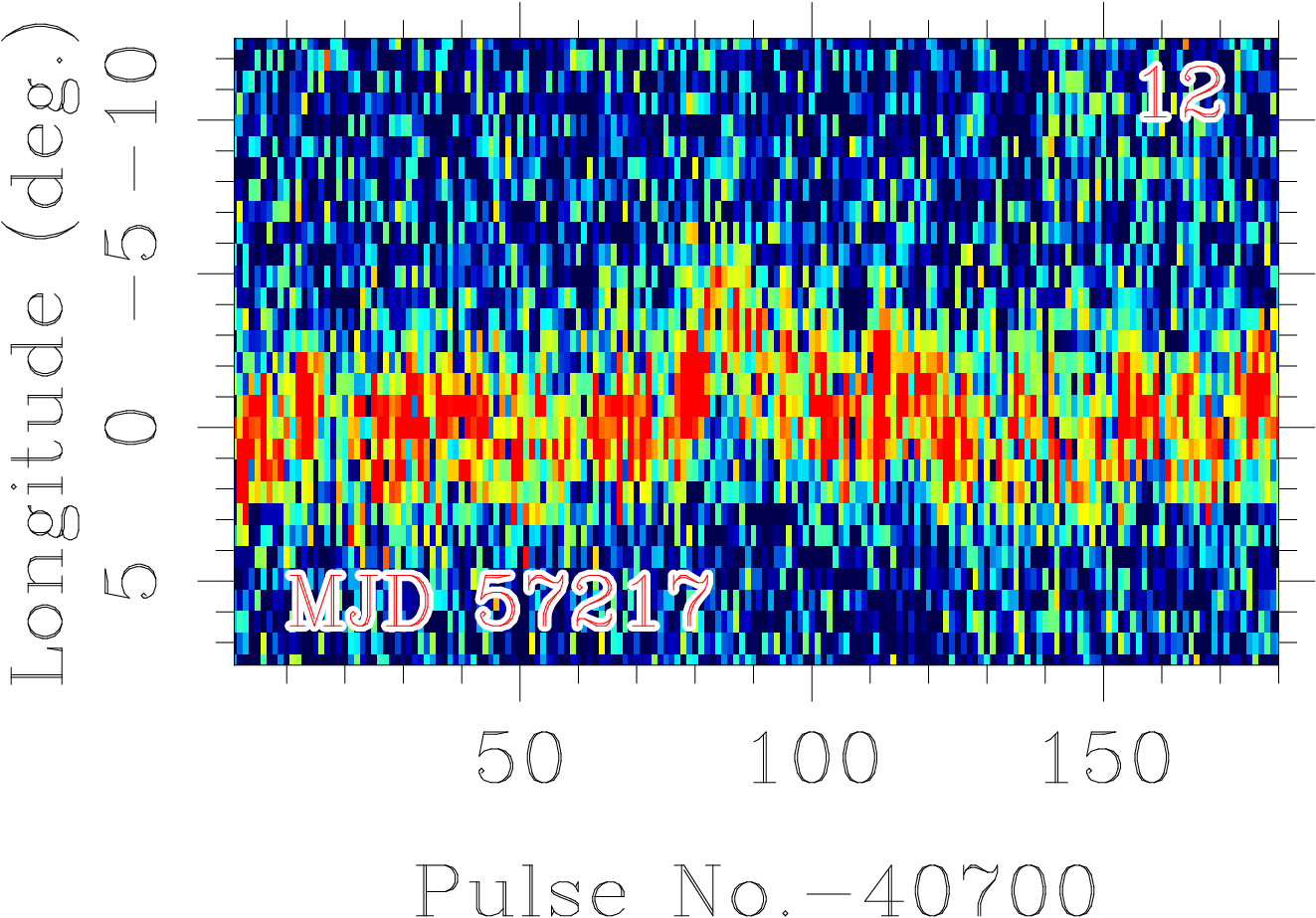}
 \includegraphics[bb = 60 45 505 685,clip,width=0.114\textwidth,angle=-90]{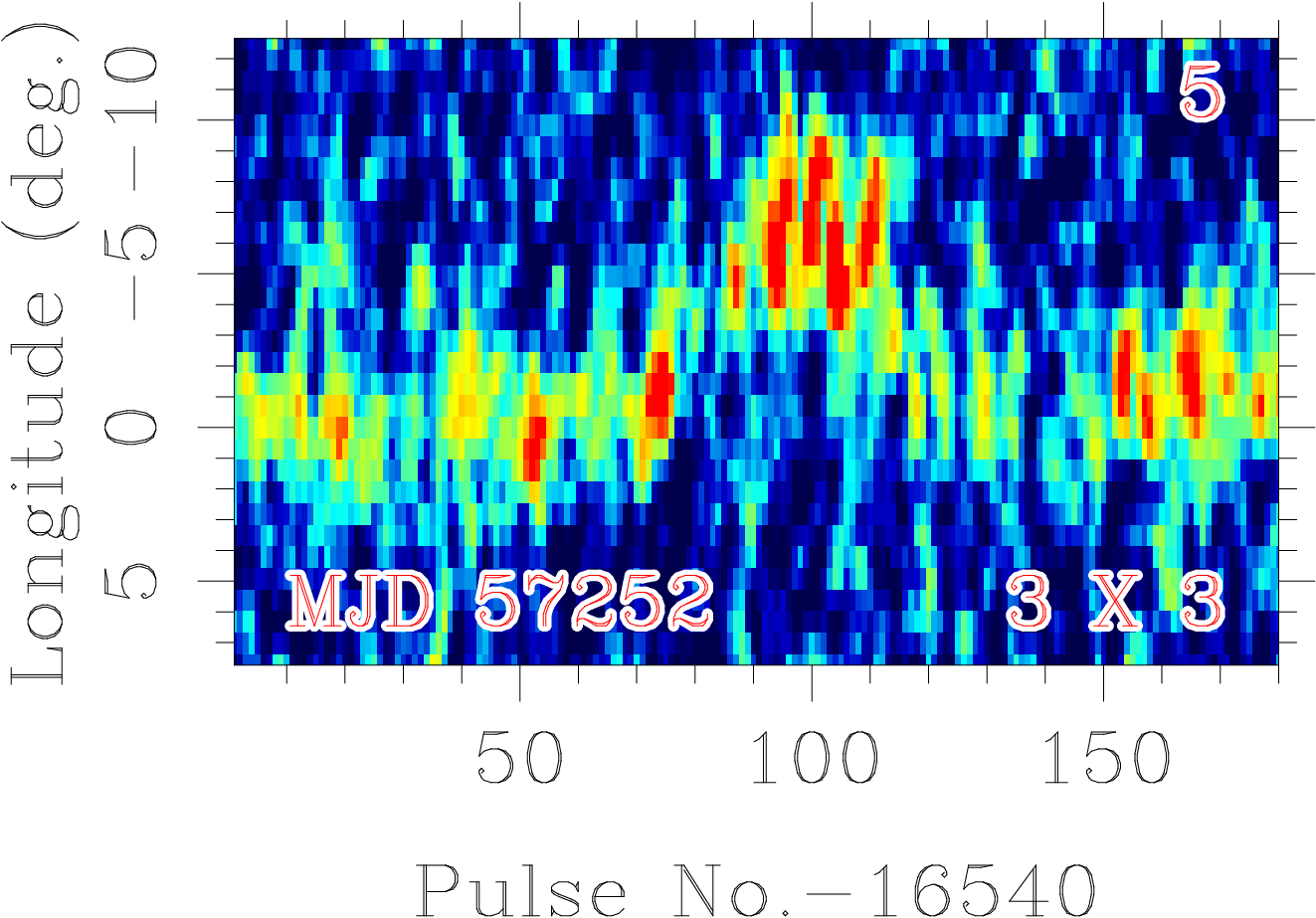}
 \includegraphics[bb = 60 95 505 685,clip,width=0.114\textwidth,angle=-90]{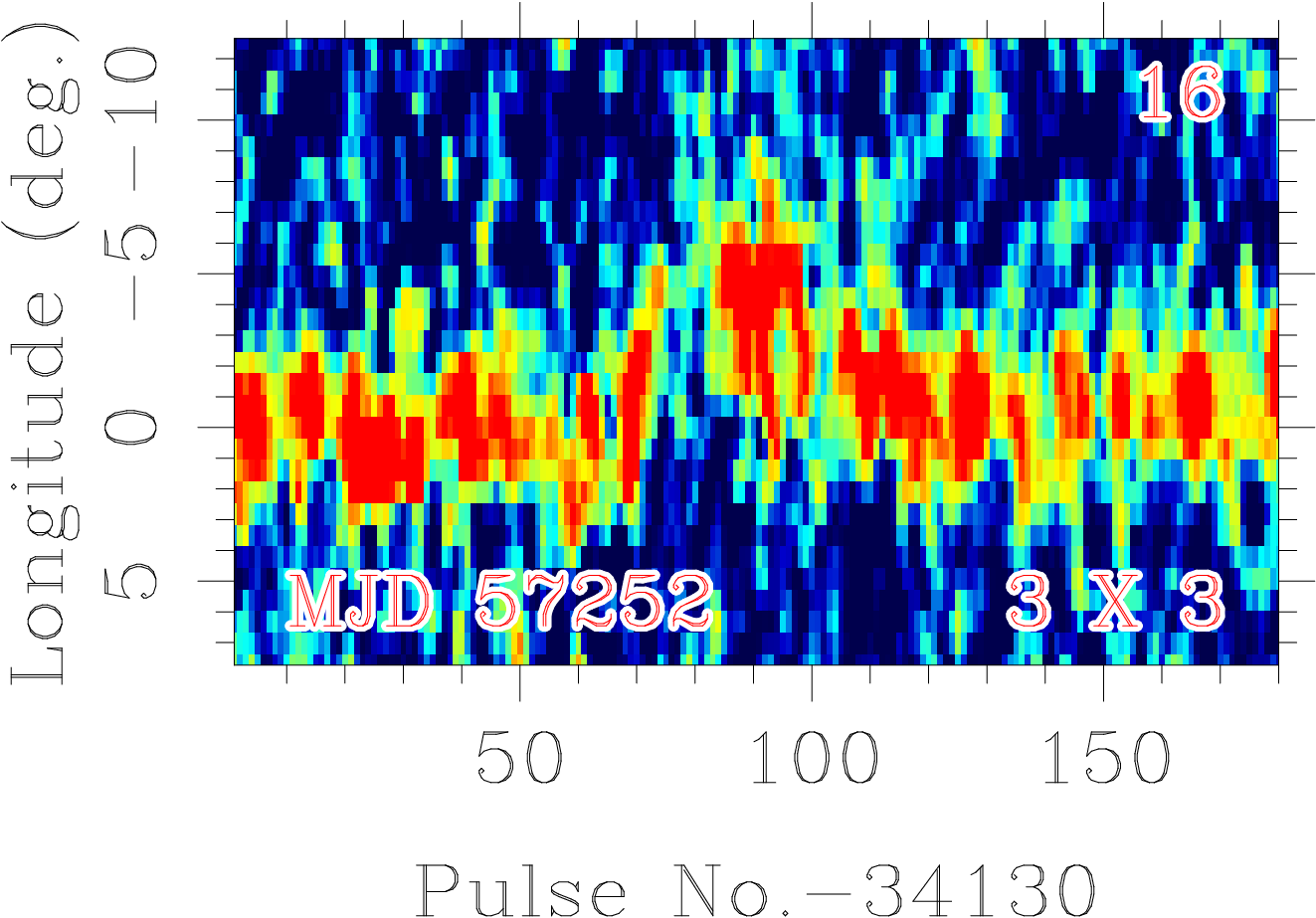} 
 \includegraphics[bb = 60 95 505 685,clip,width=0.114\textwidth,angle=-90]{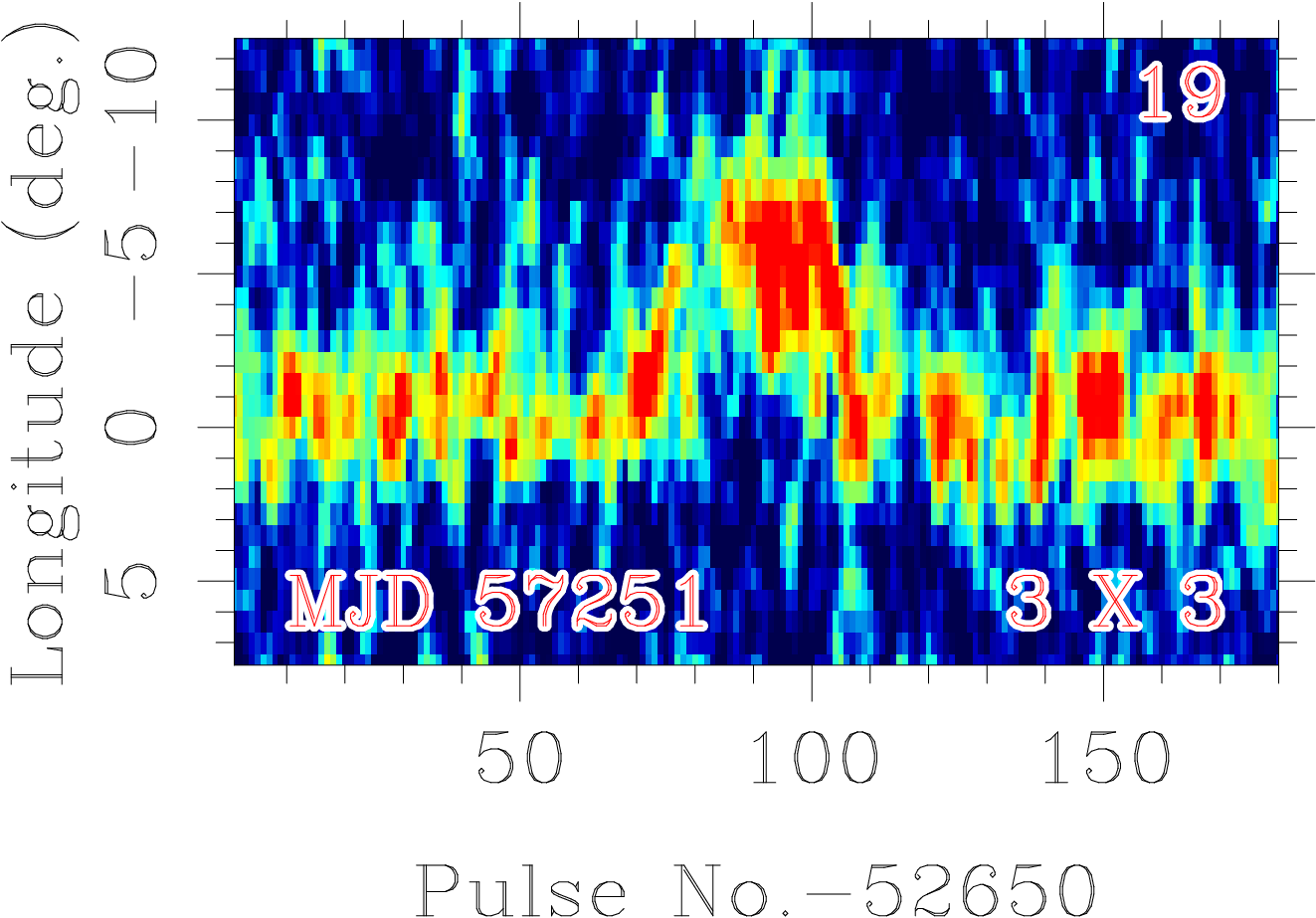}
 \includegraphics[bb = 60 45 505 685,clip,width=0.114\textwidth,angle=-90]{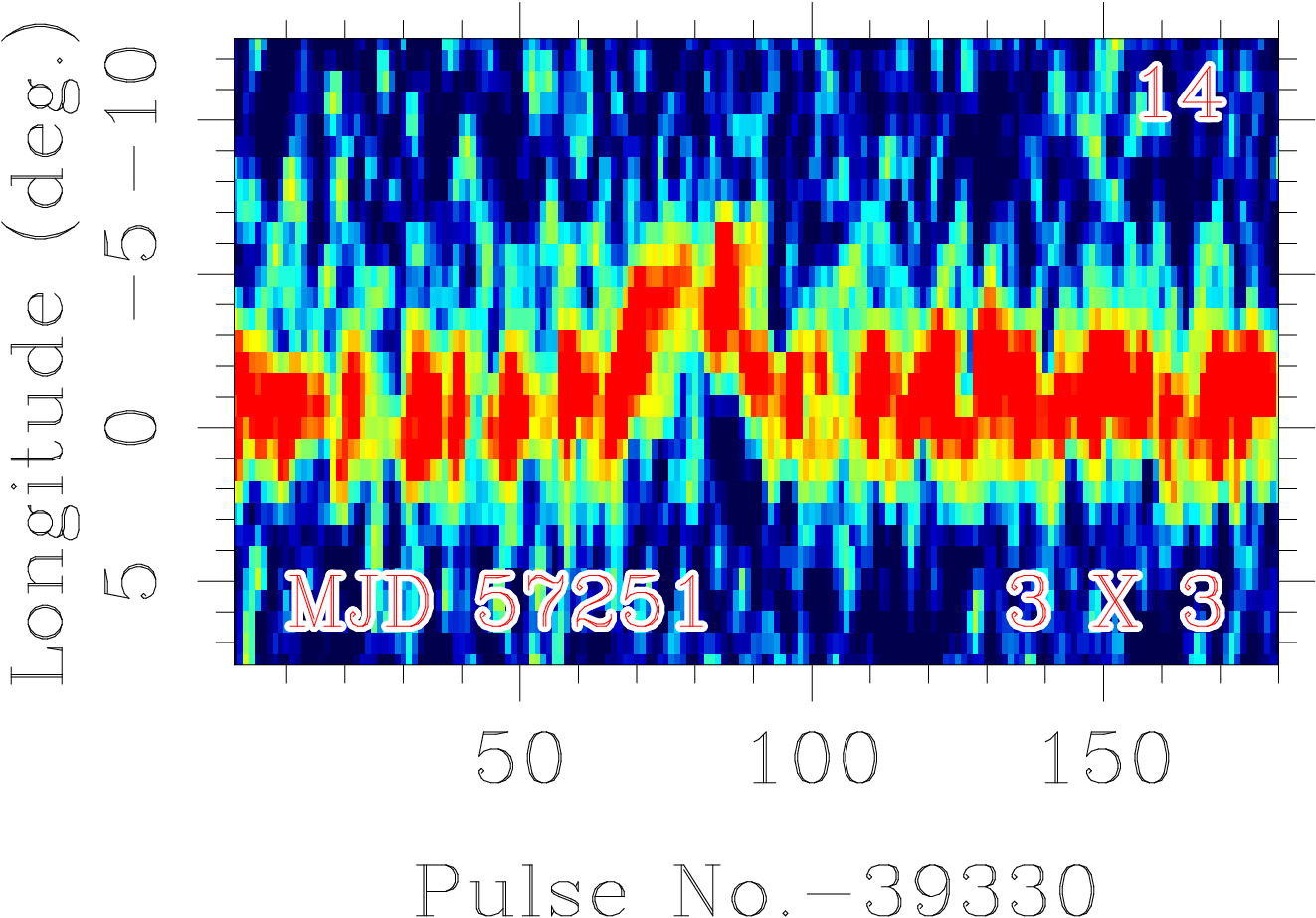} 
 \includegraphics[bb = 60 95 505 685,clip,width=0.114\textwidth,angle=-90]{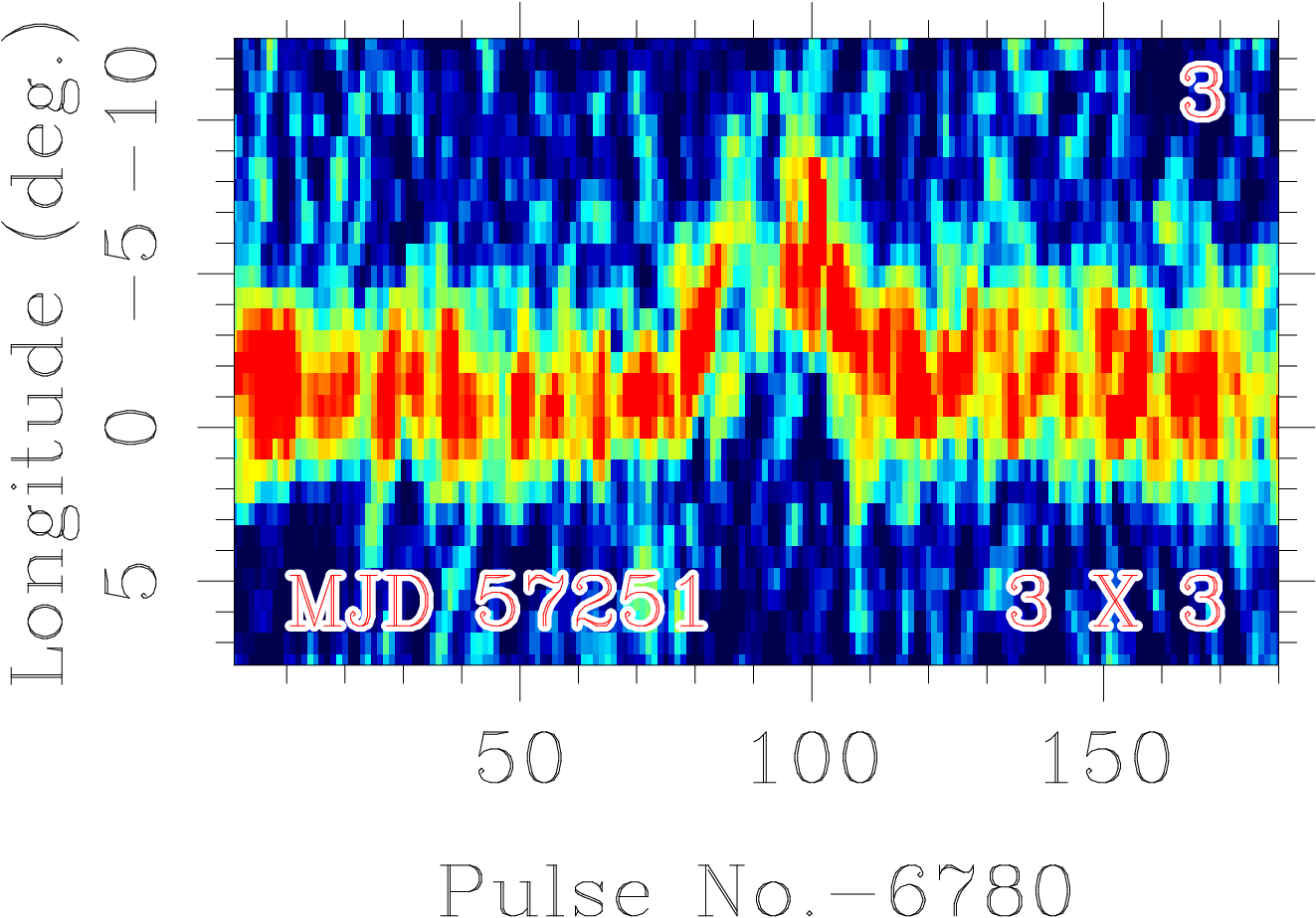}
 \includegraphics[bb = 60 95 505 685,clip,width=0.114\textwidth,angle=-90]{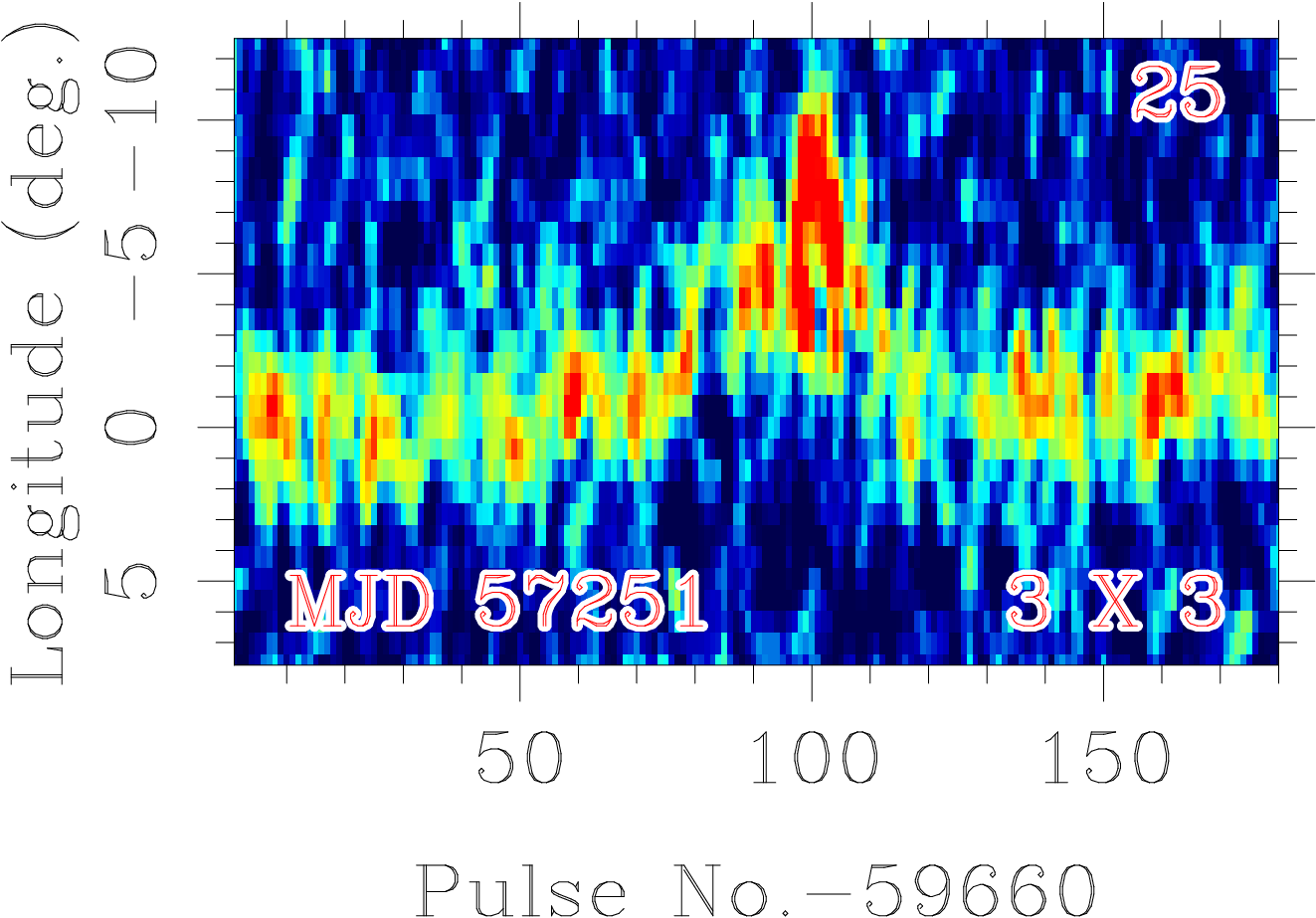}
 \includegraphics[bb = 60 45 505 685,clip,width=0.114\textwidth,angle=-90]{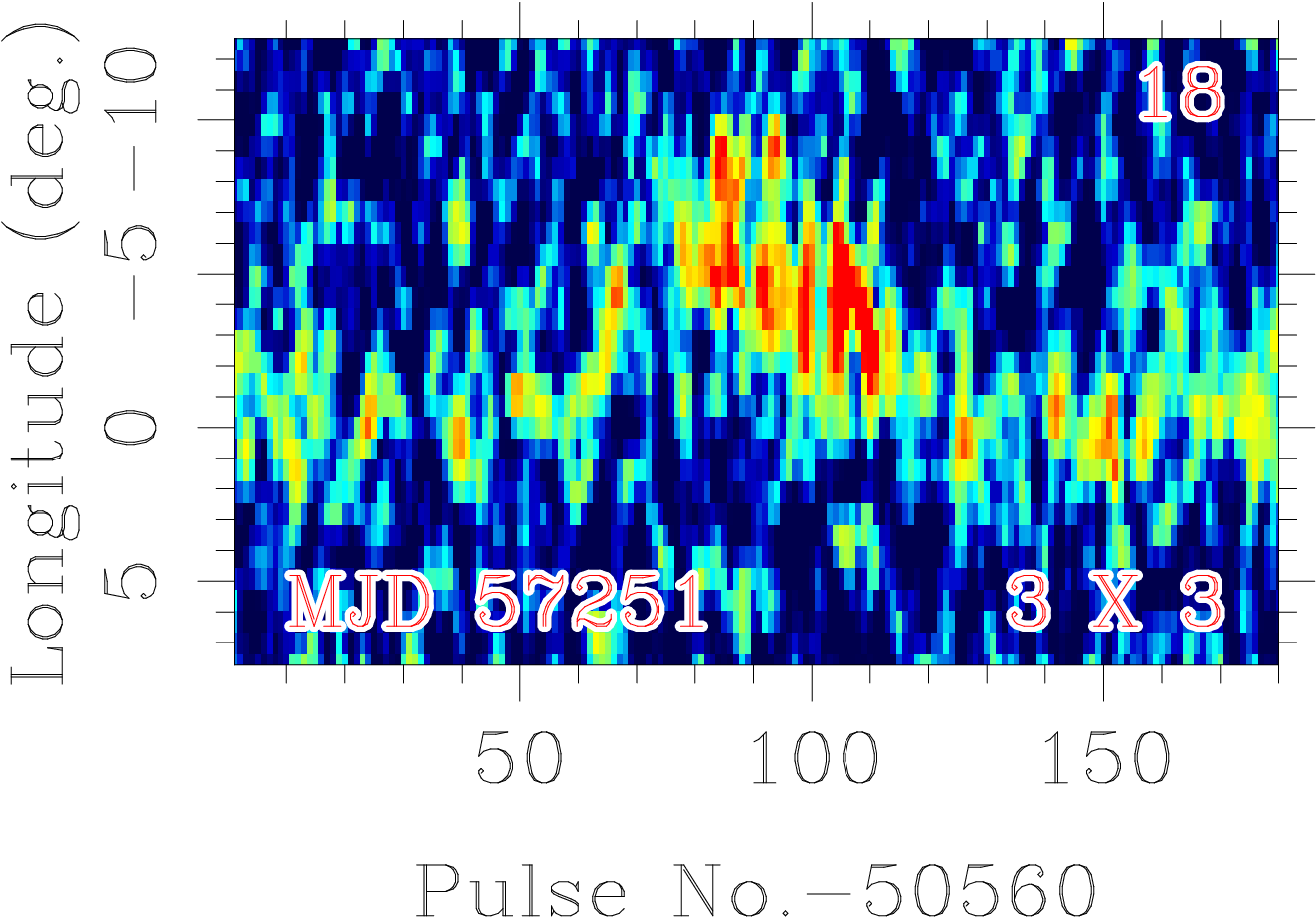}
 \includegraphics[bb = 60 95 505 685,clip,width=0.114\textwidth,angle=-90]{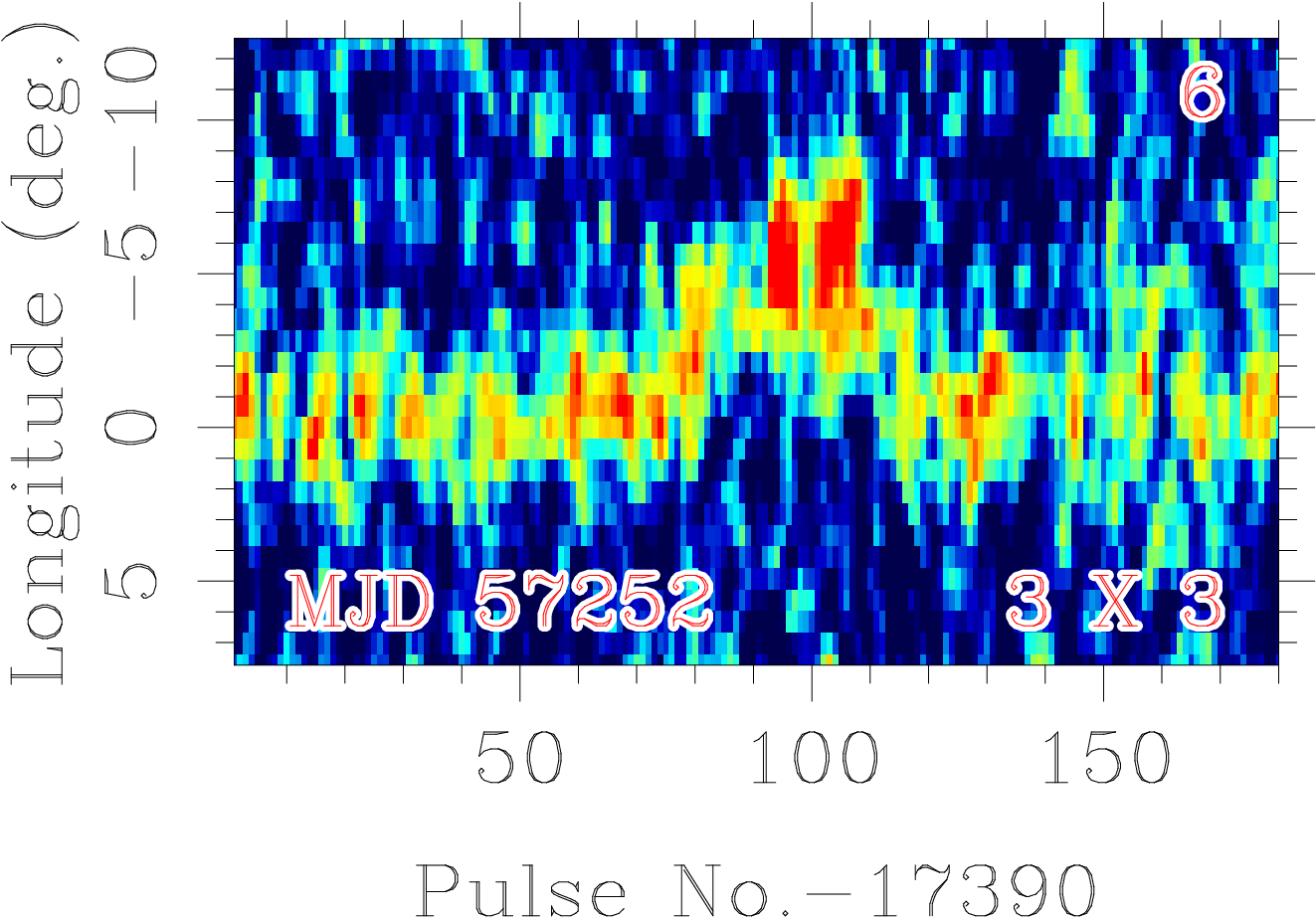}
 \includegraphics[bb = 60 95 505 685,clip,width=0.114\textwidth,angle=-90]{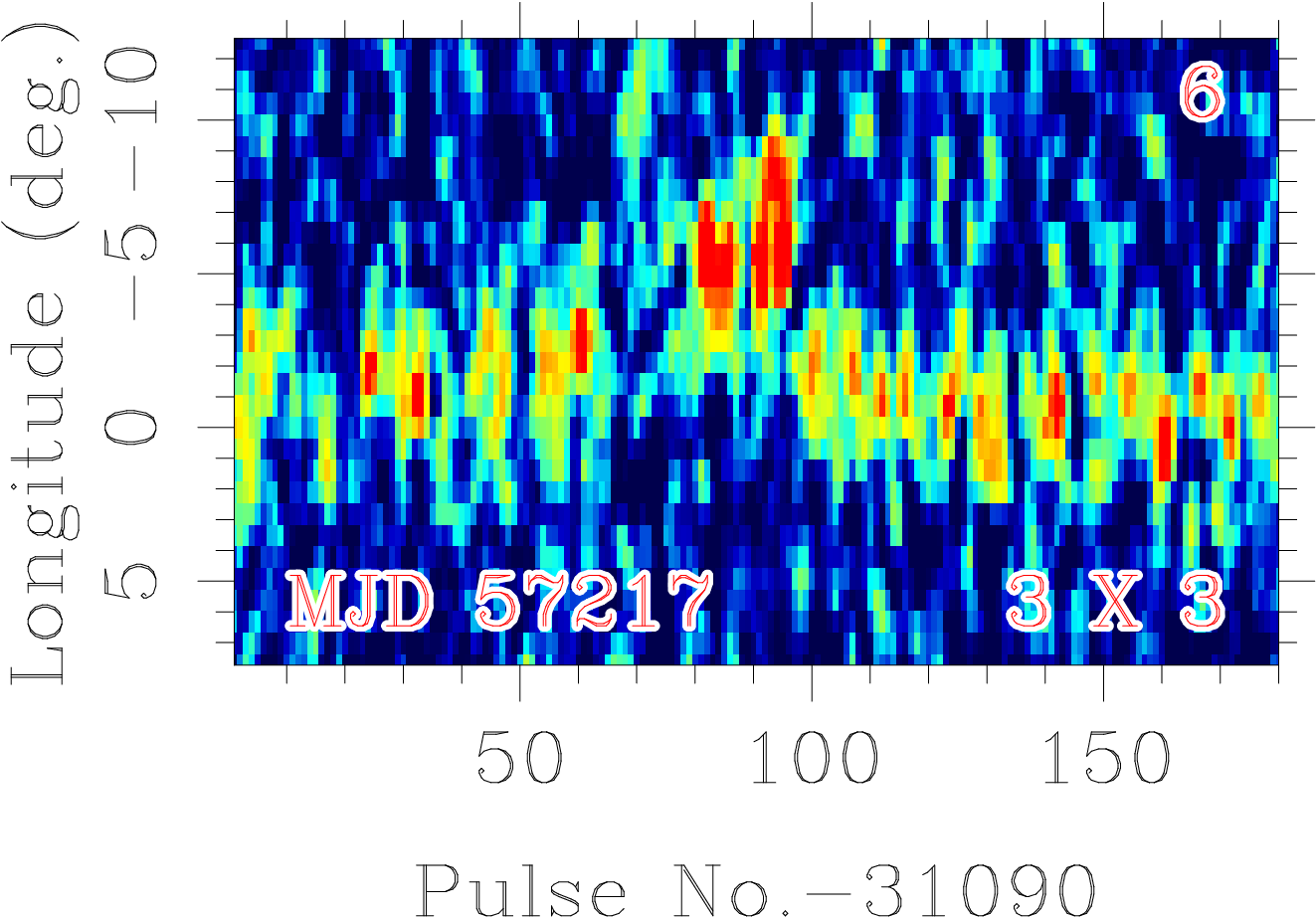}
 \includegraphics[bb = 60 45 505 685,clip,width=0.114\textwidth,angle=-90]{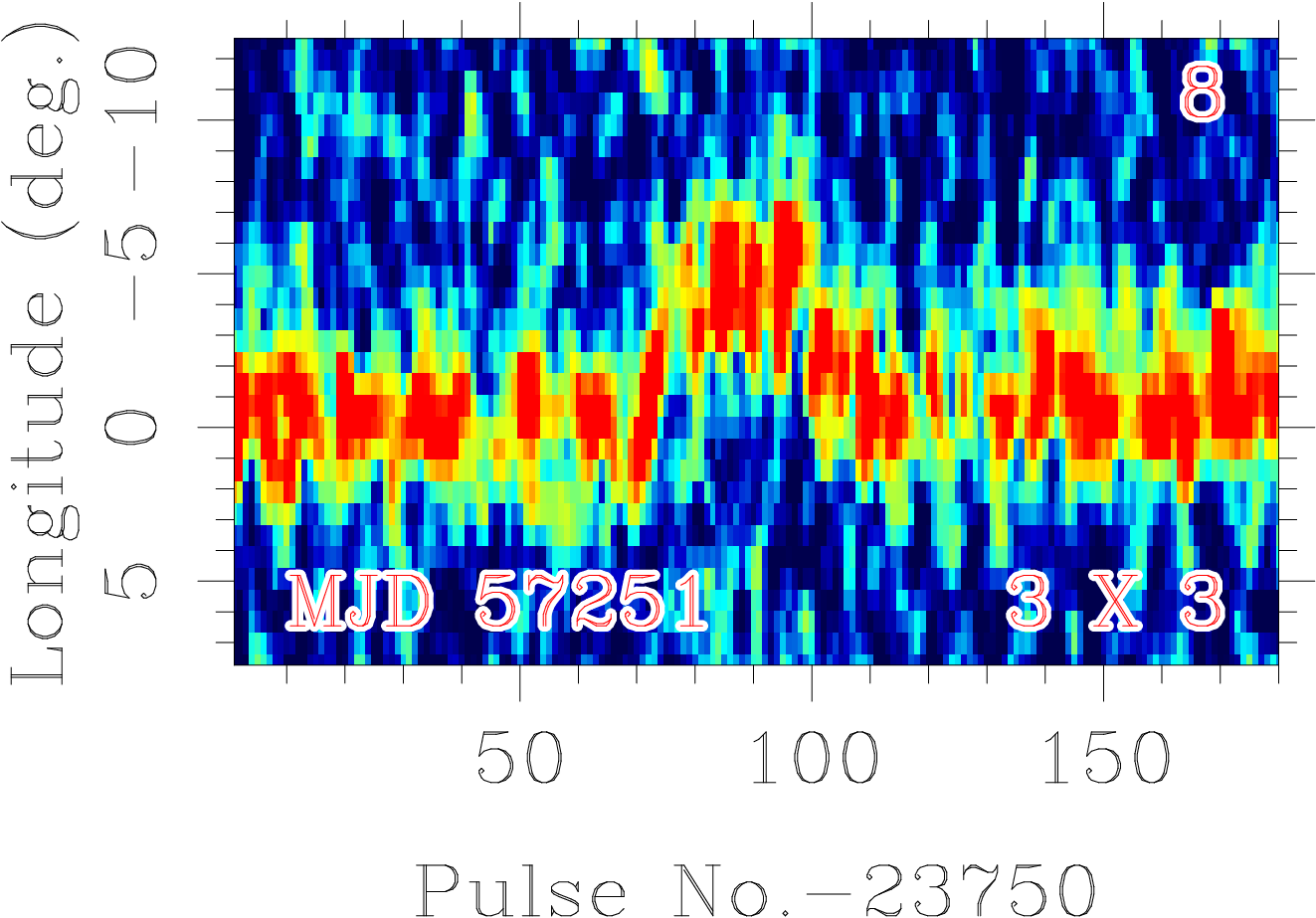}
 \includegraphics[bb = 60 95 505 685,clip,width=0.114\textwidth,angle=-90]{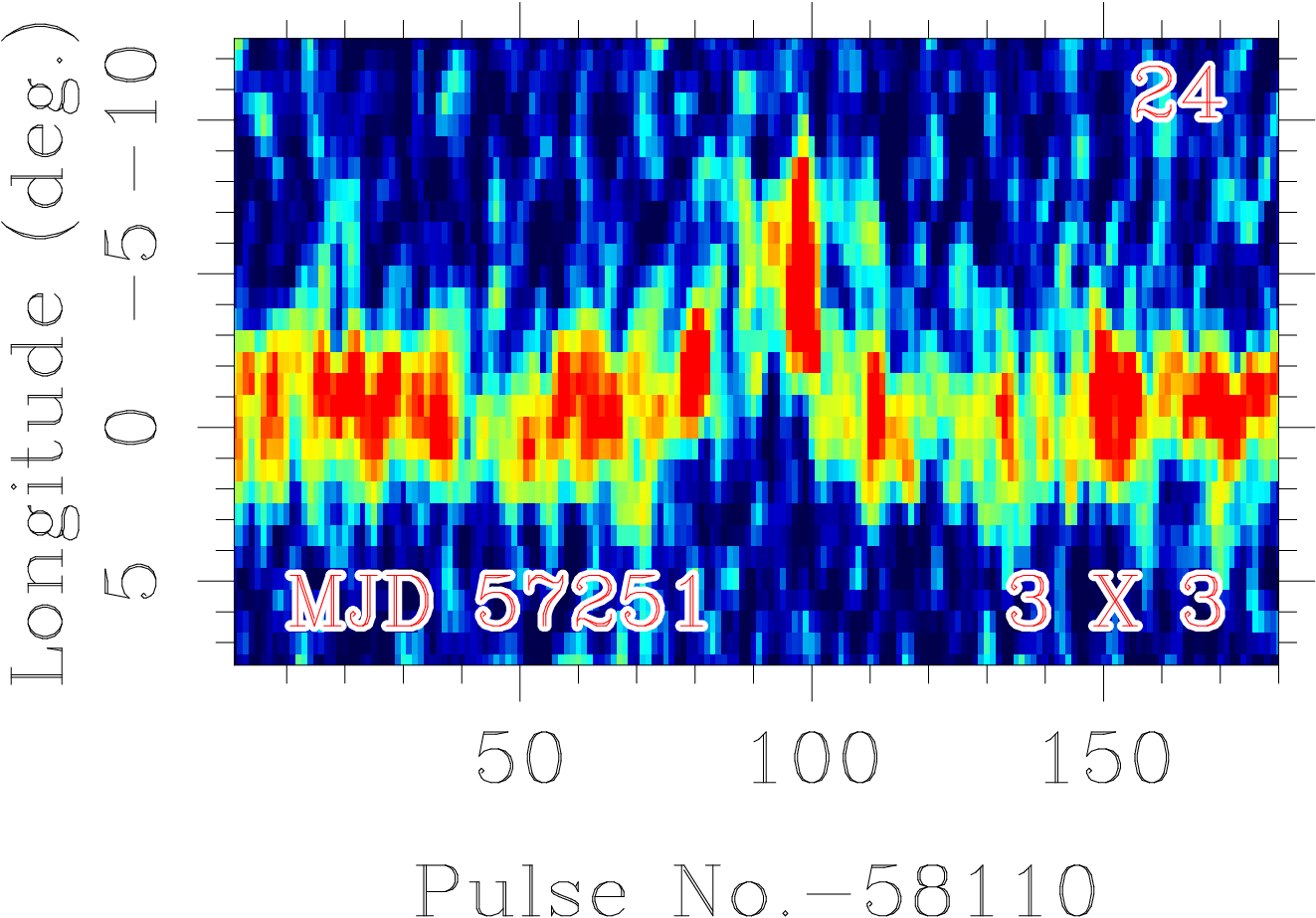}
 \includegraphics[bb = 60 95 505 685,clip,width=0.114\textwidth,angle=-90]{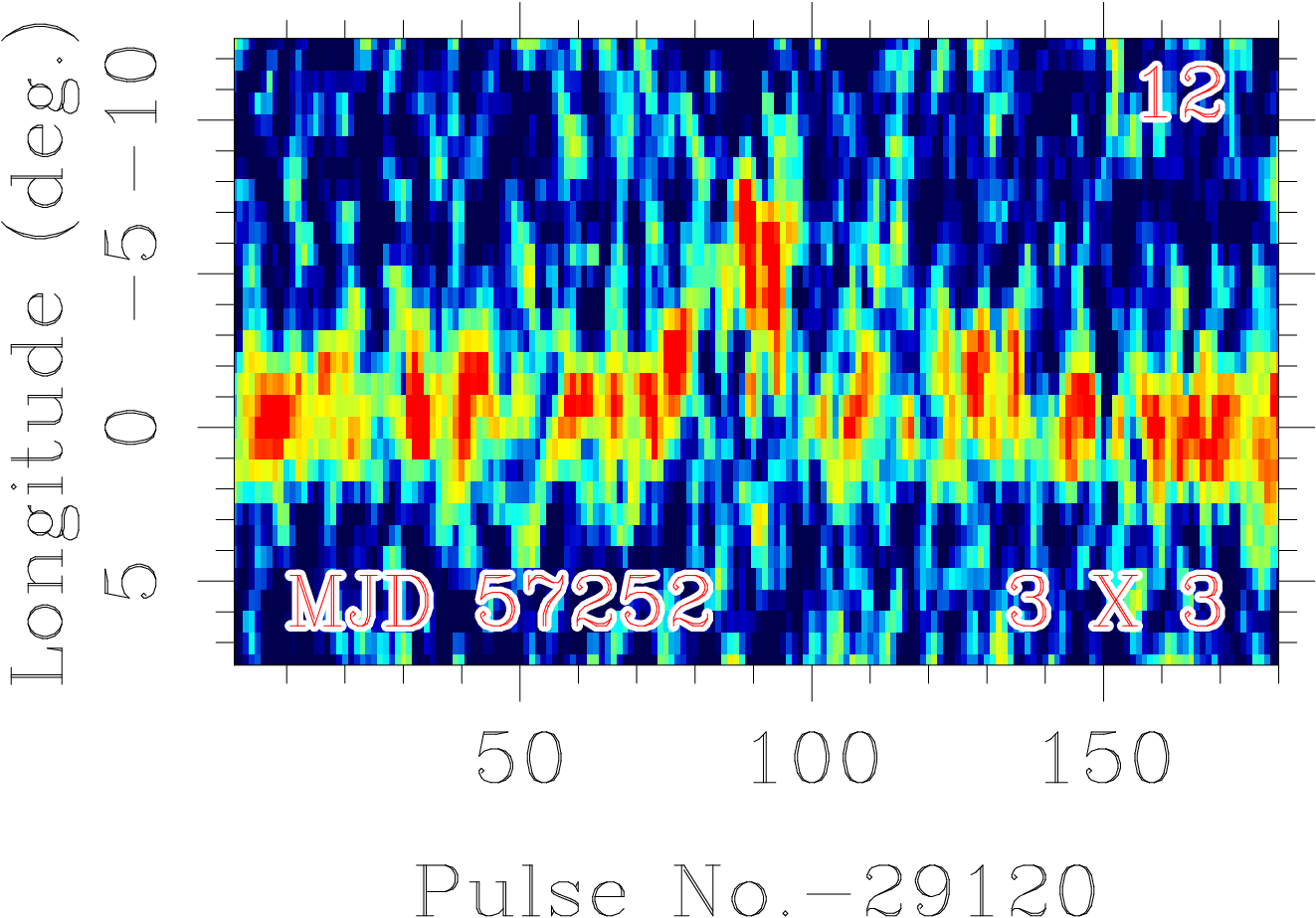} 
 \includegraphics[bb = 60 45 505 685,clip,width=0.114\textwidth,angle=-90]{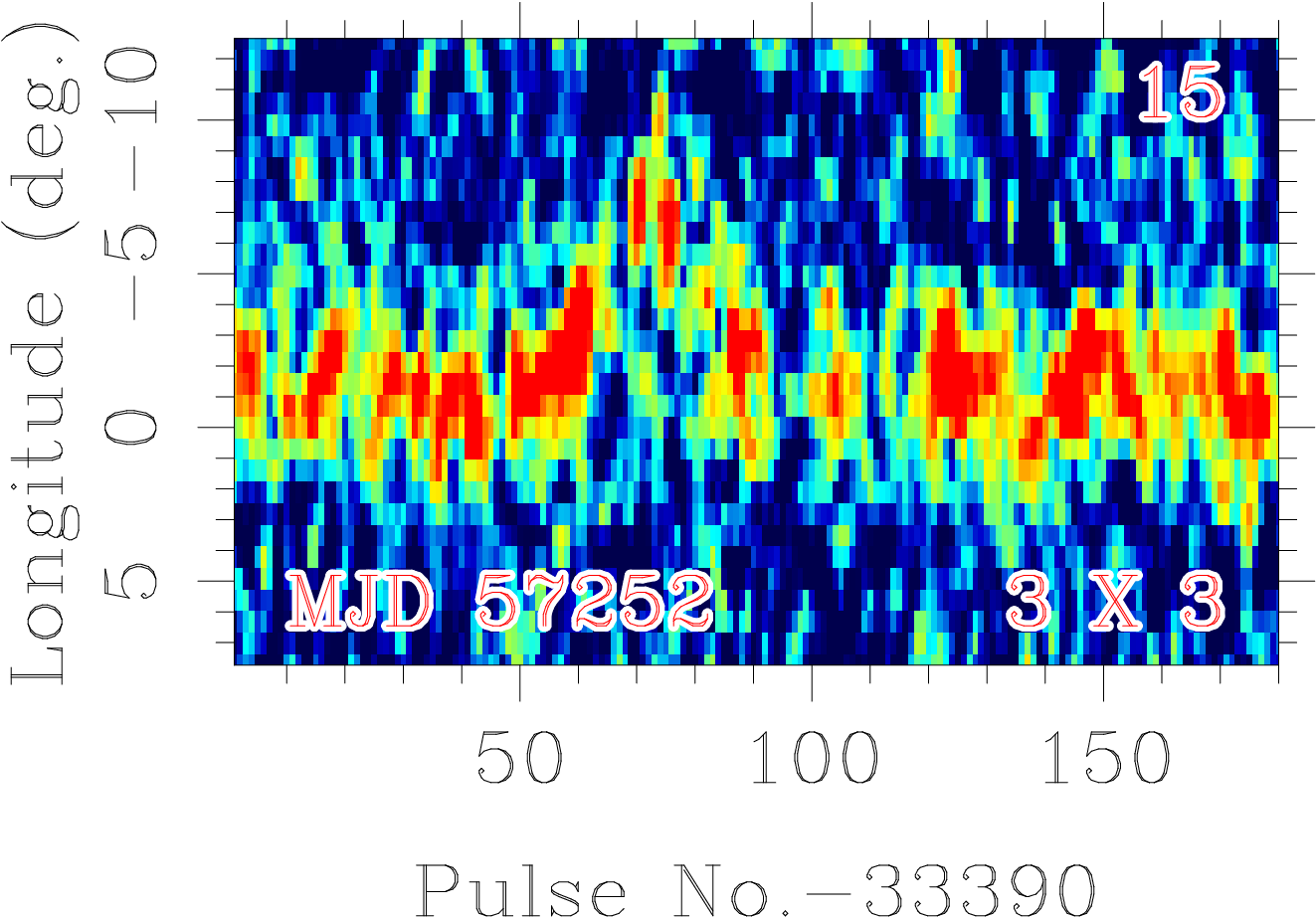} 
 \includegraphics[bb = 60 95 505 685,clip,width=0.114\textwidth,angle=-90]{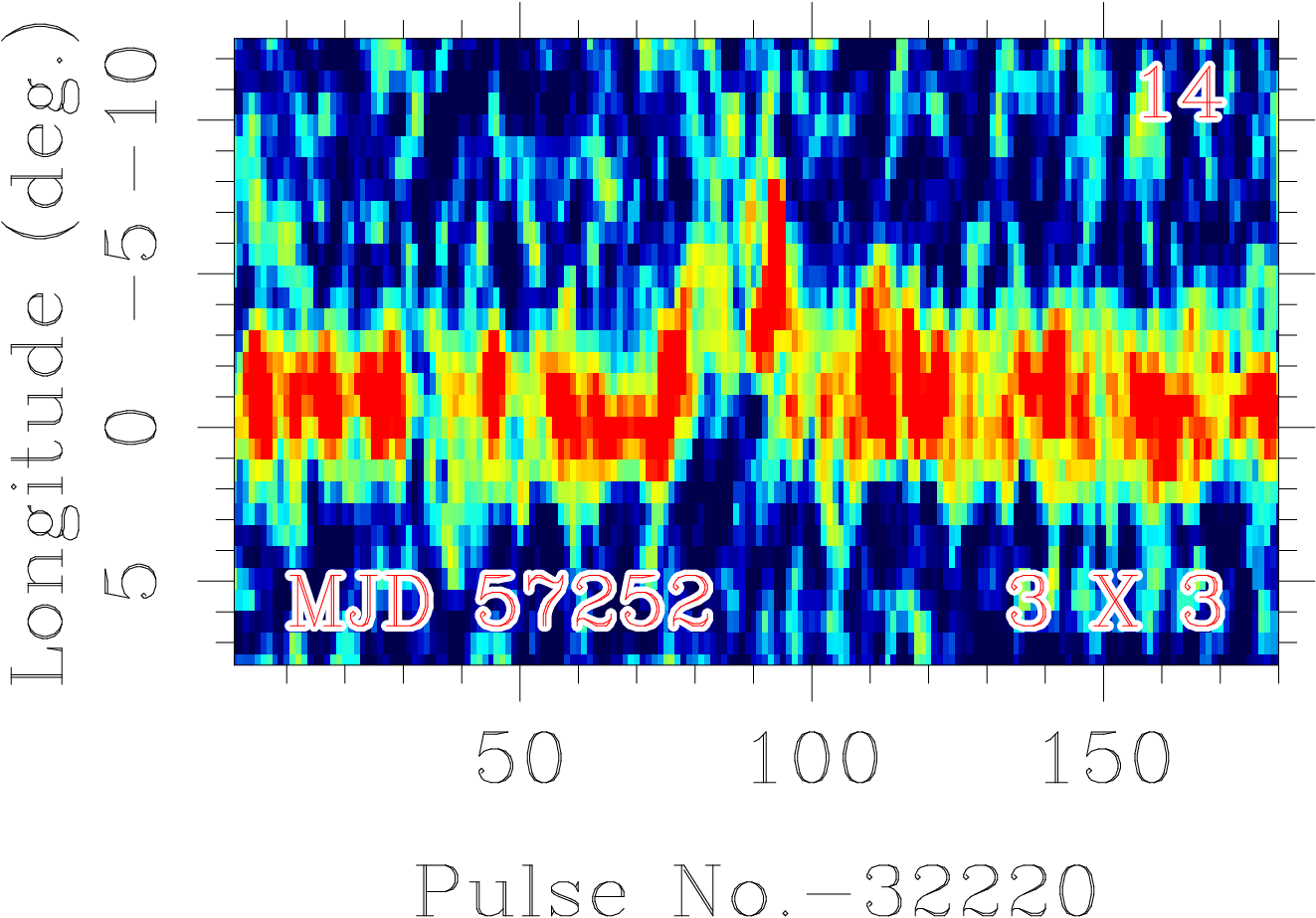}
 \includegraphics[bb = 60 95 505 685,clip,width=0.114\textwidth,angle=-90]{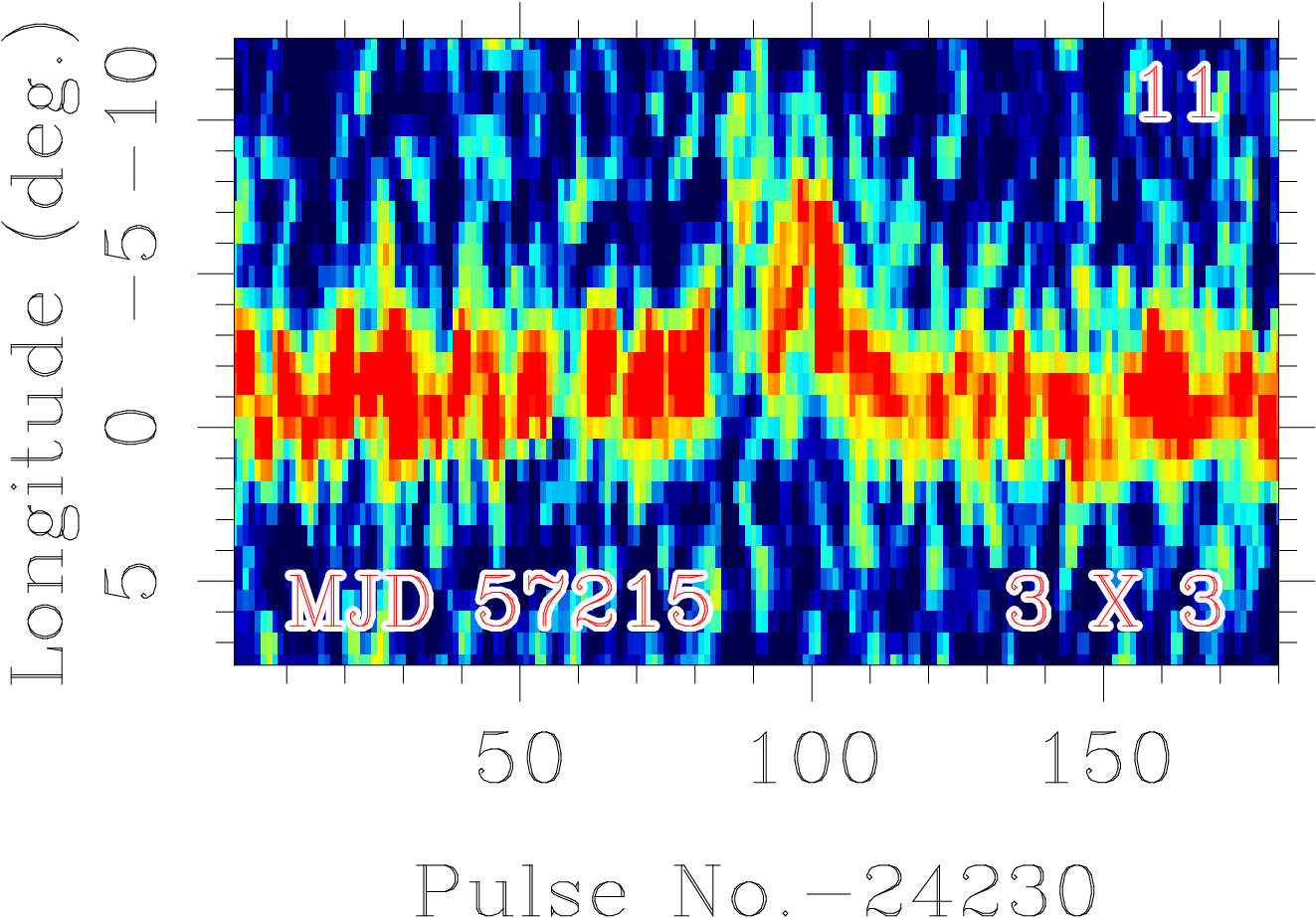}
 \includegraphics[bb = 60 45 505 685,clip,width=0.114\textwidth,angle=-90]{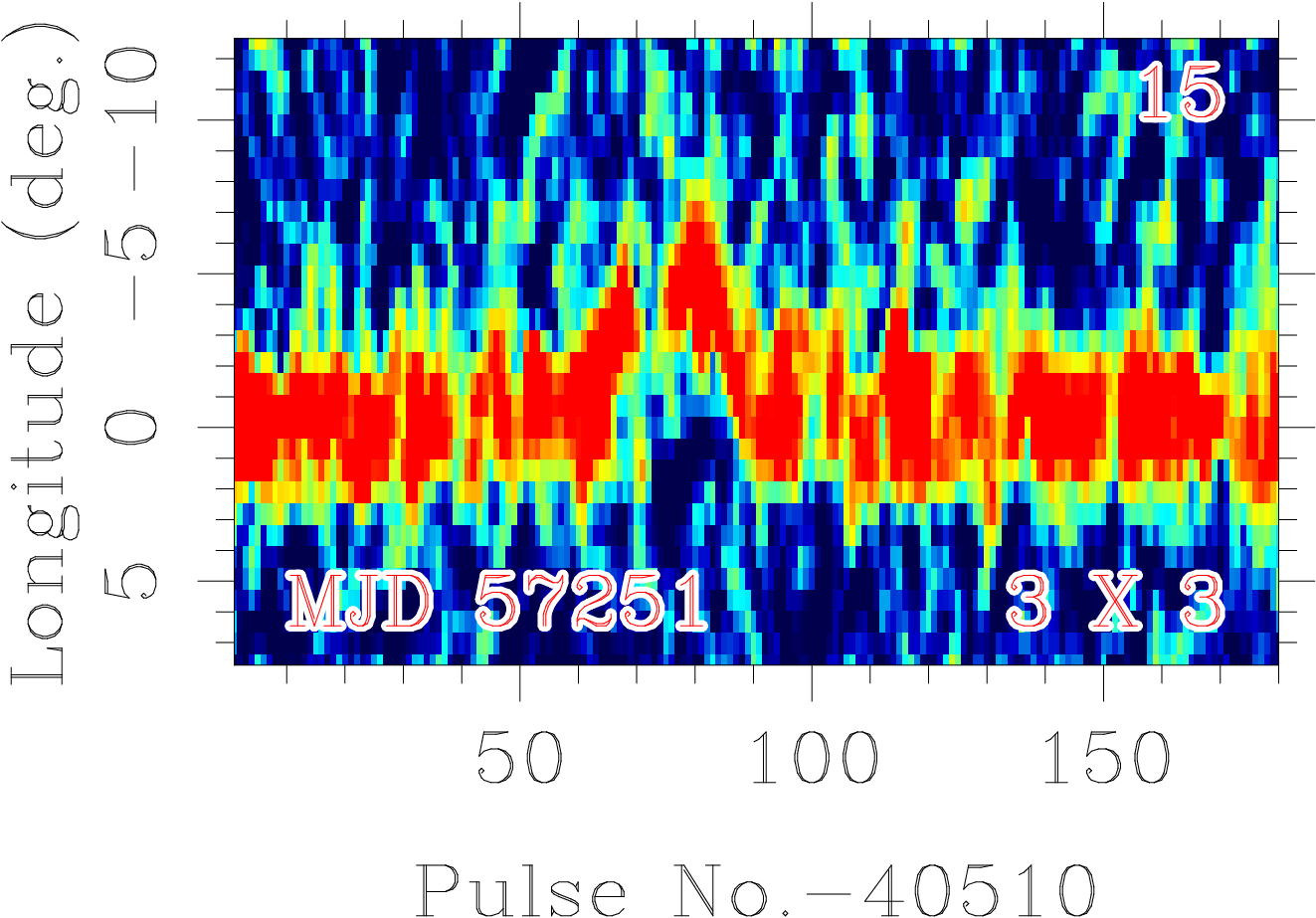} 
 \includegraphics[bb = 60 95 505 685,clip,width=0.114\textwidth,angle=-90]{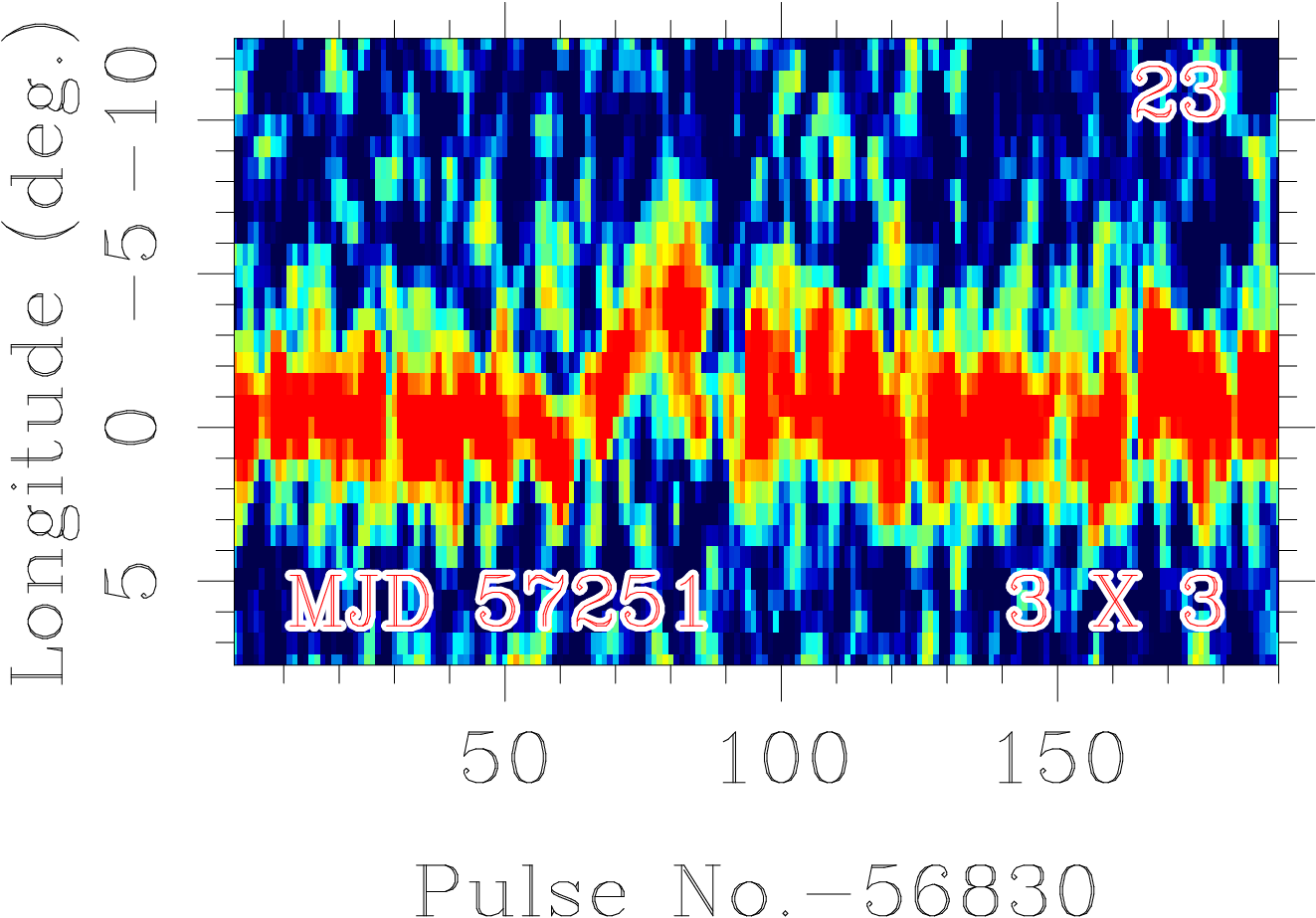} 
 \includegraphics[bb = 60 95 505 685,clip,width=0.114\textwidth,angle=-90]{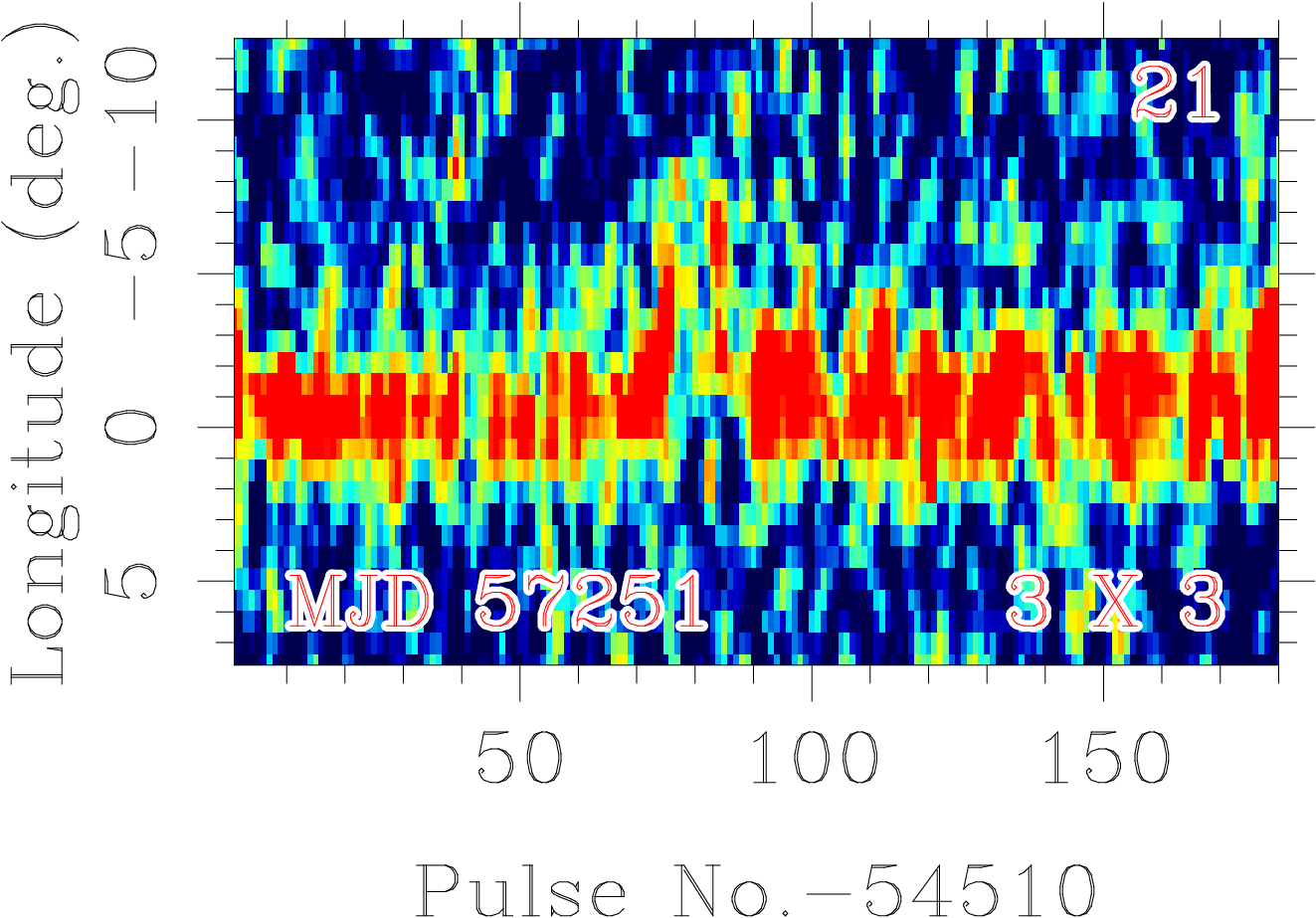} 
 \includegraphics[bb = 60 45 505 685,clip,width=0.114\textwidth,angle=-90]{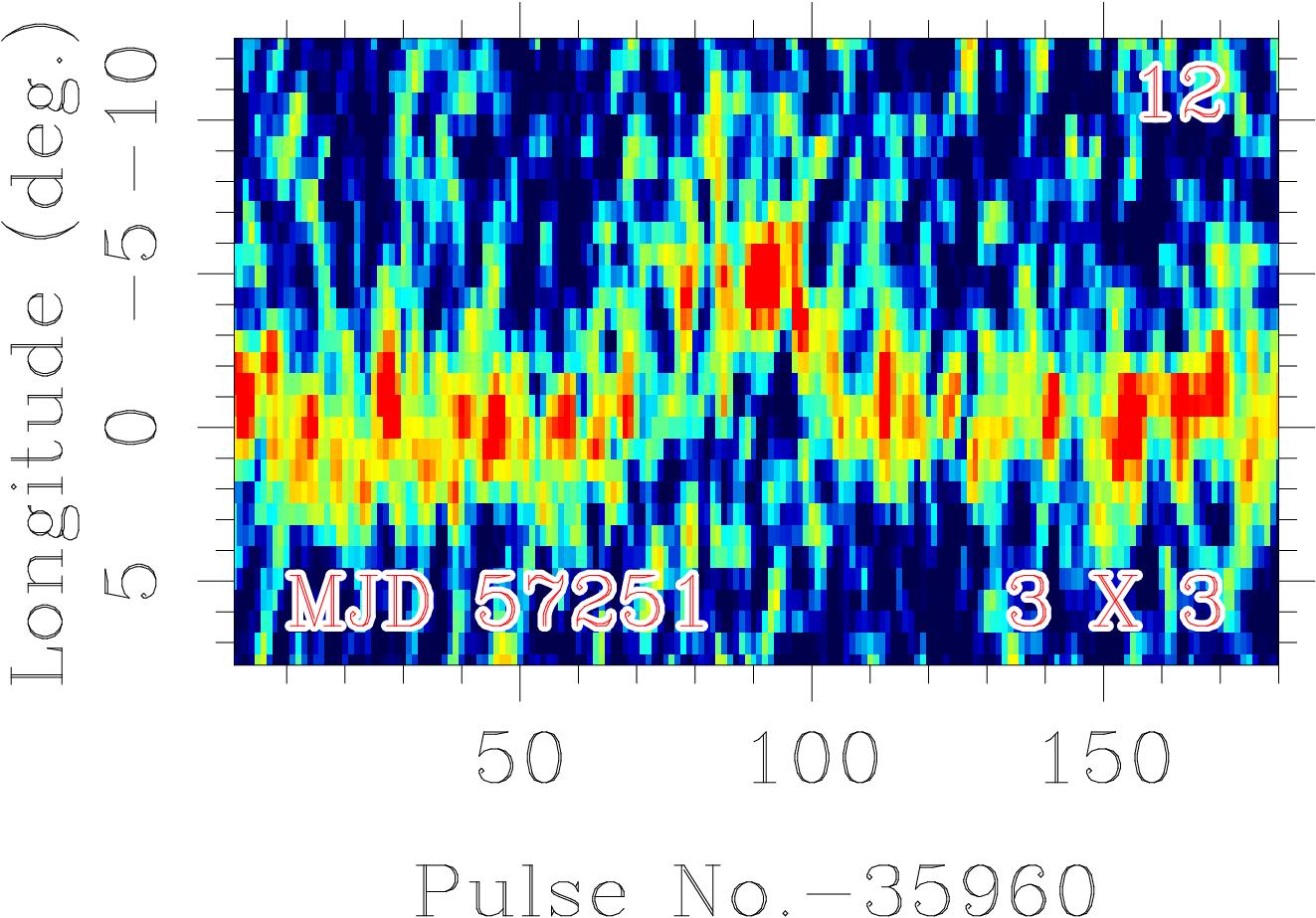}
 \caption{The detailed plots for abnormal emission events of
   ``$\Lambda$ type'', including some ``small flares''. Most ``images'' have been enhanced.}
  \label{fig10}
\end{figure}
%
\begin{figure}
 \includegraphics[bb = 60 45 505 685,clip,width=0.114\textwidth,angle=-90]{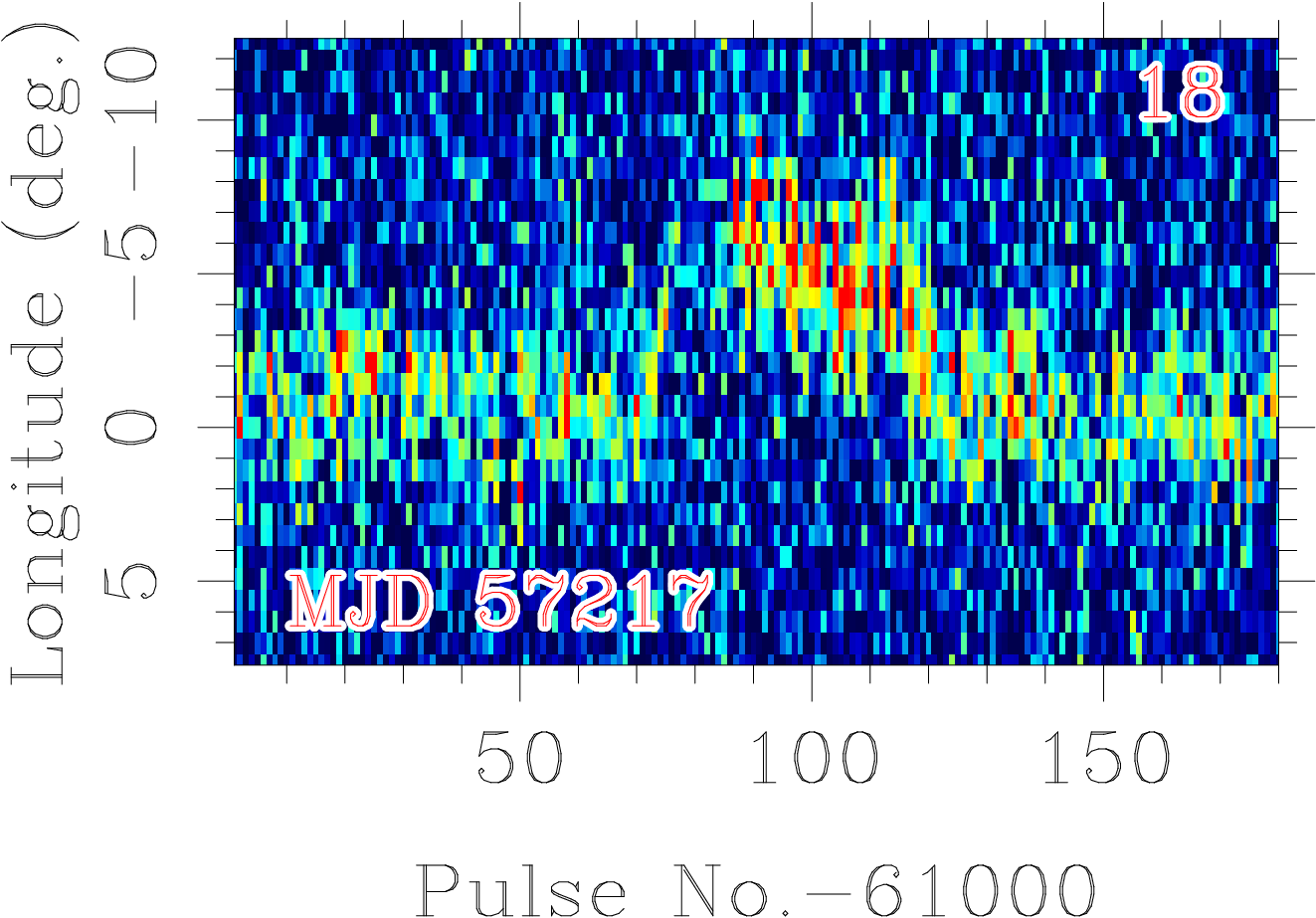}
 \includegraphics[bb = 60 95 505 685,clip,width=0.114\textwidth,angle=-90]{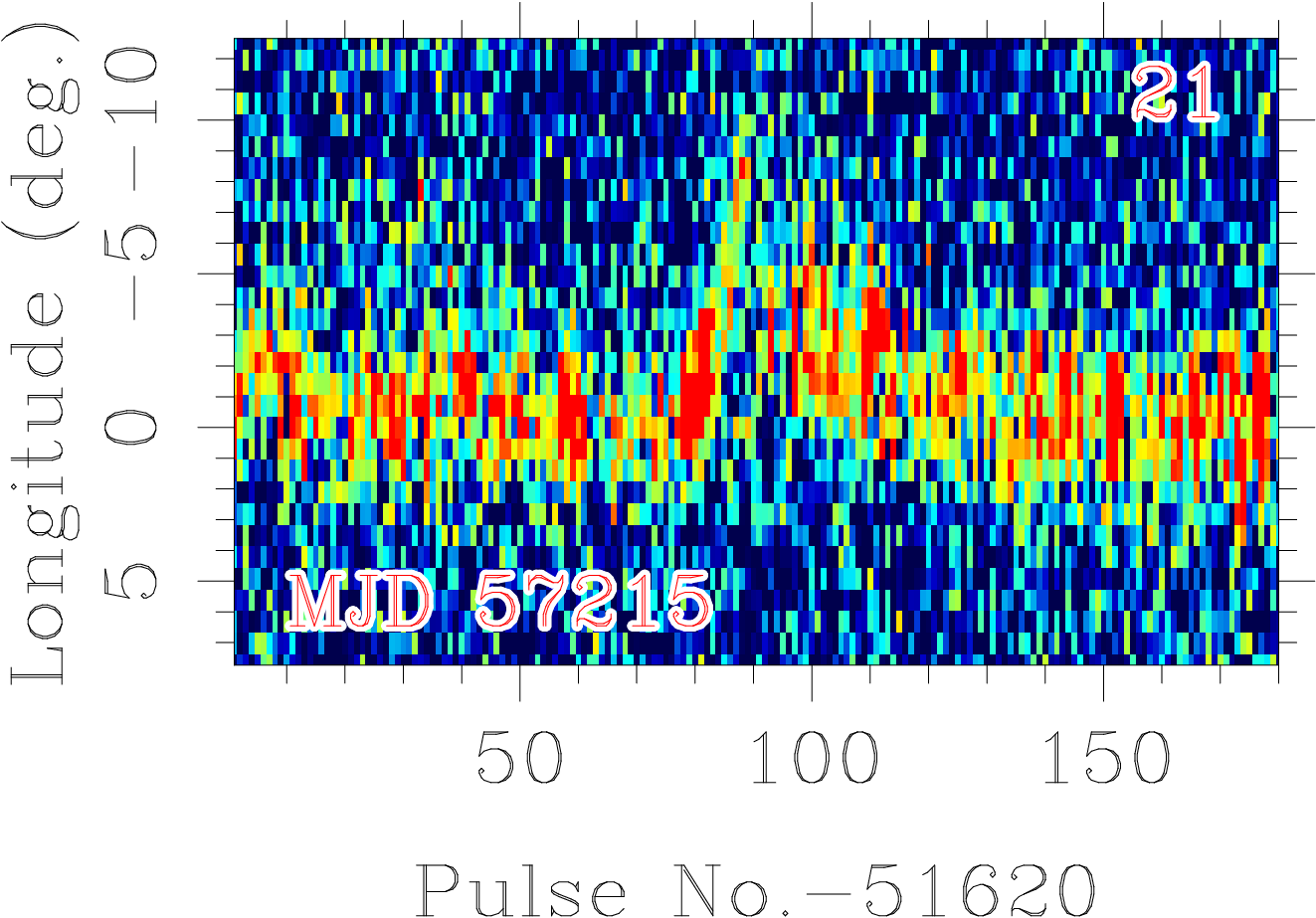}
 \includegraphics[bb = 60 95 505 685,clip,width=0.114\textwidth,angle=-90]{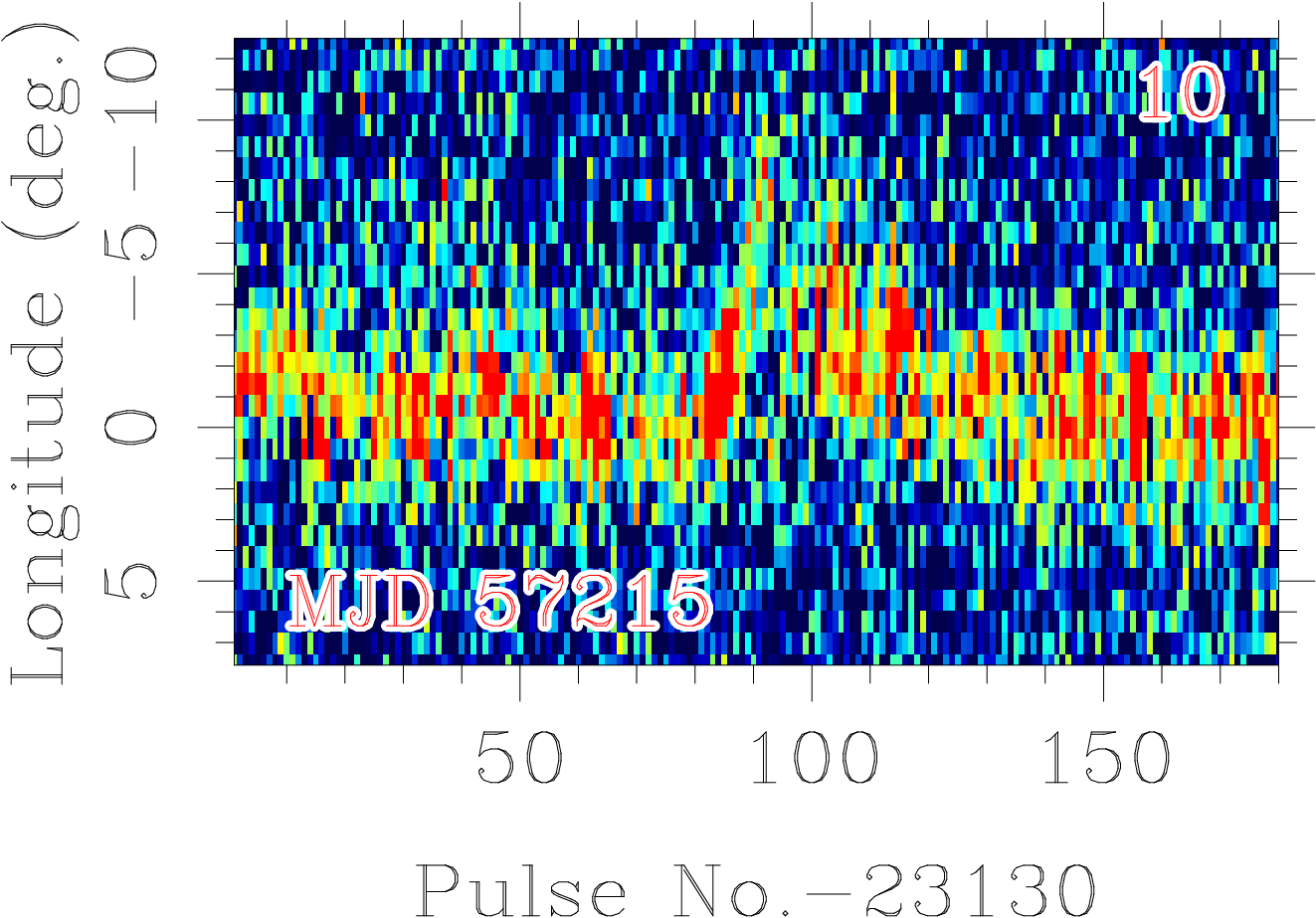}
\caption{The detailed plots for abnormal emission events of ``$\lambda$ type''. }
  \label{fig11}
\end{figure}
\begin{figure}
 \includegraphics[bb = 60 45 505 685,clip,width=0.114\textwidth,angle=-90]{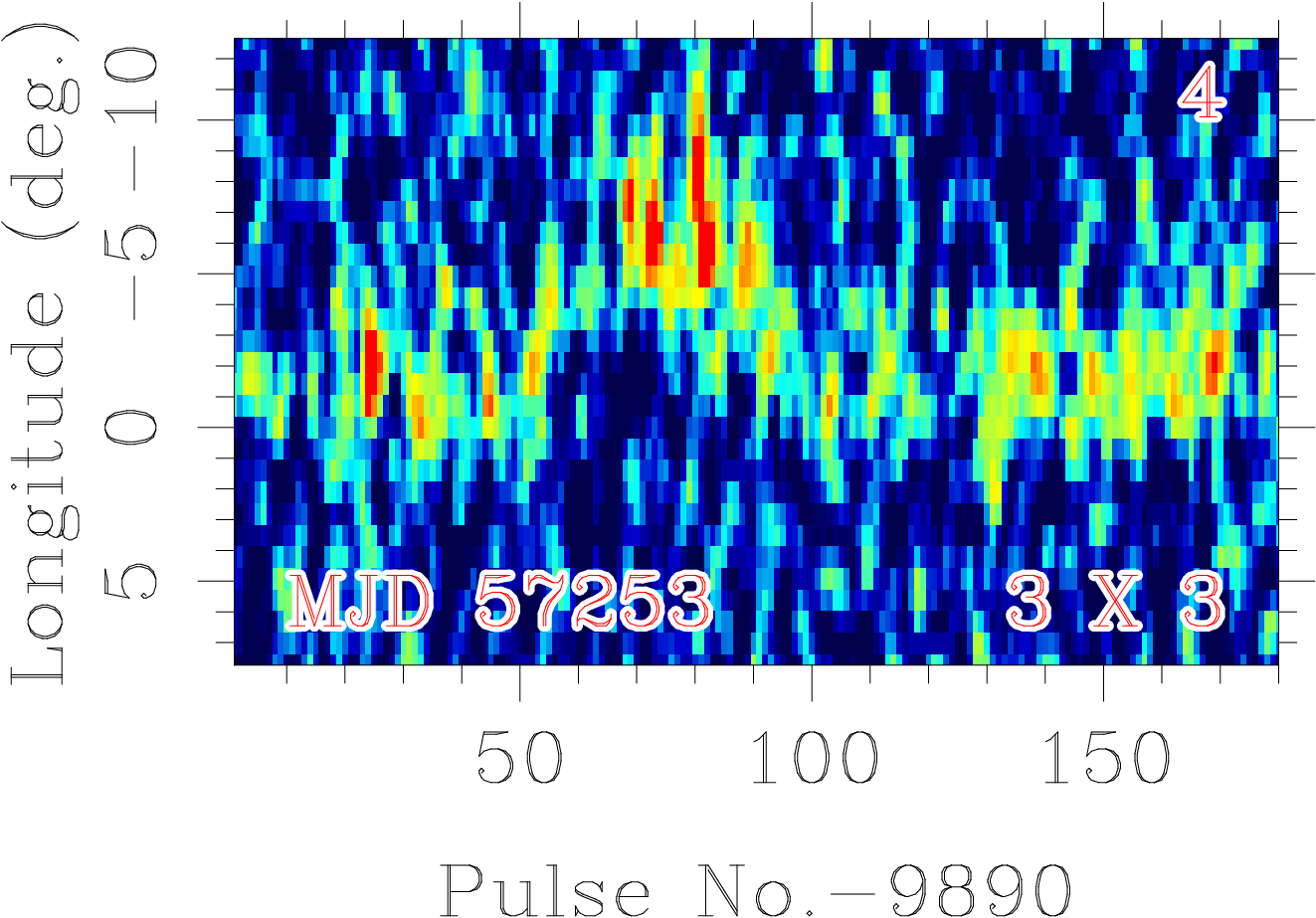}
 \includegraphics[bb = 60 95 505 685,clip,width=0.114\textwidth,angle=-90]{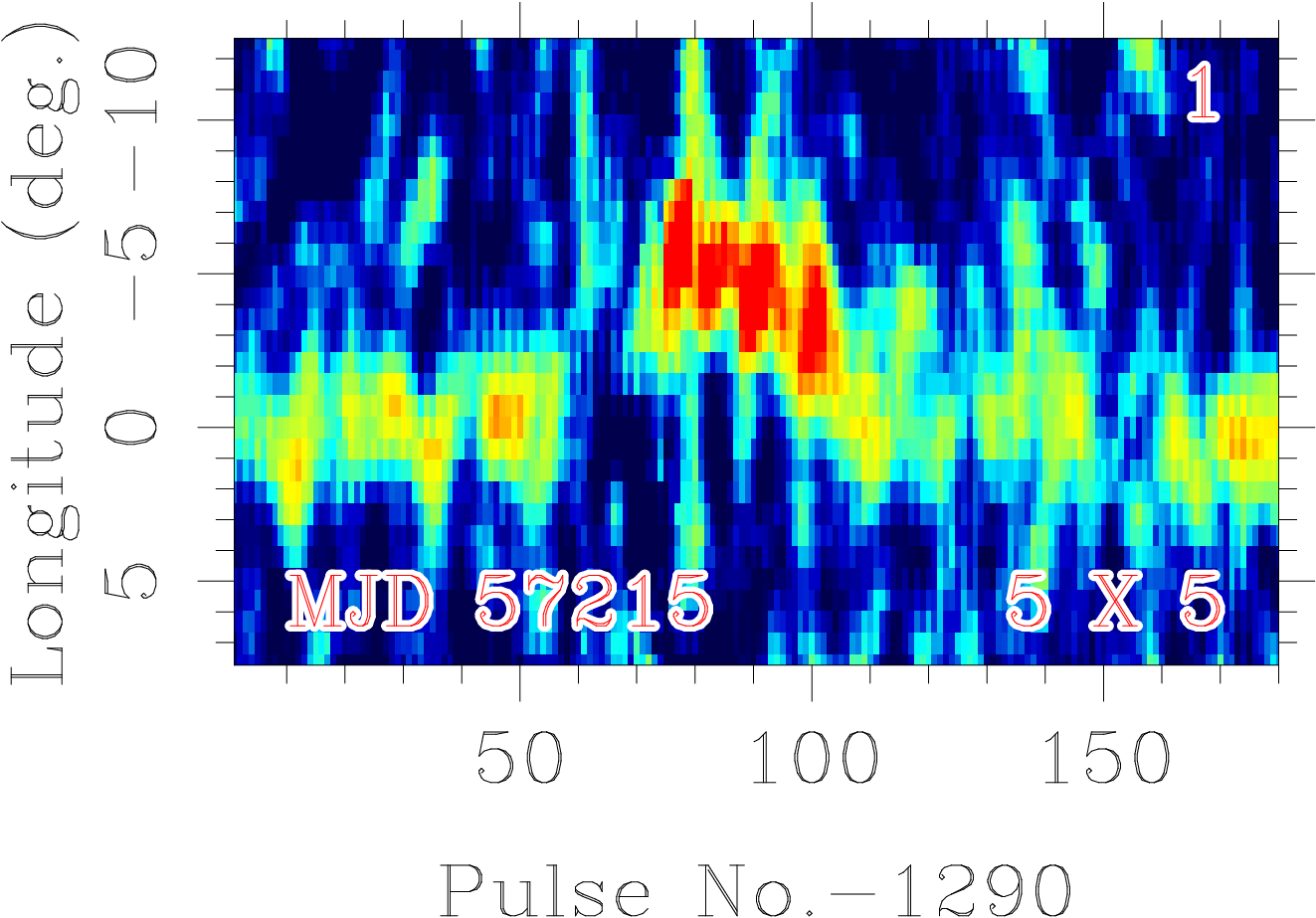}
 \includegraphics[bb = 60 95 505 685,clip,width=0.114\textwidth,angle=-90]{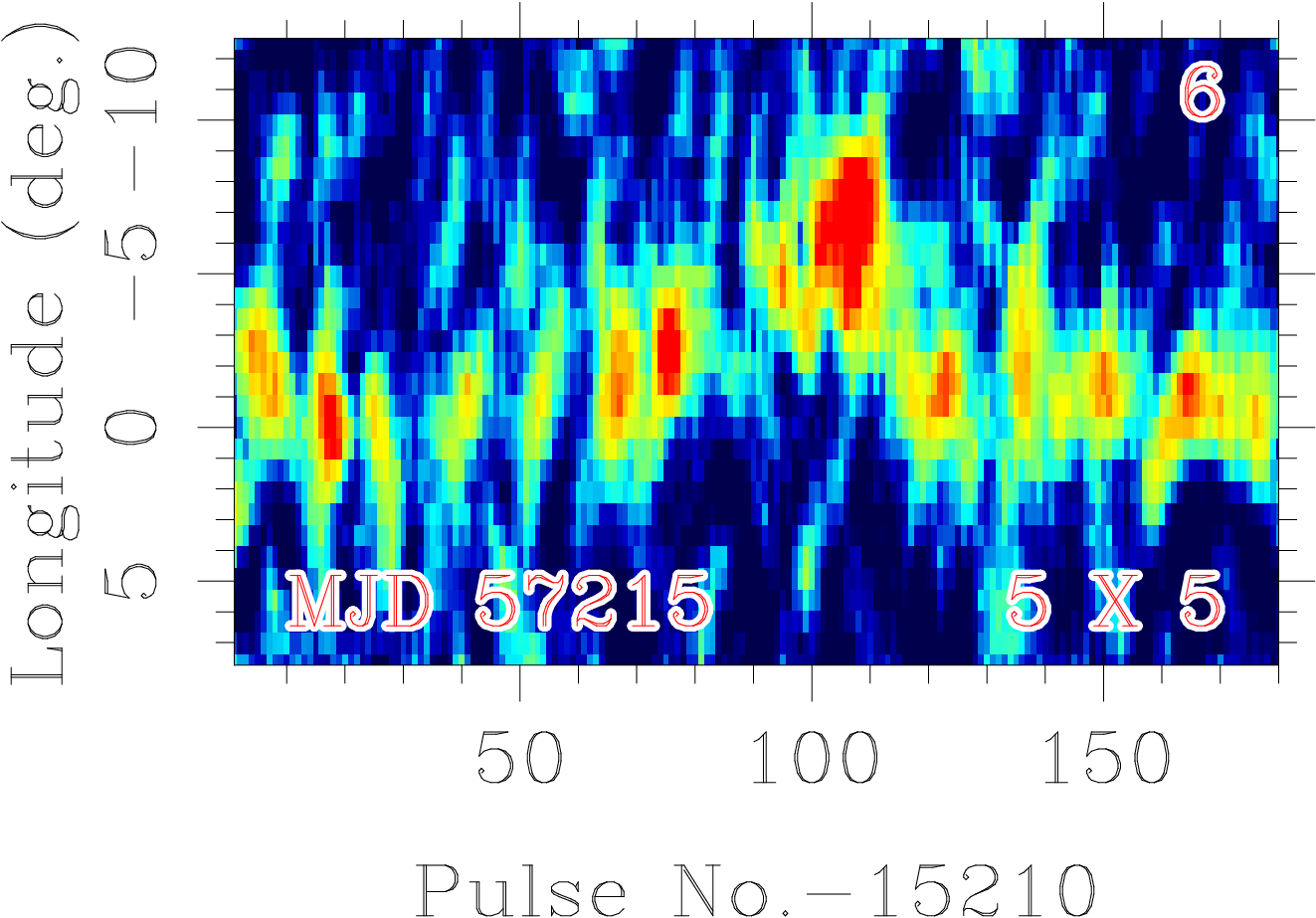}
 \includegraphics[bb = 60 45 505 685,clip,width=0.114\textwidth,angle=-90]{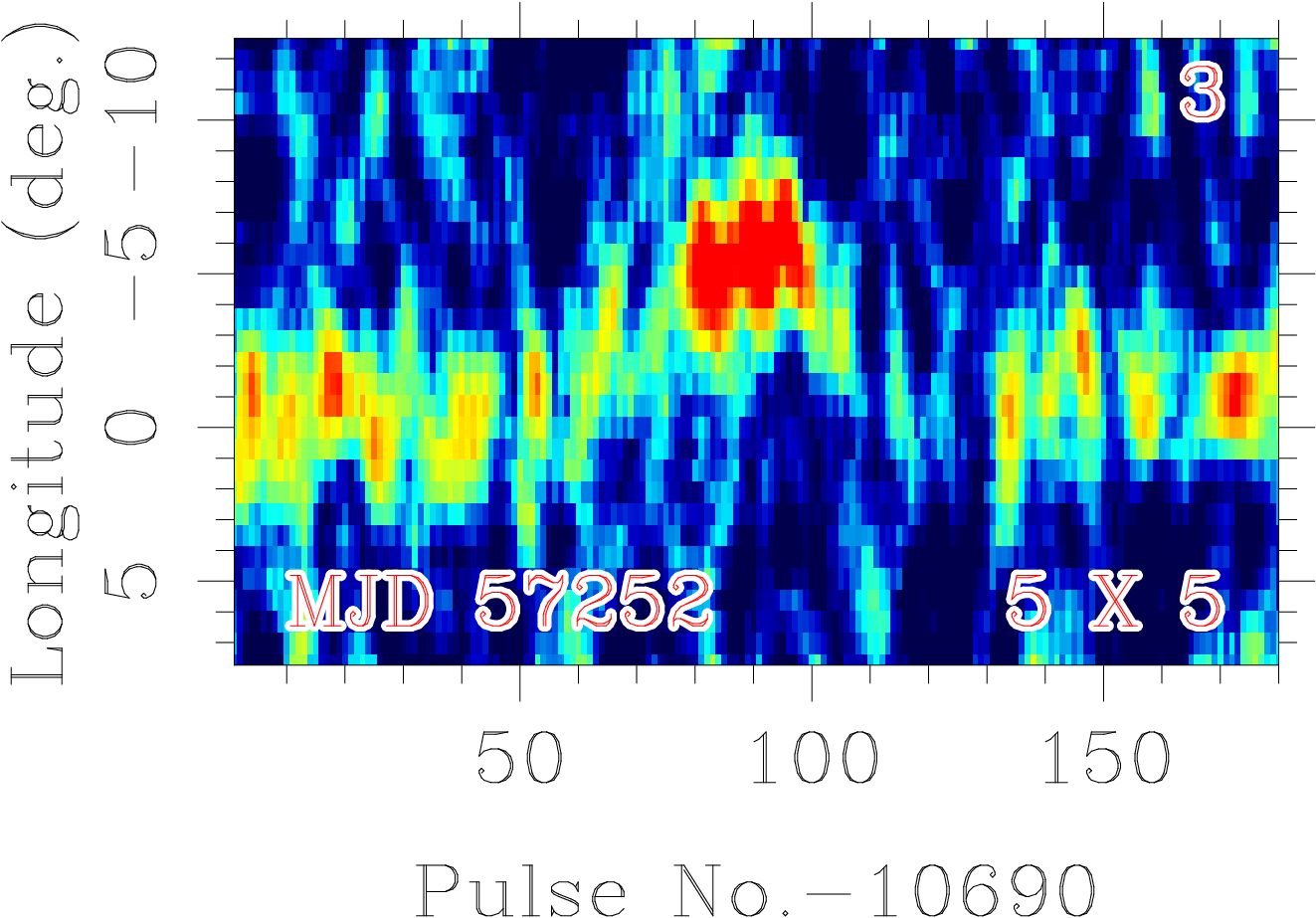}
 \includegraphics[bb = 60 95 505 685,clip,width=0.114\textwidth,angle=-90]{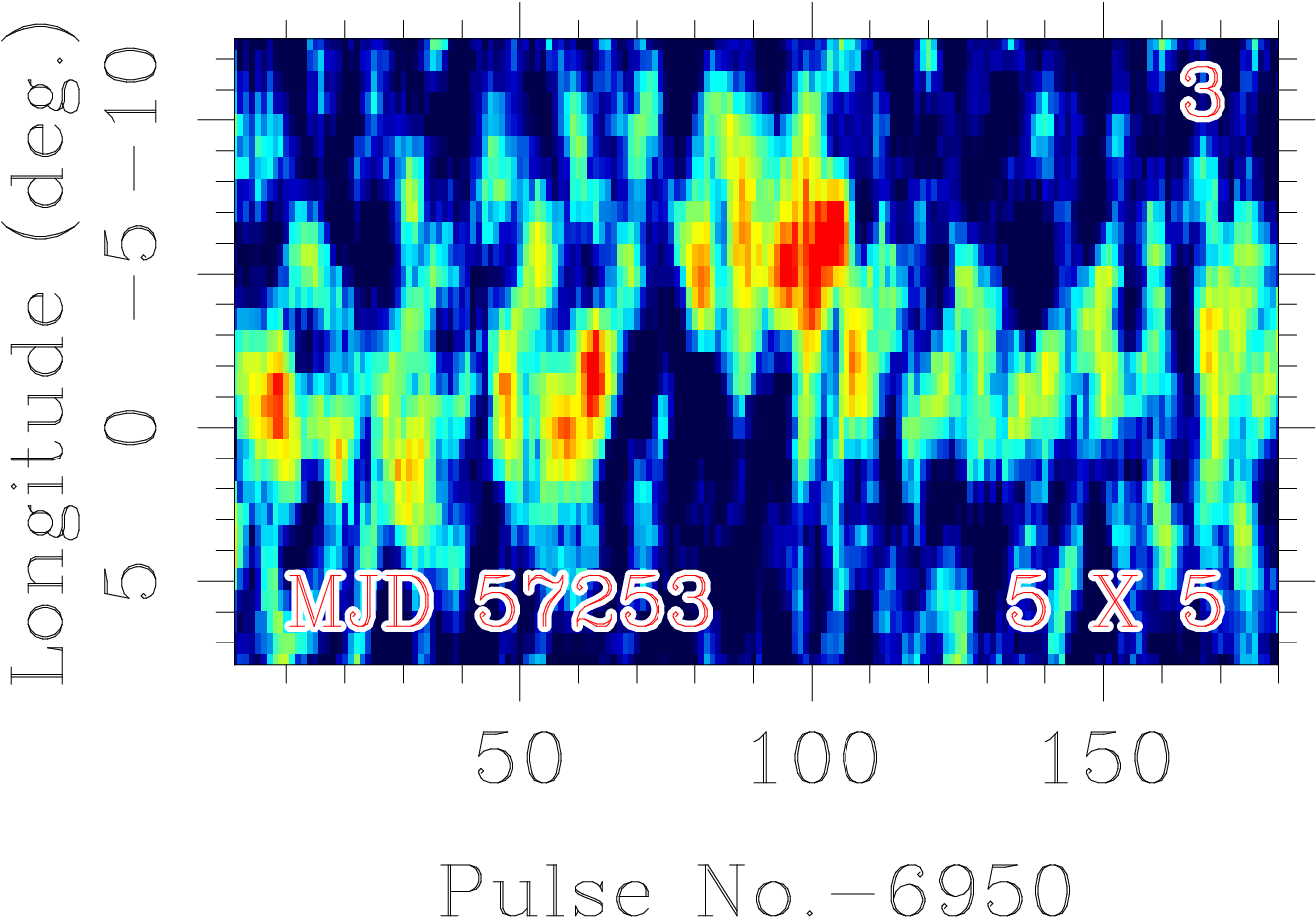}
 \includegraphics[bb = 60 95 505 685,clip,width=0.114\textwidth,angle=-90]{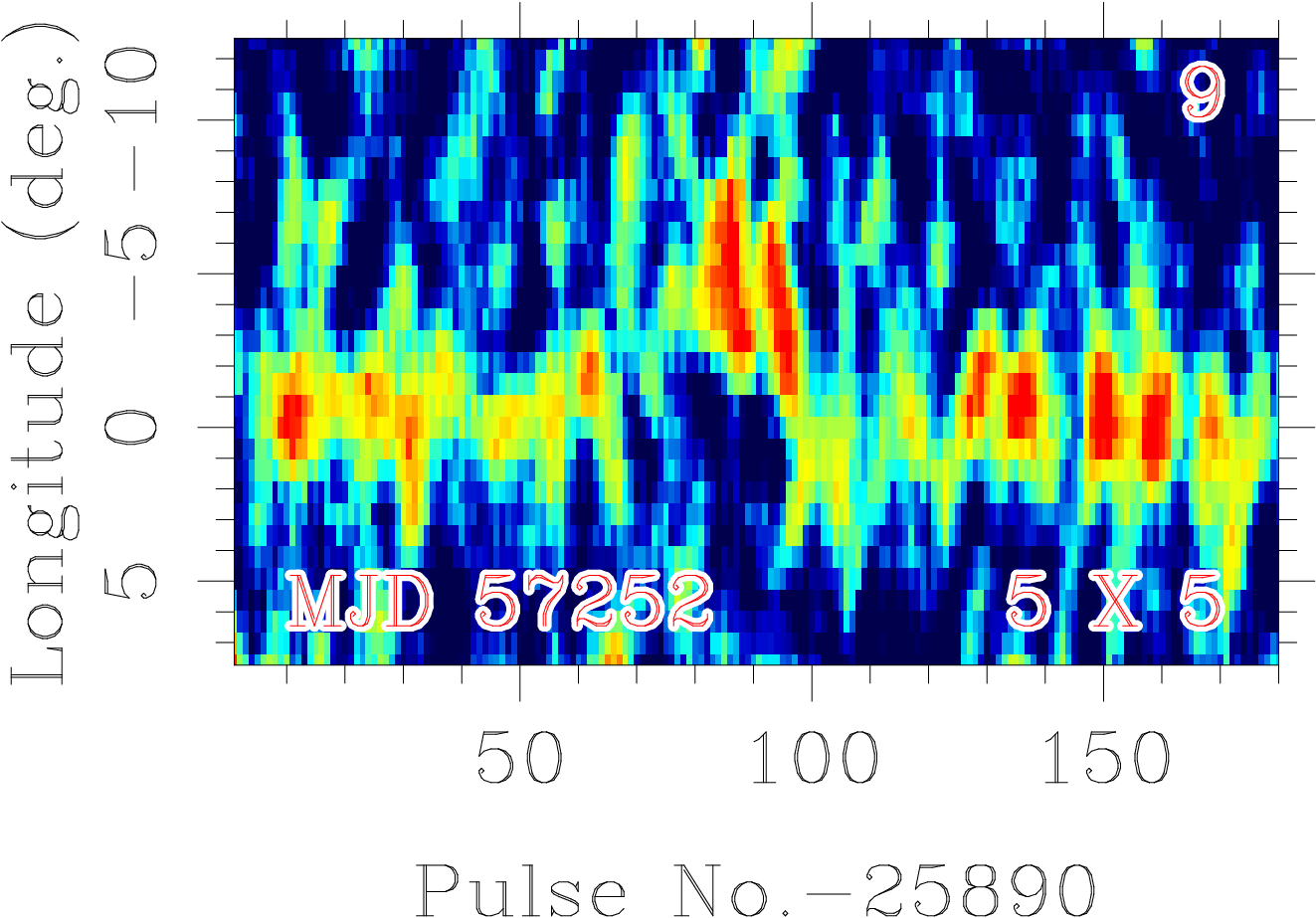}
 \includegraphics[bb = 60 45 505 685,clip,width=0.114\textwidth,angle=-90]{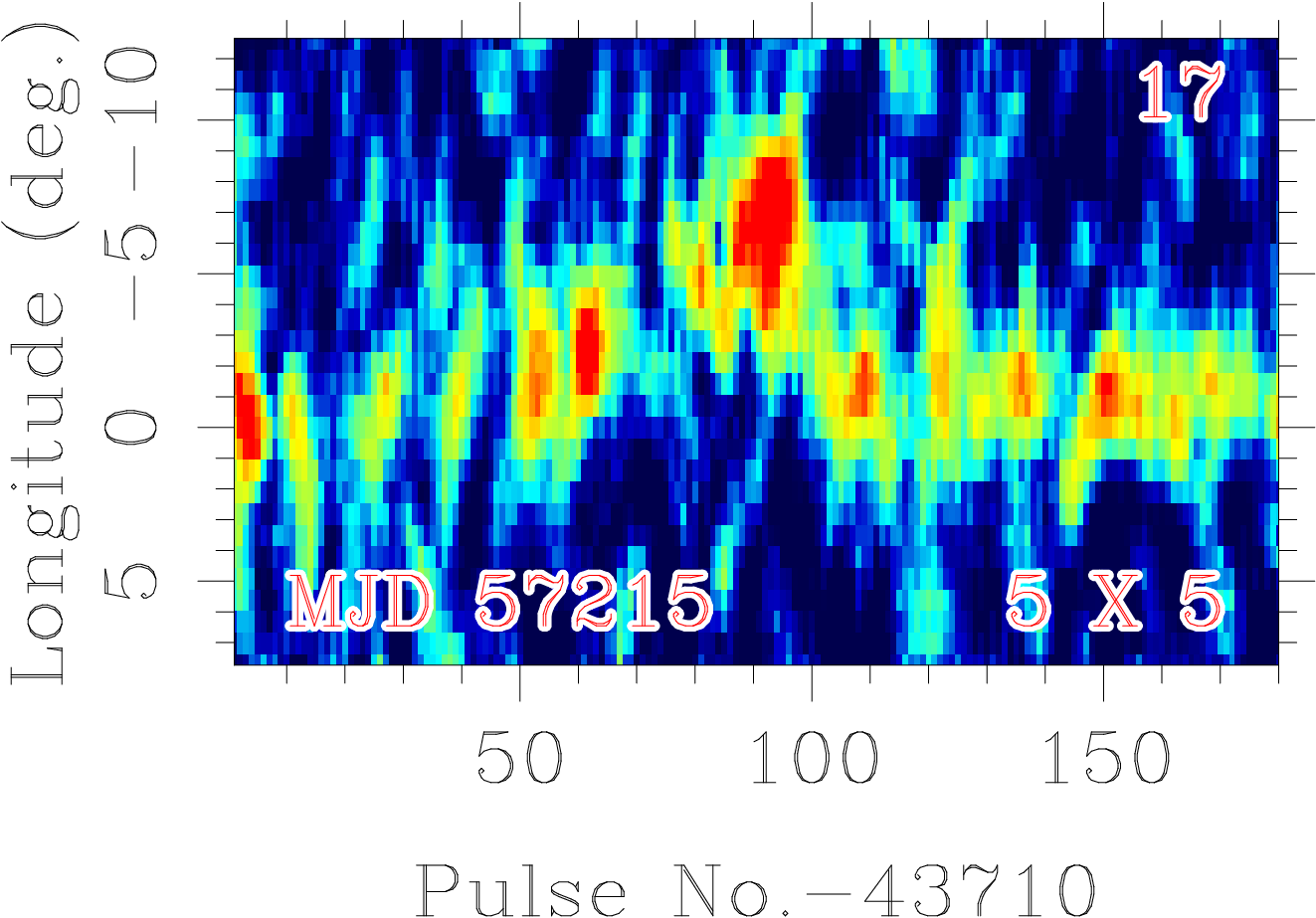}
 \includegraphics[bb = 60 95 505 685,clip,width=0.114\textwidth,angle=-90]{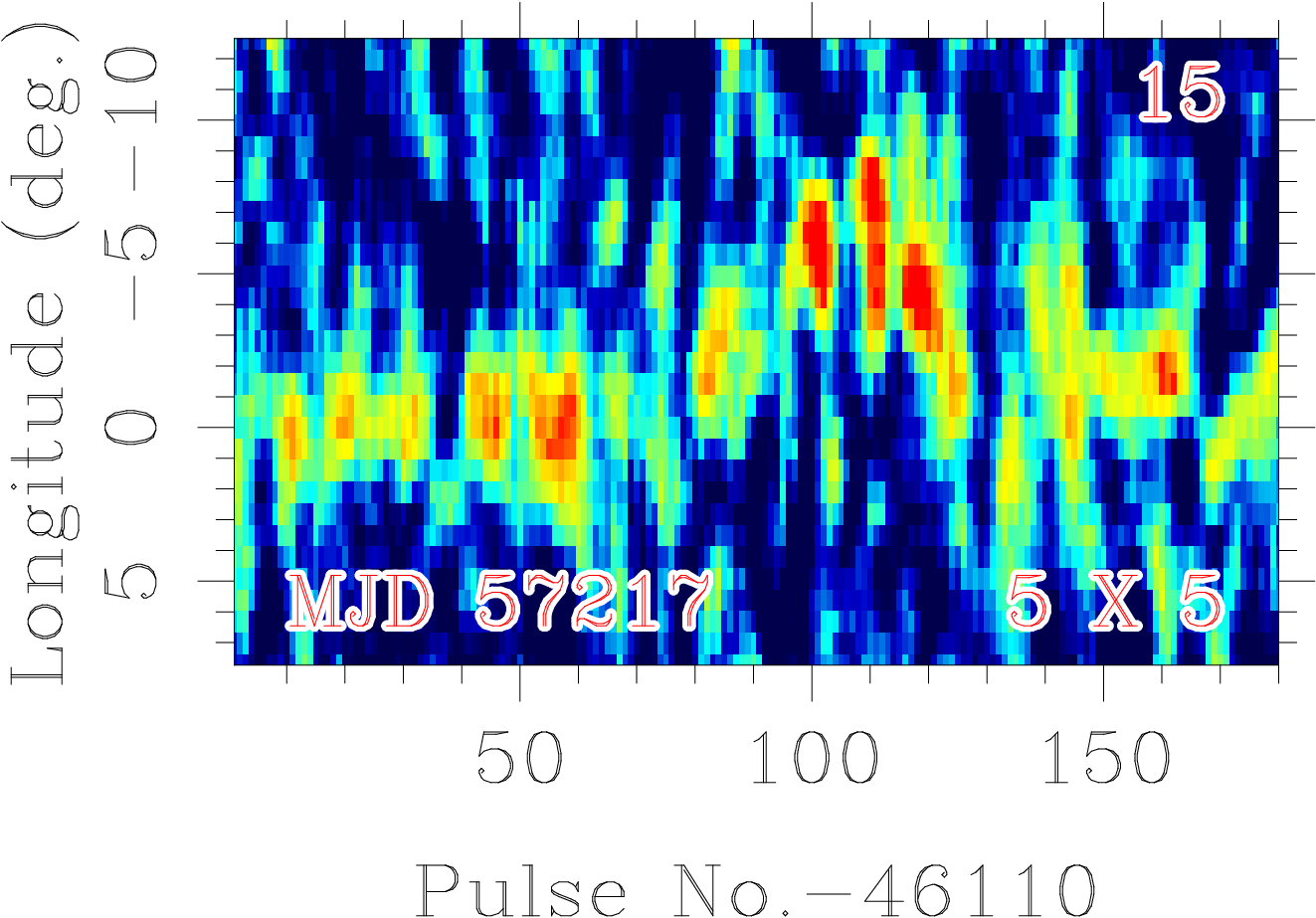}
 \includegraphics[bb = 60 95 505 685,clip,width=0.114\textwidth,angle=-90]{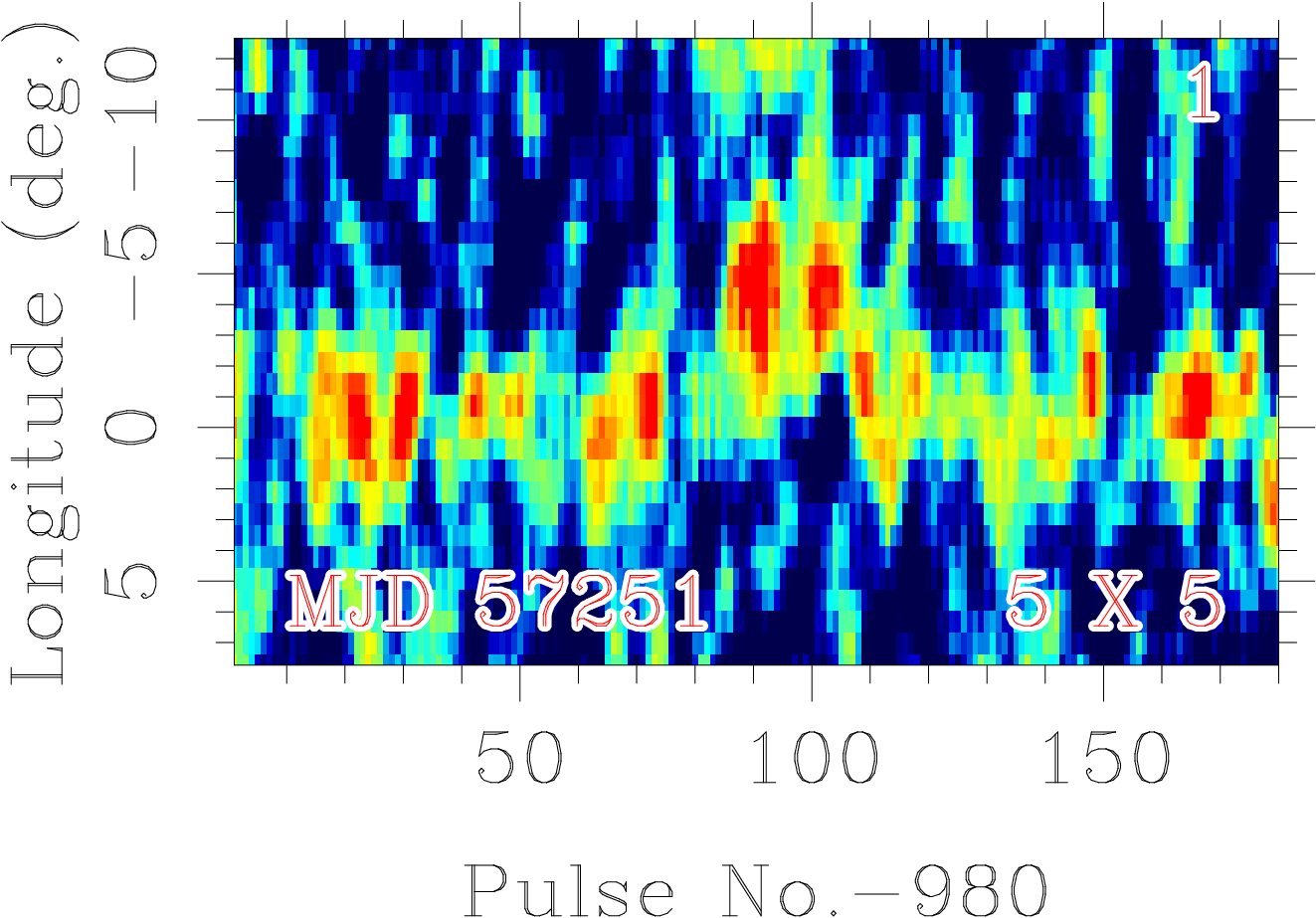}
 \includegraphics[bb = 60 45 505 685,clip,width=0.114\textwidth,angle=-90]{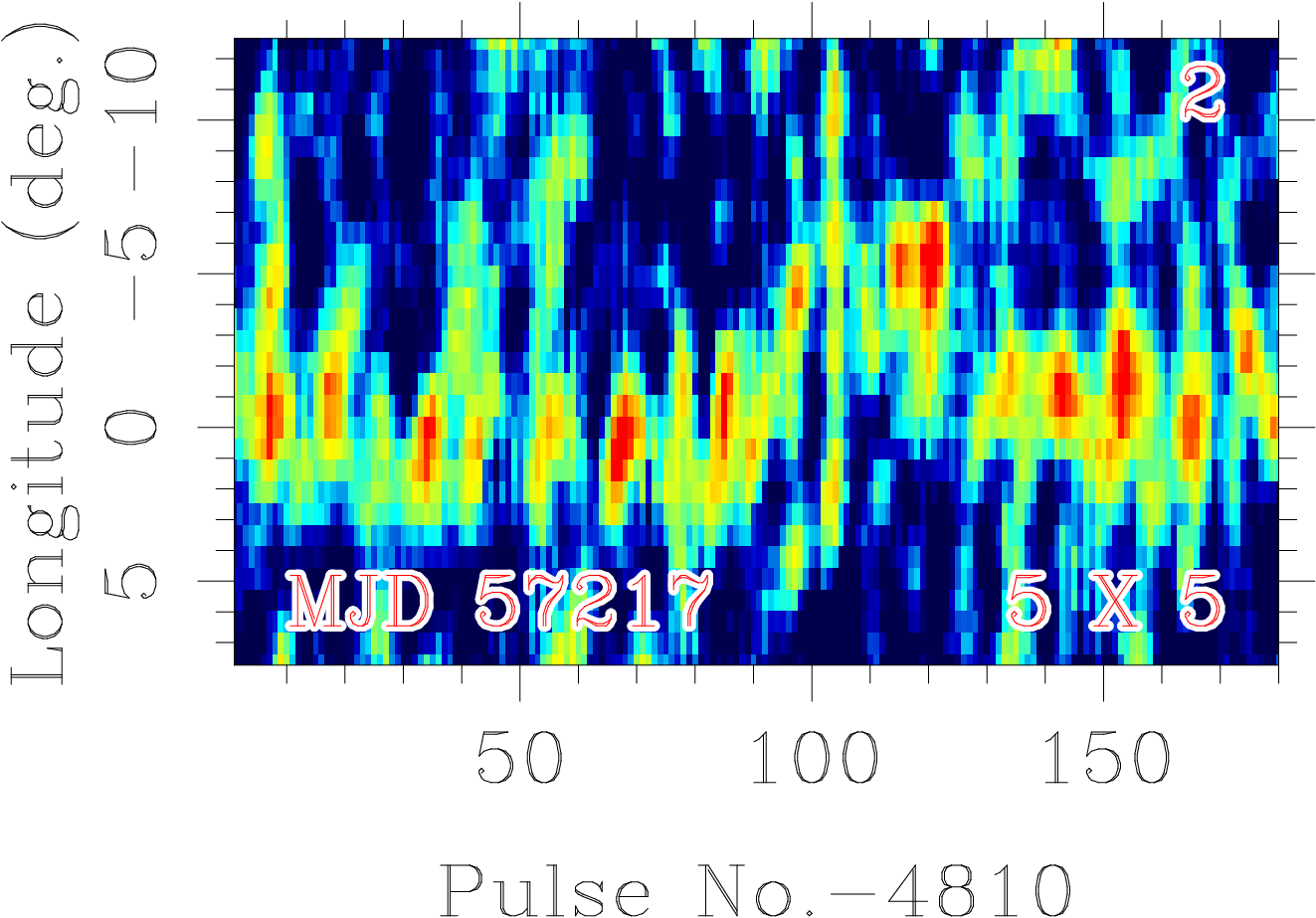}
 \includegraphics[bb = 60 95 505 685,clip,width=0.114\textwidth,angle=-90]{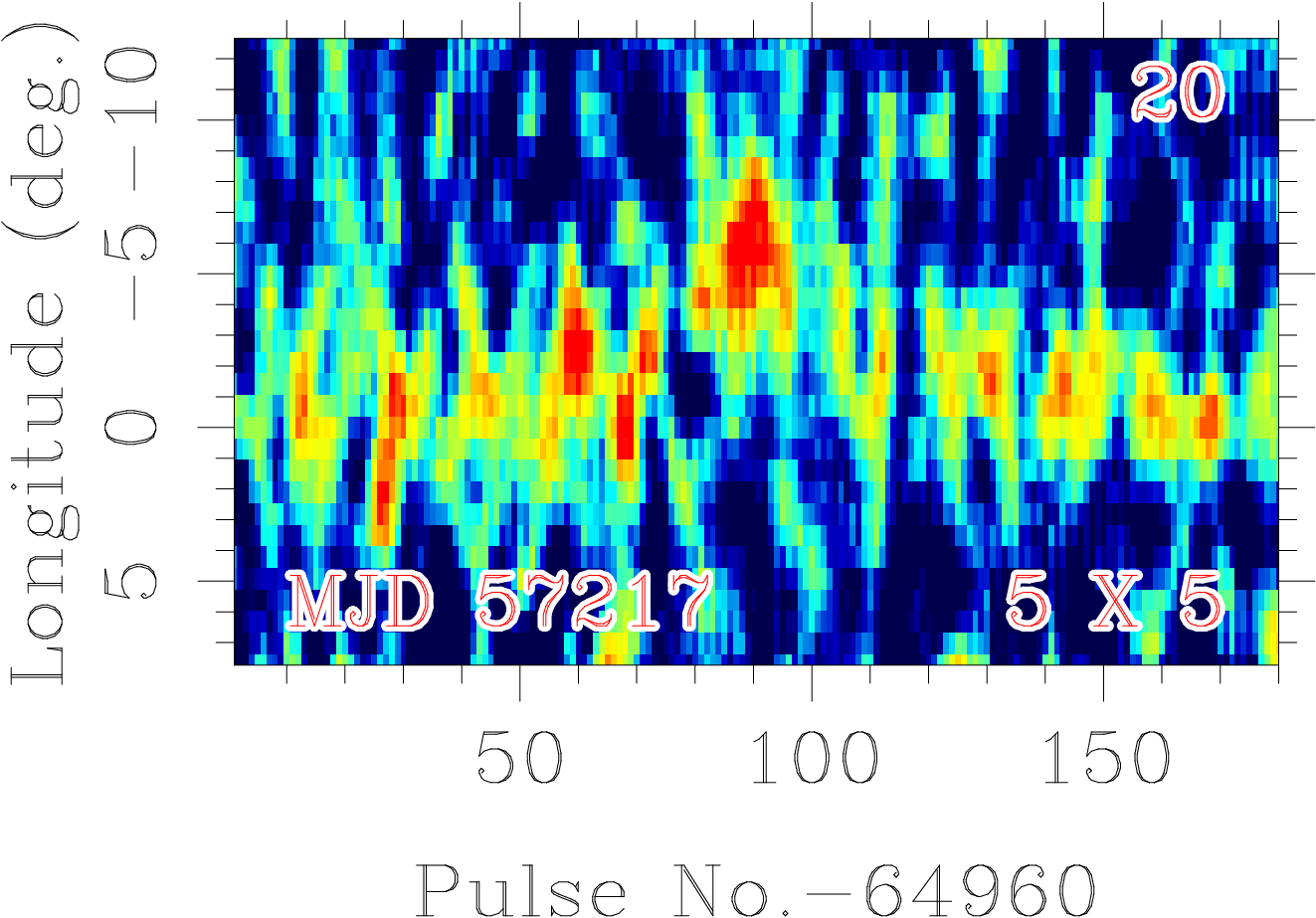}
 \includegraphics[bb = 60 95 505 685,clip,width=0.114\textwidth,angle=-90]{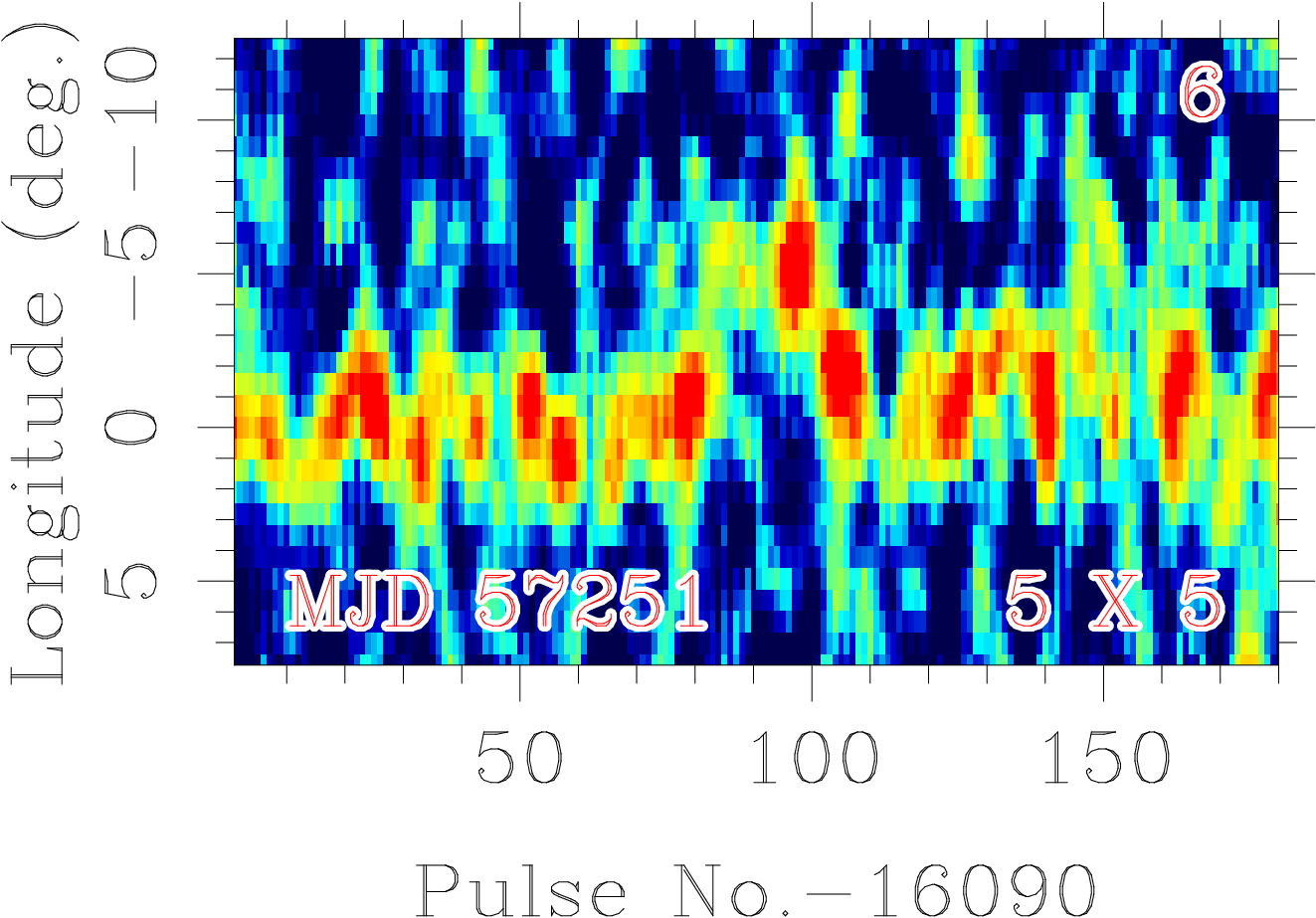}
 \includegraphics[bb = 60 45 505 685,clip,width=0.114\textwidth,angle=-90]{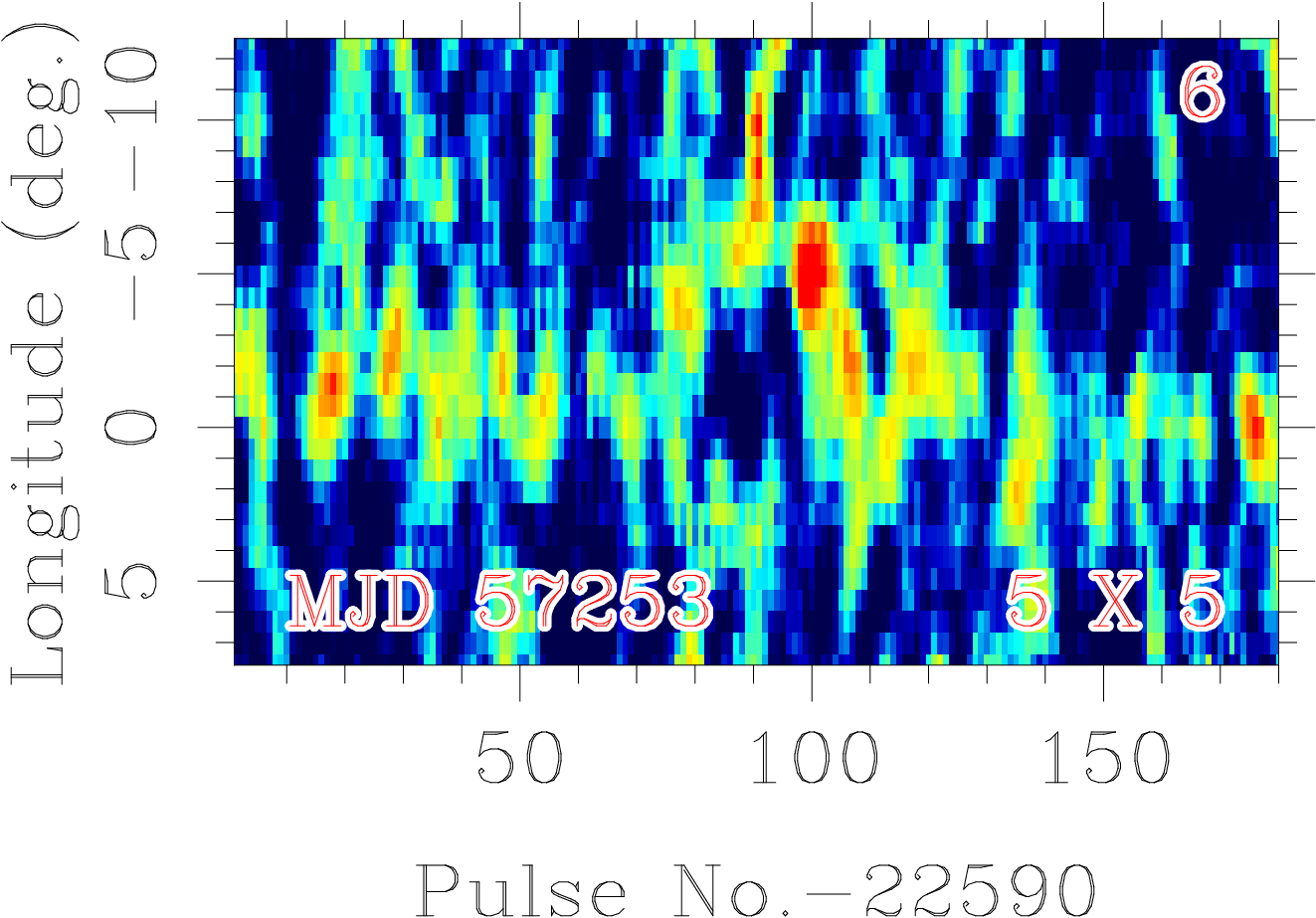}
 \includegraphics[bb = 60 95 505 685,clip,width=0.114\textwidth,angle=-90]{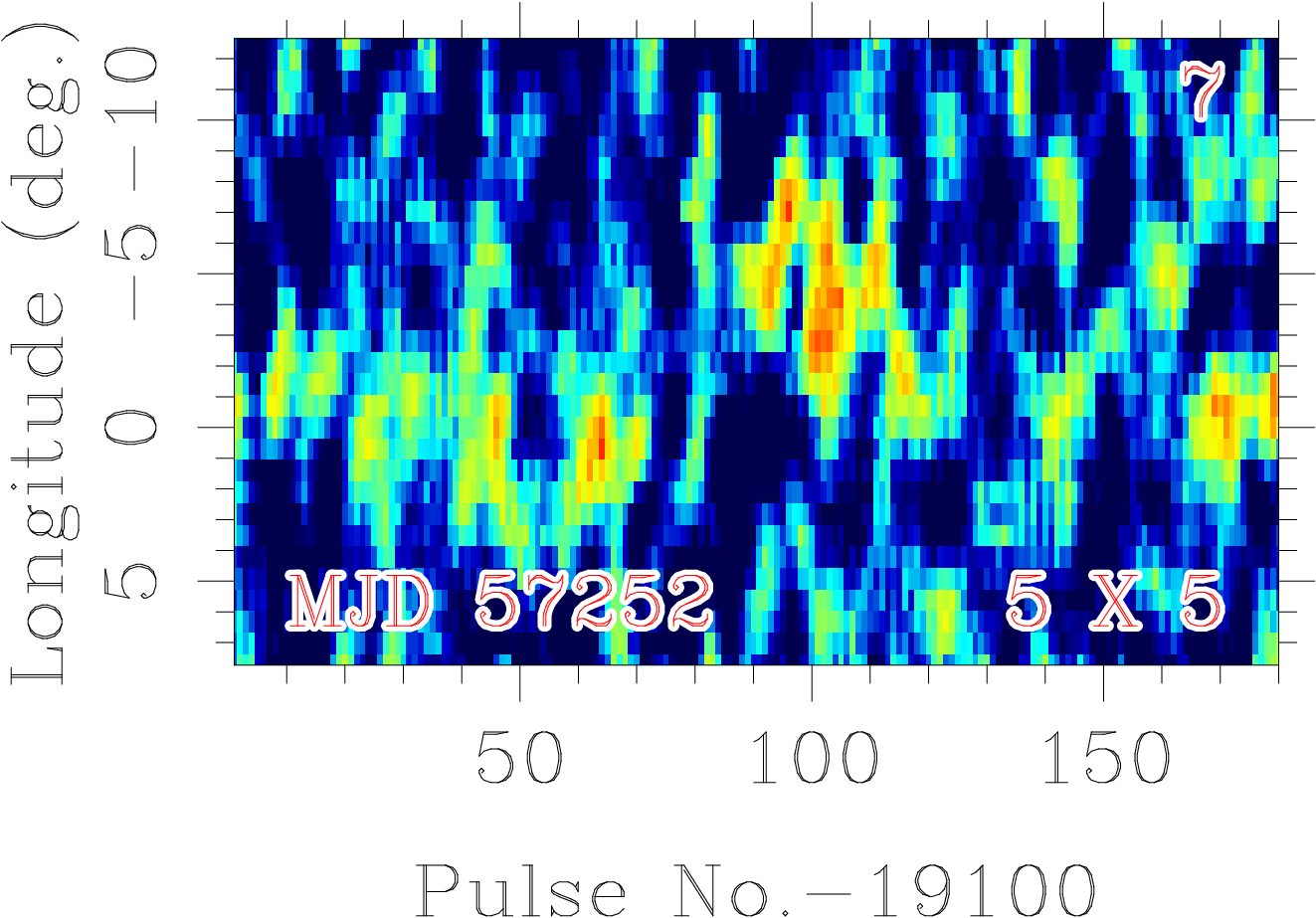}
 \includegraphics[bb = 60 95 505 685,clip,width=0.114\textwidth,angle=-90]{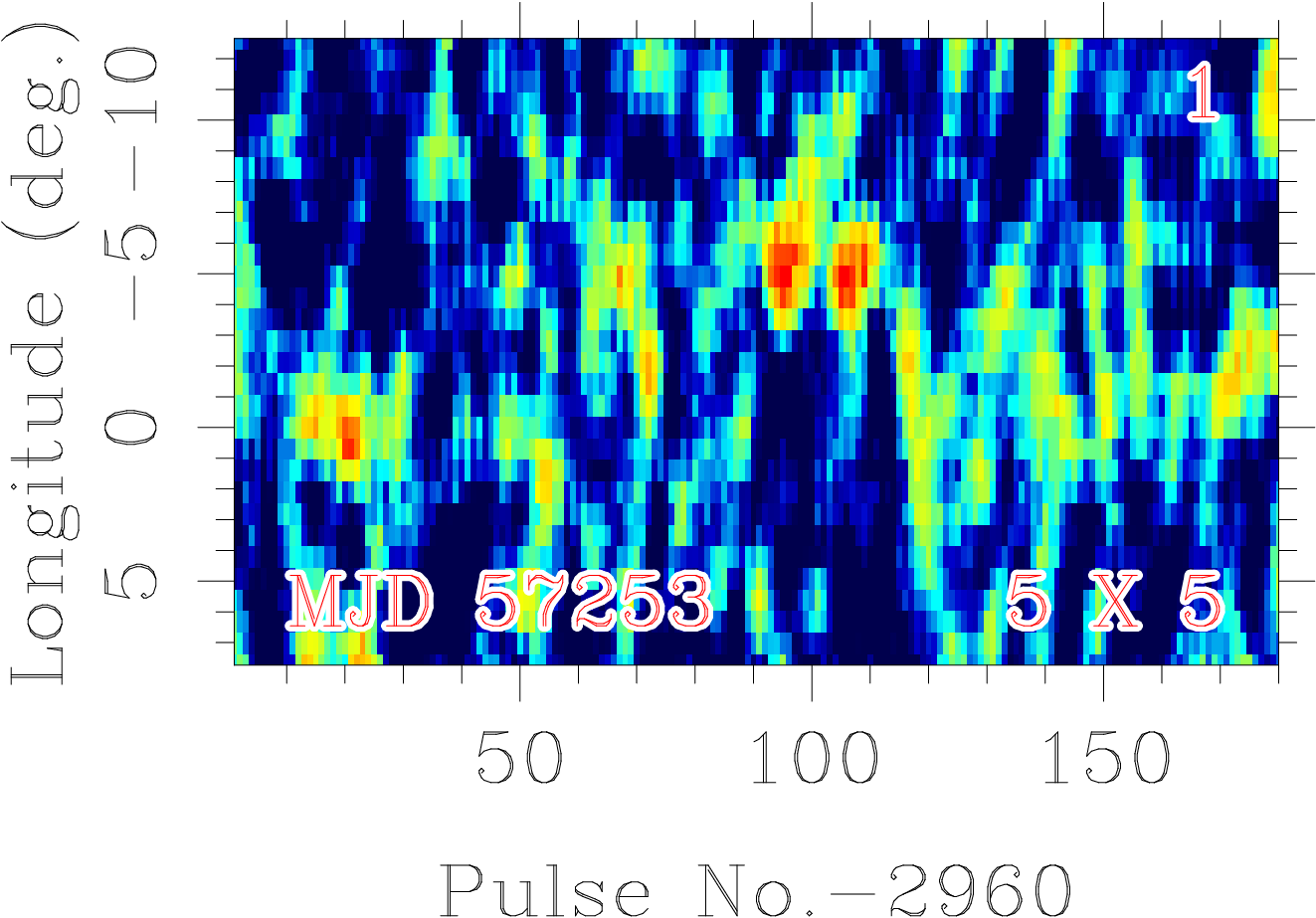}
 \includegraphics[bb = 60 45 505 685,clip,width=0.114\textwidth,angle=-90]{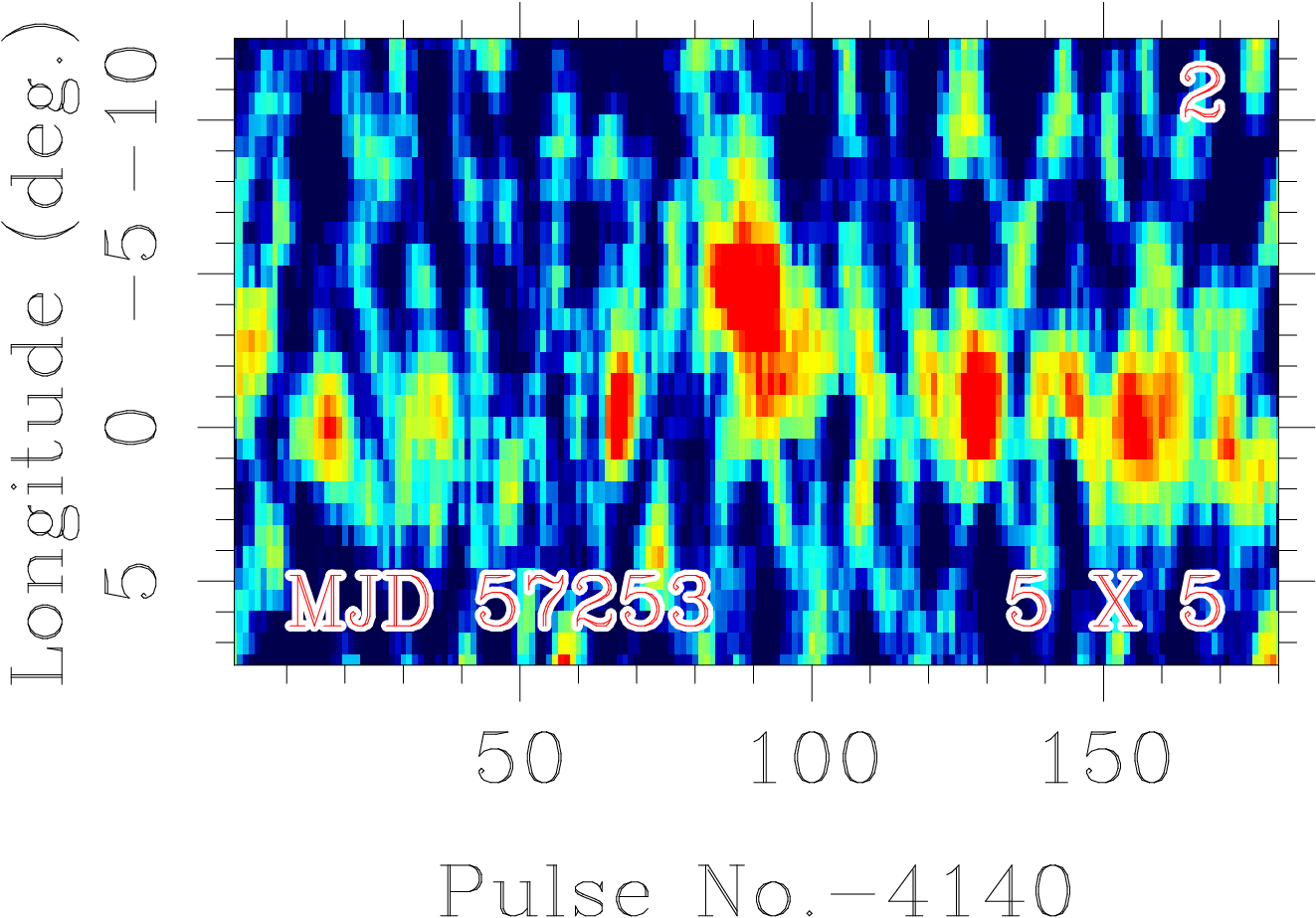}
 \includegraphics[bb = 60 95 505 685,clip,width=0.114\textwidth,angle=-90]{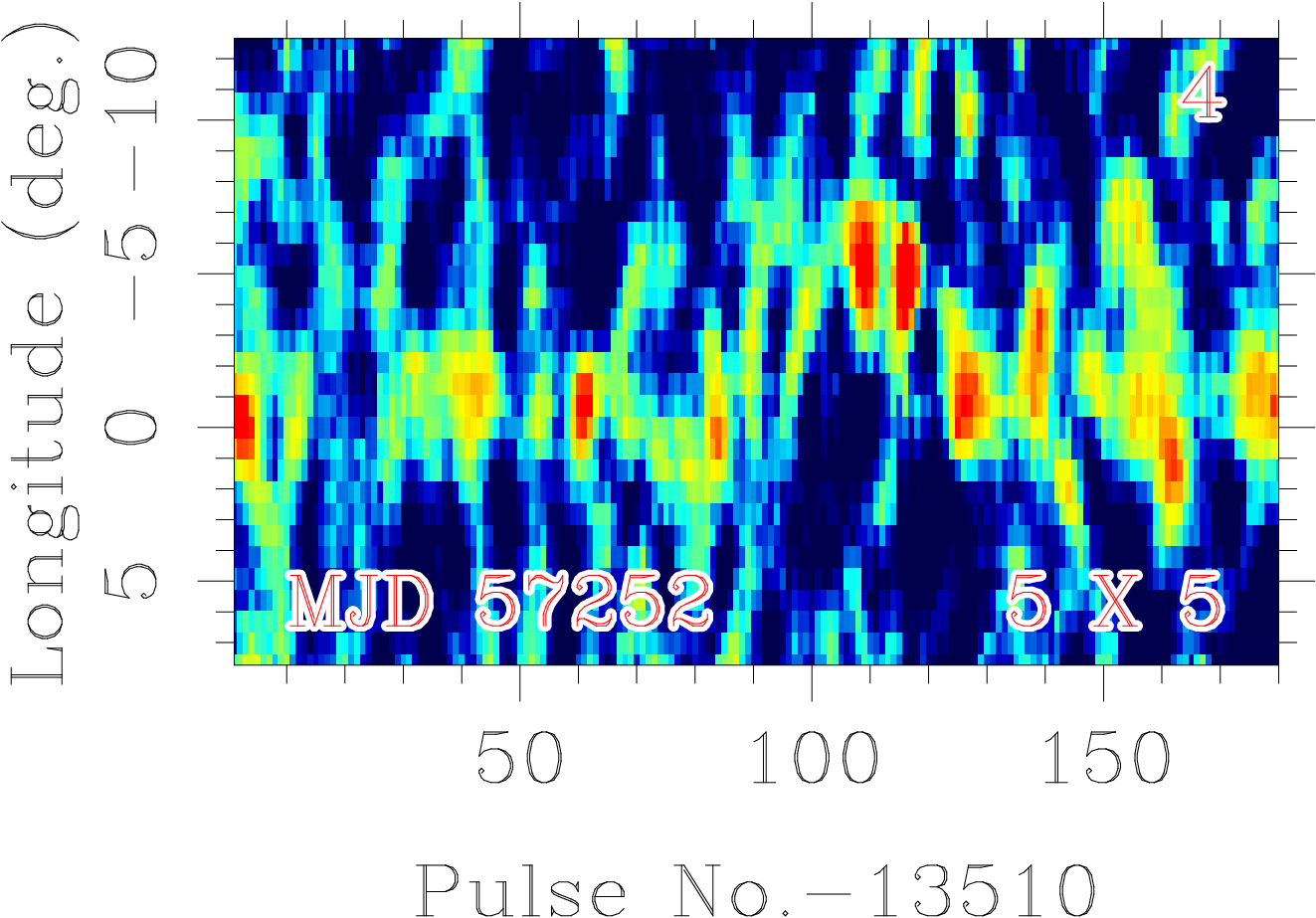}
 \includegraphics[bb = 60 95 505 685,clip,width=0.114\textwidth,angle=-90]{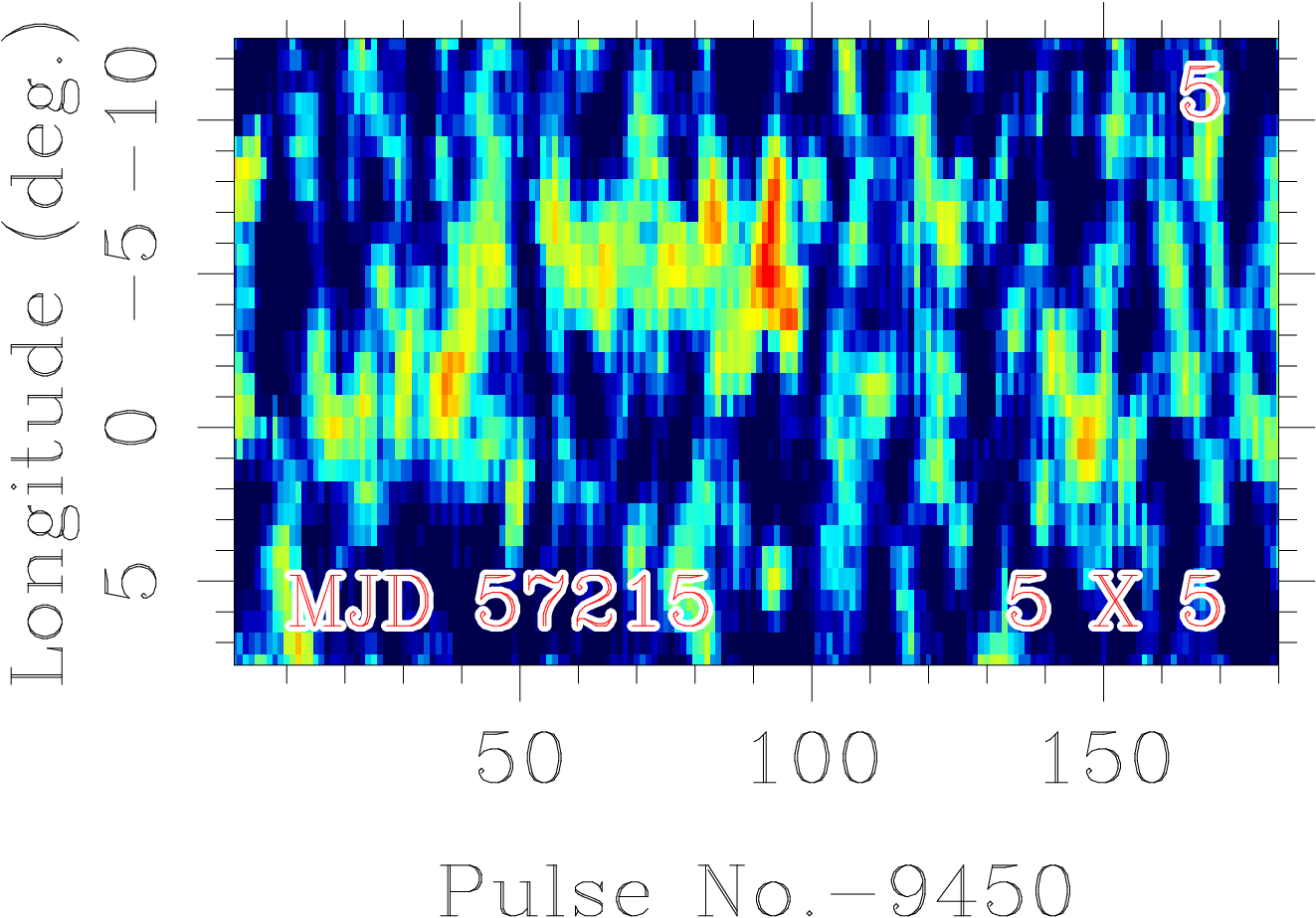}
 \includegraphics[bb = 60 45 505 685,clip,width=0.114\textwidth,angle=-90]{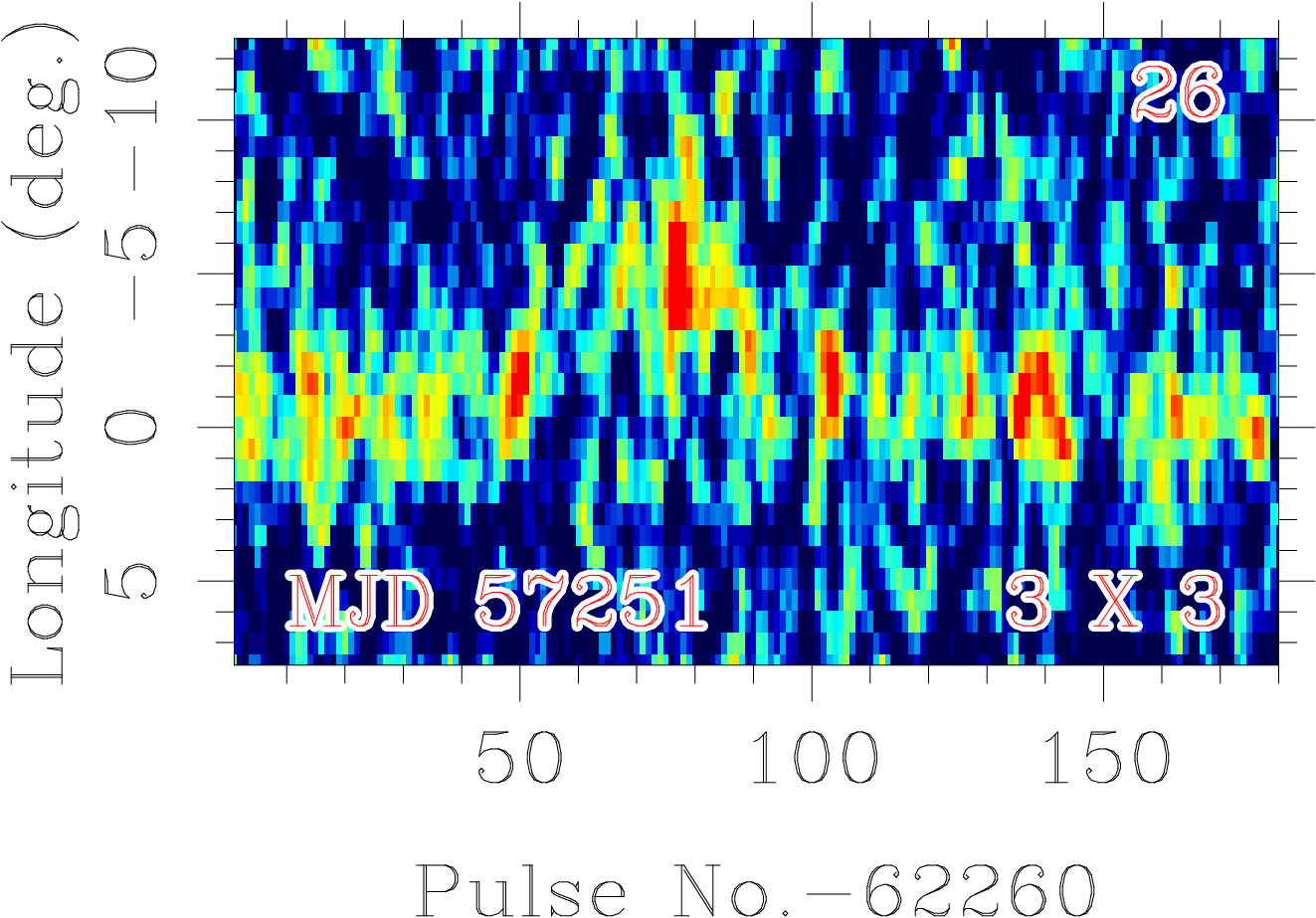} 
 \includegraphics[bb = 60 95 505 685,clip,width=0.114\textwidth,angle=-90]{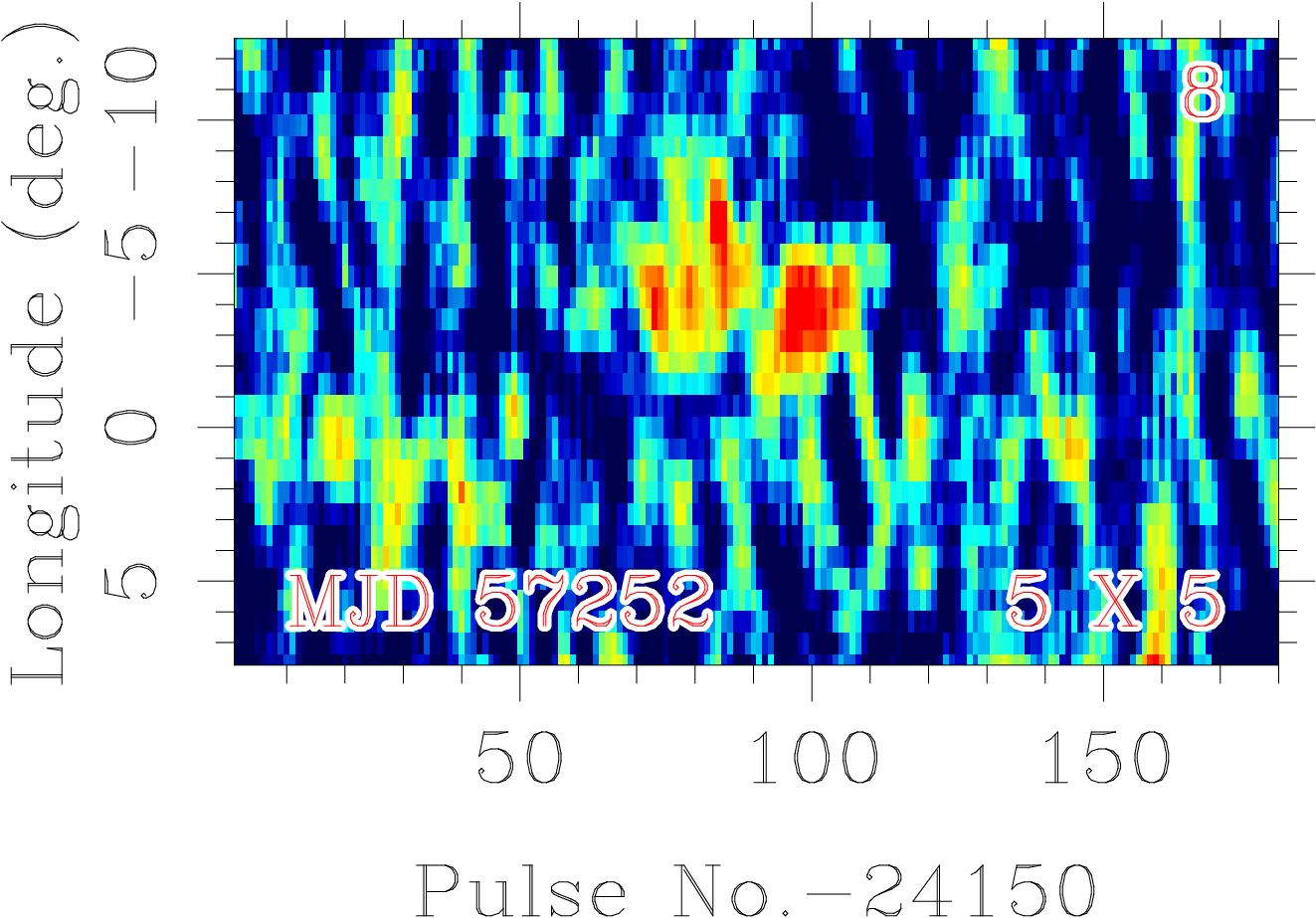}
 \includegraphics[bb = 60 95 505 685,clip,width=0.114\textwidth,angle=-90]{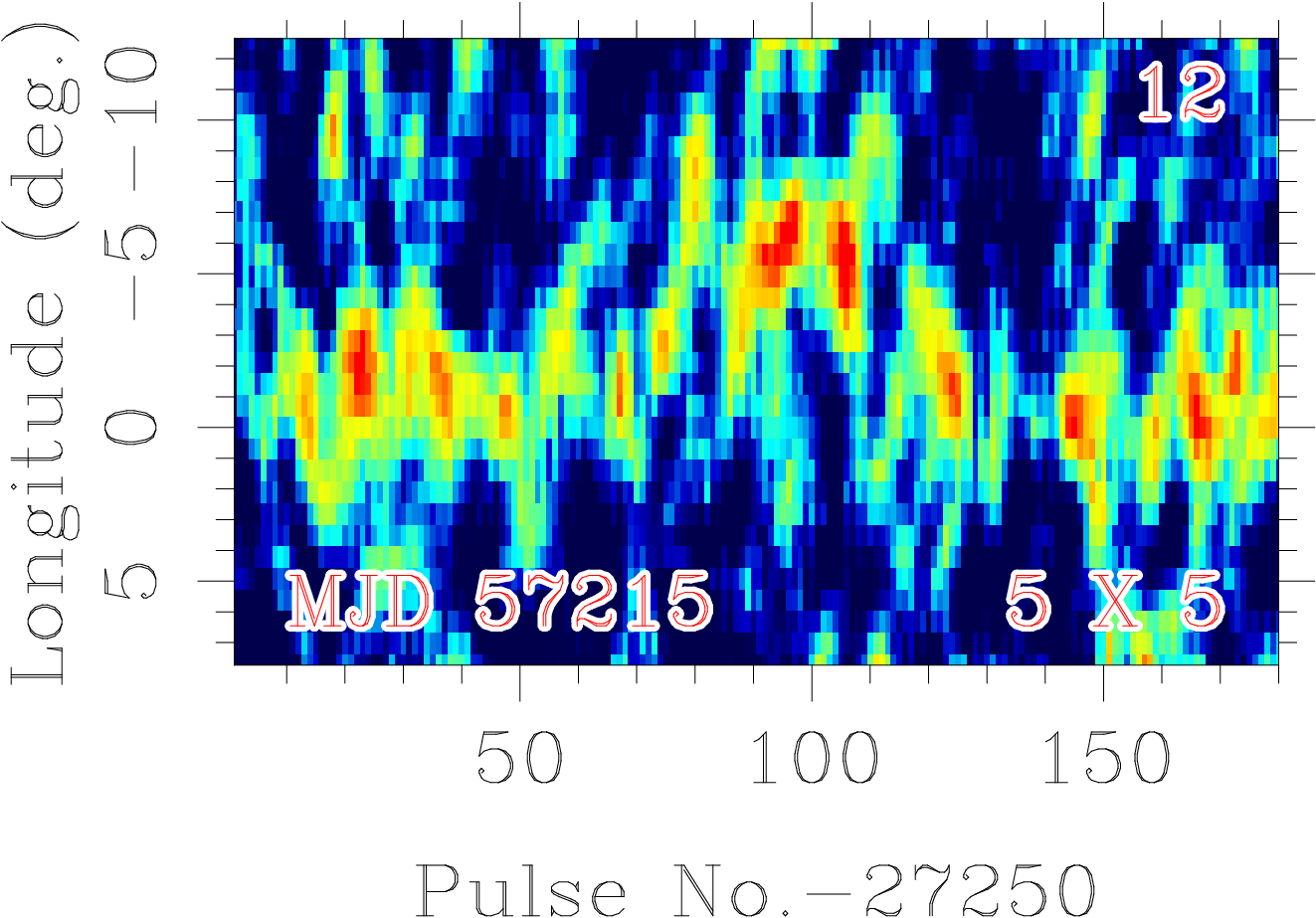}
 \includegraphics[bb = 60 45 505 685,clip,width=0.114\textwidth,angle=-90]{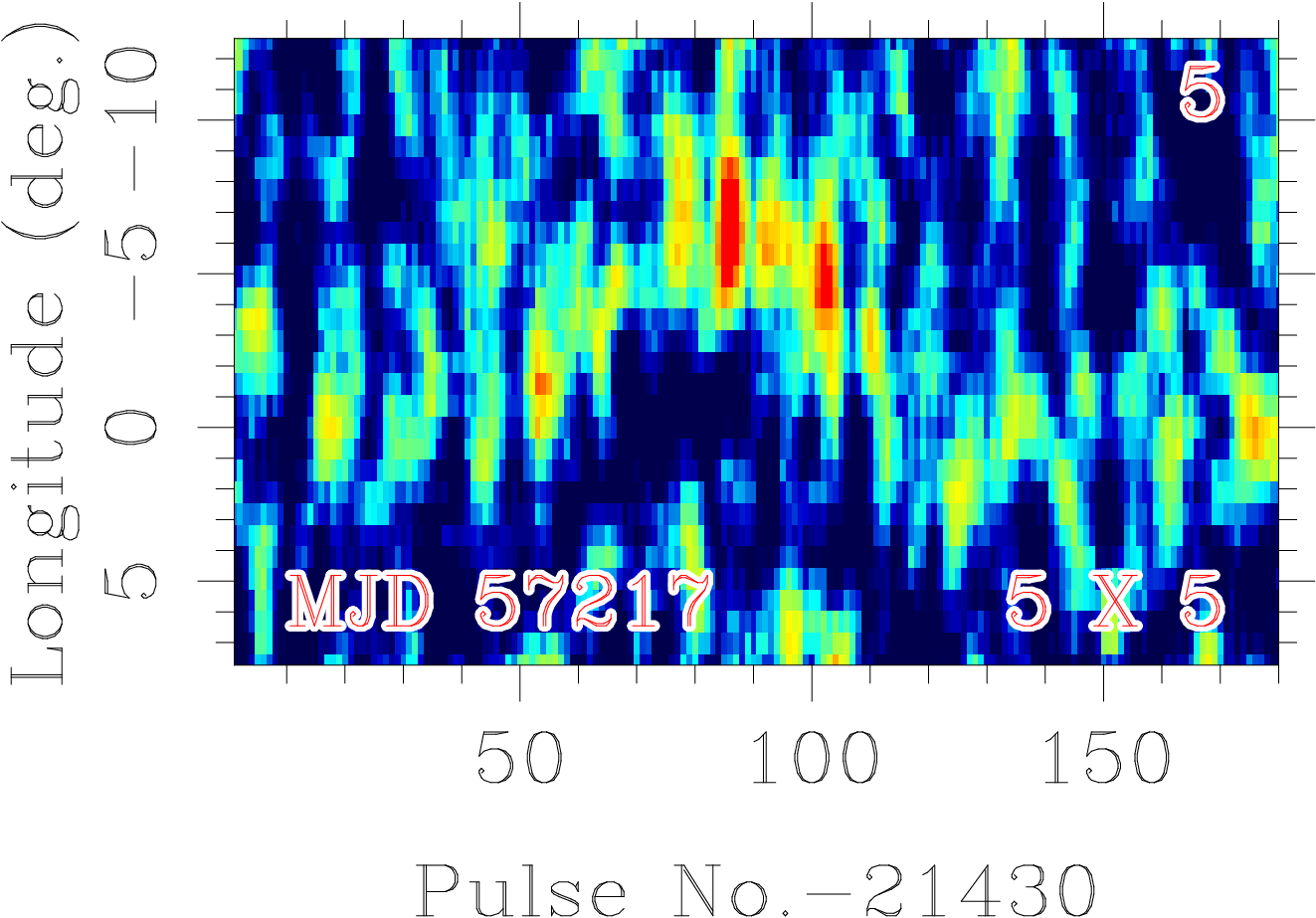}
 \includegraphics[bb = 60 95 505 685,clip,width=0.114\textwidth,angle=-90]{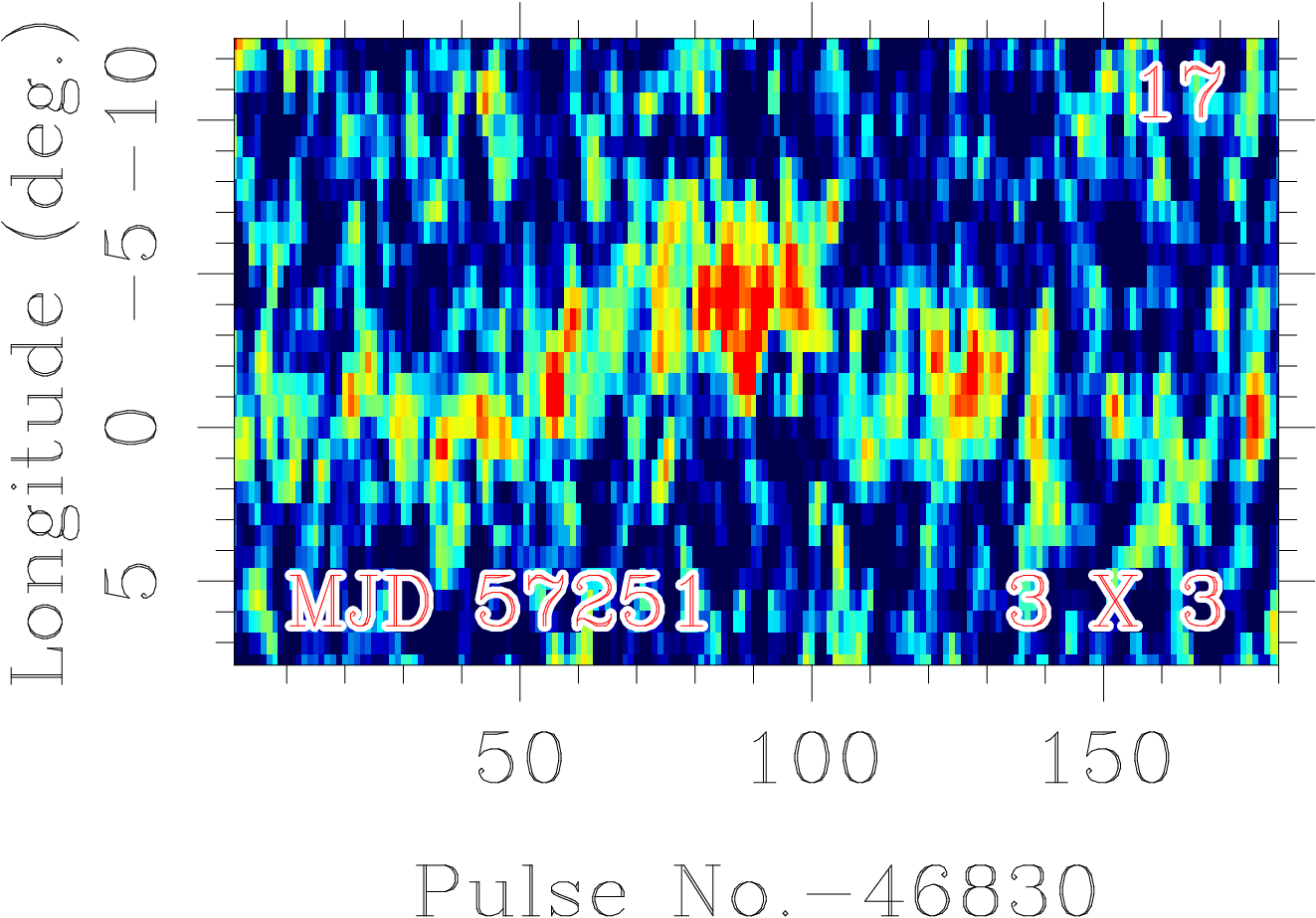} 
 \includegraphics[bb = 60 95 505 685,clip,width=0.114\textwidth,angle=-90]{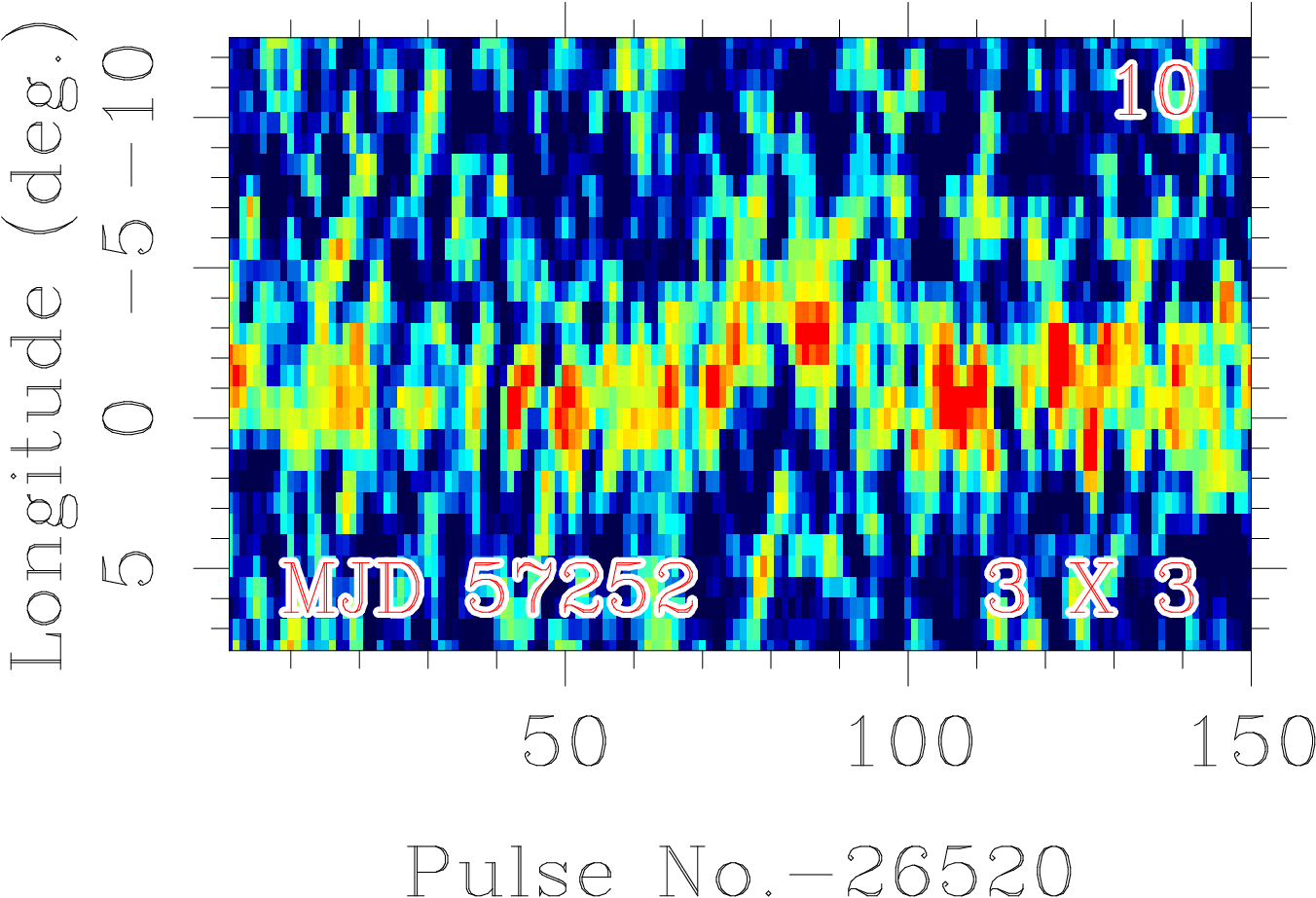} 
 \includegraphics[bb = 60 45 505 685,clip,width=0.114\textwidth,angle=-90]{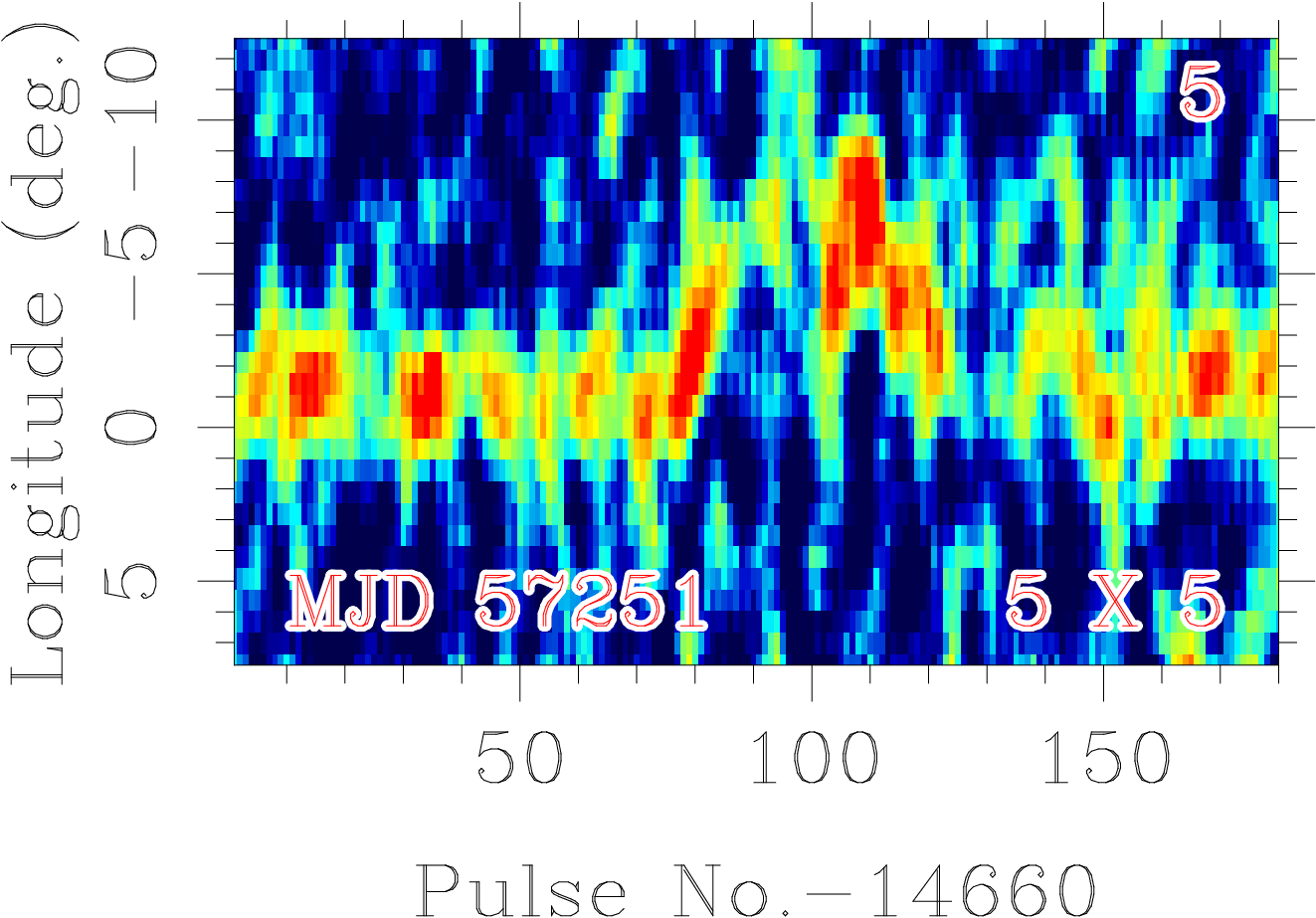}
 \includegraphics[bb = 60 95 505 685,clip,width=0.114\textwidth,angle=-90]{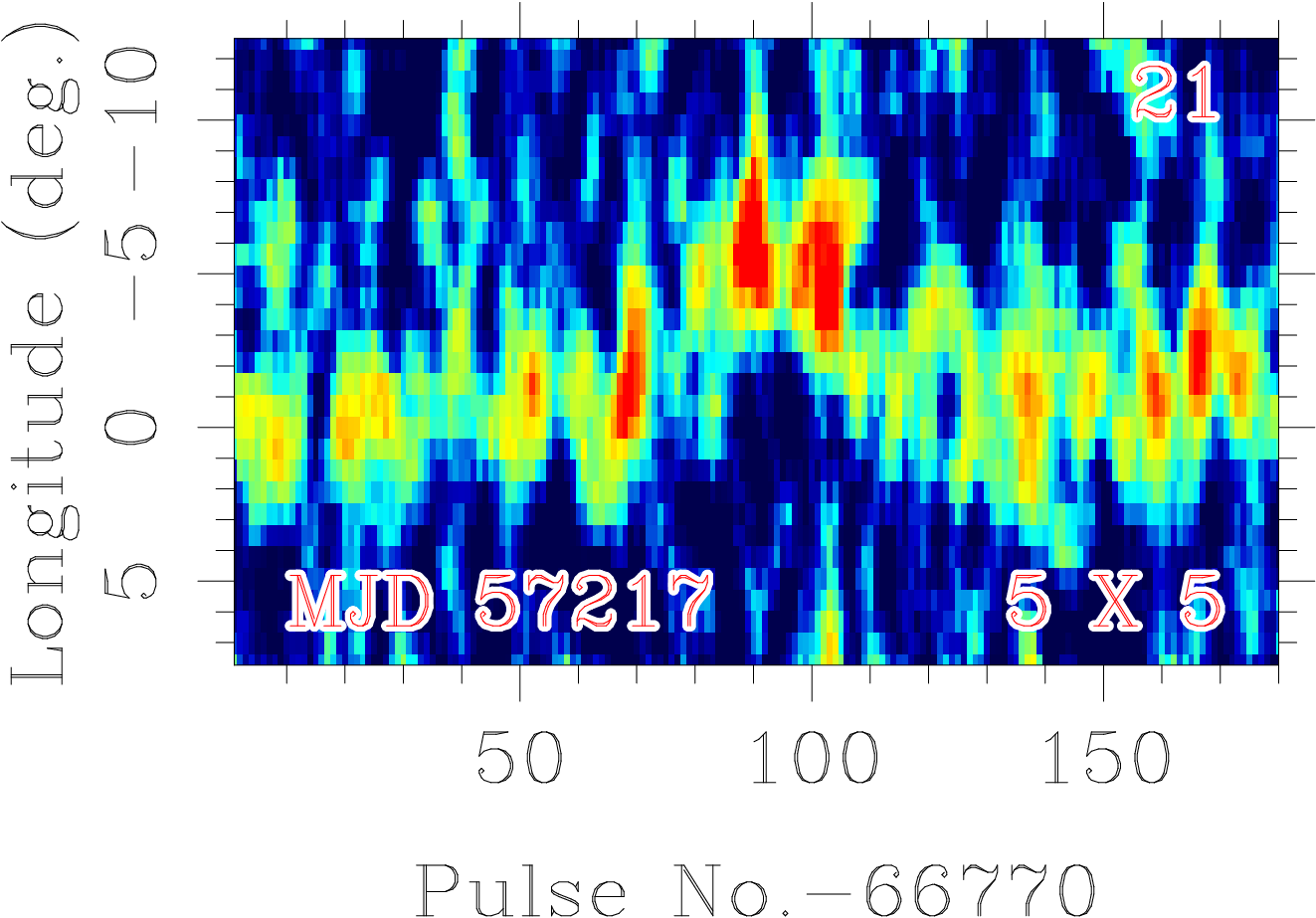}
 \includegraphics[bb = 60 95 505 685,clip,width=0.114\textwidth,angle=-90]{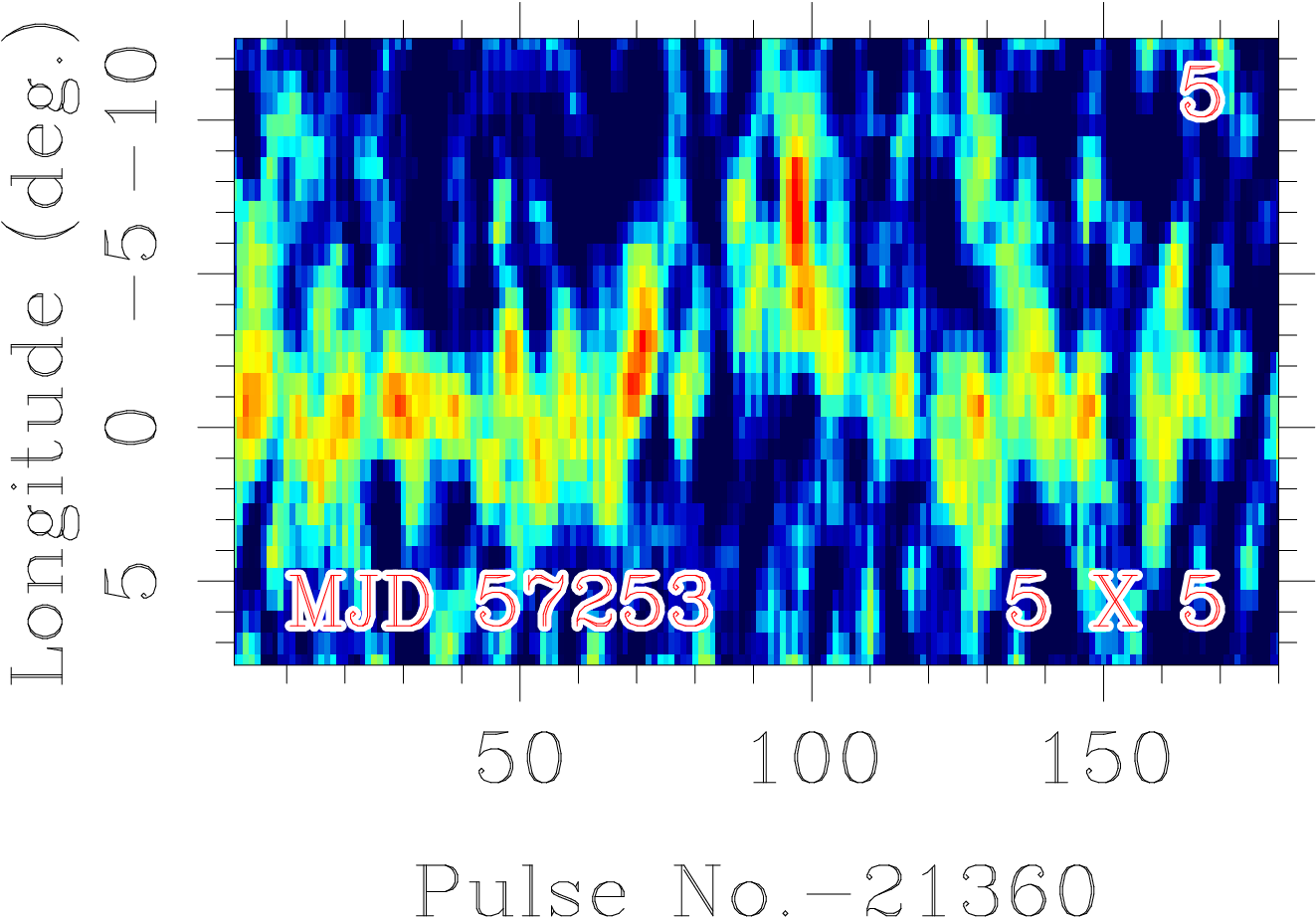}
\caption{The detailed plots for unclassified weak abnormal emission
  events. All ``images'' have been enhanced.}
  \label{fig12}
\end{figure}